\documentclass[12pt, draftclsnofoot, onecolumn]{IEEEtran}

\usepackage{cite}
\usepackage{comment}
\usepackage{amsfonts}
\usepackage{amssymb}
\usepackage{amsmath}
\usepackage{xcolor}
\DeclareMathOperator*{\minimize}{minimize}

\DeclareMathOperator*{\argmin}{arg\,min}
\newtheorem{prop}{Proposition}

\newtheorem{lemma}{Lemma}
\newtheorem{corollary}{Corollary}

\usepackage{graphicx}
\usepackage{bigdelim}
\usepackage{caption,setspace}
\usepackage[ruled,vlined]{algorithm2e}
\usepackage{booktabs}
\usepackage{subfig}

\ifCLASSINFOpdf
\else
\fi
\begin{document}
%
\title{Energy-Efficient Node Deployment in Static and Mobile Heterogeneous Multi-Hop Wireless Sensor Networks}
%
%
%

\author{Saeed Karimi-Bidhendi,
        Jun Guo,
        and Hamid Jafarkhani
\thanks{Authors are with the Center for Pervasive Communications \& Computing, University of California, Irvine, Irvine CA, 92697 USA (e-mail: \{skarimib, guoj4, hamidj\}@uci.edu). This work was presented in part in 2020 IEEE International Conference on Communications \cite{guo2020energy}. This work has been submitted to the IEEE for possible publication.  Copyright may be transferred without notice, after which this version may no longer be accessible. This work was supported in part by the NSF Award CCF-1815339.}
}

%
%

\markboth{}%
{Shell \MakeLowercase{\textit{et al.}}: Bare Demo of IEEEtran.cls for IEEE Journals}
%



\maketitle

\begin{abstract}

We study a heterogeneous wireless sensor network (WSN) where $N$ heterogeneous access points (APs) gather data from densely deployed sensors and transmit their sensed information to $M$ heterogeneous fusion centers (FCs) via multi-hop wireless communication. This heterogeneous node deployment problem is modeled as an optimization problem with total wireless communication power consumption of the network as its objective function. We consider both static WSNs, where nodes retain their deployed position, and mobile WSNs where nodes can move from their initial deployment to their optimal locations. Based on the derived necessary conditions for the optimal node deployment in static WSNs, we propose an iterative algorithm to deploy nodes. In addition, we study the necessary conditions of the optimal movement-efficient node deployment in mobile WSNs with constrained movement energy, and present iterative algorithms to find such deployments, accordingly. Simulation results show that our proposed node deployment algorithms outperform the existing methods in the literature, and achieves a lower total wireless communication power in both static and mobile WSNs, on average.

\end{abstract}


\IEEEpeerreviewmaketitle

\section{Introduction}\label{introduction}

Wireless sensor networks (WSNs) consist of small and low-cost sensor devices used to monitor the environment and transfer the sensed information through wireless channels to dedicated fusion centers. WSNs can be classified into either homogeneous WSNs \cite{guo2020energy, guo2018source, cortes2005spatially, chunawale2014minimization, benaddy2017mutlipath}, in which sensors share the same characteristics such as storage, antennas, sensitivity etc., or heterogeneous WSNs where sensor nodes have different characteristics \cite{karimi2019using, hou2005energy, karimi2020energy, noori2012design, guo2019quantizers, guo2016sensor}. Based on the network architecture, WSNs can be divided into either hierarchical WSNs, where sensors are often grouped into clusters with some of them chosen to be cluster heads, or non-hierarchical WSNs where sensors have identical functionality and multi-hop wireless communications is used to maintain the connectivity of the network. Sensor nodes can also be classified as either static \cite{karimi2020energy, guo2016sensor, zafar2009improved, shams2008fast}, in which each node remains at its deployed position, or mobile where nodes can move to their optimal locations to improve the energy efficiency and sensing quality of the WSNs \cite{guo2019movement, song2013distributed, zou2004sensor, yong2009sensor, senouci2013localized, chellappan2007deploying}.

Energy efficiency is a key determinant in longevity of the WSNs since sensors have limited energy resources and it is difficult or infeasible to recharge the batteries of densely deployed sensors. In general, many factors contribute to the energy consumption of the WSNs, e.g. communication energy, movement energy, sensing energy and computation energy \cite{yousefi2004power, heinzelman2000application}. Empirical measurements in many applications have shown that the data processing and computation energy is negligible compared to communication energy \cite{anastasi2009energy, razzaque2014energy}. Moreover, the sensing energy for passive sensors, such as light or thermal sensors, is considerably small. Therefore, wireless communication dominates the energy consumption in static sensors in practice while movement energy dominates the energy consumption in mobile wireless sensor networks \cite{karimi2020energy, dantu2005robomote, wu2007optimal}. According to the study in \cite{wang2005optimizing}, for the optimal angular velocity and acceleration, the movement energy consumption is approximately linear to the distance that the sensor has to travel, and this linear model is a widely adopted assumption in the literature \cite{liao2012minimizing, liao2014minimizing, wichmann2015analysis, chen2016ptas}.

Several methods have been proposed in the literature to reduce the energy consumption of wireless communication in WSNs. Topology control has been adopted in \cite{li2008survey, xinlian2014sensor} to circumvent excess energy consumption by appropriately switching sensors between awake and asleep states. Energy efficient routing protocols have been established in \cite{benaddy2017mutlipath, hao2013energy} to find optimal paths to transfer data from sensors to fusion centers. Power control protocols reduce the energy consumption of WSNs by calibrating the transmission power of sensors while a reliable communication is maintained \cite{kawadia2003power, narayanaswamy2002power}. Clustering methods \cite{kawadia2003power, younis2004distributed} iterate among cluster heads to balance the energy consumption among the sensor nodes. The common assumption of these approaches is that the node deployment is assumed to be known and fixed while a proper deployment can significantly affect the energy consumption of the WSNs. Furthermore, the above MAC protocols require a large number of message exchanges since the geometry and energy of the network is needed for the operation \cite{younis2004distributed, koyuncu2018asynchronous}. Inspired by dynamics of swarm behavior, a population based iterative algorithm called particle swarm optimization (PSO) is proposed in \cite{dandekar2013energy} to find optimal node locations. An iterative algorithm is proposed in \cite{vincze2007deploying} to  determine the position of nodes such that the average sensors' distance to the nearest fusion center is minimized. Given a maximum length on the cluster diameter, a cluster formation algorithm is proposed in \cite{chatterjee2015multiple} to enhance the network lifetime and reduce the average number of hops for data packets to reach fusion centers. An efficient routing scheme is proposed in \cite{jain2015lifetime} to minimize the maximum energy consumed by each fusion center. The optimal node deployment in two-tier WSNs has been studied for heterogeneous networks in \cite{karimi2020energy}; however, the WSN is restricted to a two-tier architecture while a multi-hop model can provide more degrees of freedom to optimize the data routing. The optimal deployment and trajectory of UAVs with a fixed altitude is discussed in \cite{koyuncu2018deployment} to maintain a reliable communication with ground terminal stations.

Many methods have been developed for mobile WSNs, where movement energy dominates the energy consumption of nodes, to find the optimal deployment given a constraint on available movement energy. The Lloyd$-\alpha$ and DEED algorithms proposed in \cite{song2013distributed} use a movement-dependent penalty term to implement centroidal Voronoi tessellation for sensor deployment. In Lloyd$-\alpha$ algorithm, each movement iteration is scaled by a factor of $\alpha \in [0, 1]$ to compensate for limited movement energy resources. In DEED, the gradient and Hessian matrix of the objective function are used to optimize the sensor movements. Several virtual force based algorithms are proposed in \cite{zou2004sensor, yong2009sensor, senouci2013localized} to determine virtual motion trajectories and the rate of sensor movements using a combination of attractive and repulsive forces. A minimum cost maximum weighted flow based algorithm is developed to determine the optimal movement plan of sensors and encourage a minimum number of sensor nodes in each region of a square field \cite{chellappan2007deploying}; however, the proposed algorithm regards sensor movements as hops between neighboring grid points and lacks a rigorous formulation of movement distance. Similarly, other methods also lack a theoretical framework for a movement efficient sensor deployment that prolongs the network lifetime, e.g., the scaling parameter $\alpha \in [0, 1]$ in the Lloyd$-\alpha$ algorithm has to be specified empirically to meet the movement energy constraints given an initial node deployment. Furthermore, many of the existing work explore the one-tier network architecture while a two-tier or multi-hop protocol provides more flexibility on how sensory data from the physical environment is transferred to the virtual information world.

In this paper, we study the node deployment in heterogeneous multi-hop WSNs consisting of homogeneous densely deployed sensors, heterogeneous APs and heterogeneous FCs, to minimize the total wireless communication power consumption with and without movement energy constraints. The energy efficient node deployment is studied in \cite{karimi2020energy} for heterogeneous WSNs; however, the network is restricted to a two-tiered architecture. In \cite{guo2020energy}, we studied the necessary conditions for an optimal node deployment in homogeneous multi-hop WSNs; however, the homogeneous setting in \cite{guo2020energy} does not address many challenges that is inherent in heterogeneous WSNs, e.g., non-convexity or discontinuity of cells in the optimal partitioning of the sensing environment. To the best of our knowledge, the energy efficient node deployment in heterogeneous multi-hop WSNs is still an open problem. By deriving the necessary conditions of the optimal deployments that minimizes the total wireless communication power consumption of such heterogeneous multi-hop WSNs, we design iterative algorithms to deploy nodes. In addition, we study the optimal node deployment in such networks with limited movement energy for mobile nodes.

The rest of this paper is organized as follows: In Section \ref{system-model}, we provide the system model and problem formulation. In Section \ref{optimal_deployment-without-movement-energy-constraint}, we study the optimal node deployment in static heterogeneous multi-hop WSNs, and propose an iterative algorithm based on the derived necessary conditions. The analysis of optimal node deployment with network's total movement energy constraint is provided in Section \ref{Total-Energy-Constraint}. In Section \ref{Network-Lifetime-Constraint}, we study an energy efficient node deployment that guarantees a given network's lifetime in mobile WSNs. Experimental results are provided in Section \ref{Experiments} and Section \ref{Conclusion} concludes the paper.

\section{System Model and Problem Formulation}\label{system-model}

\begin{figure}[!htb]
\setlength\abovecaptionskip{0pt}
\setlength\belowcaptionskip{0pt}
\centering
\includegraphics[width=4.0in]{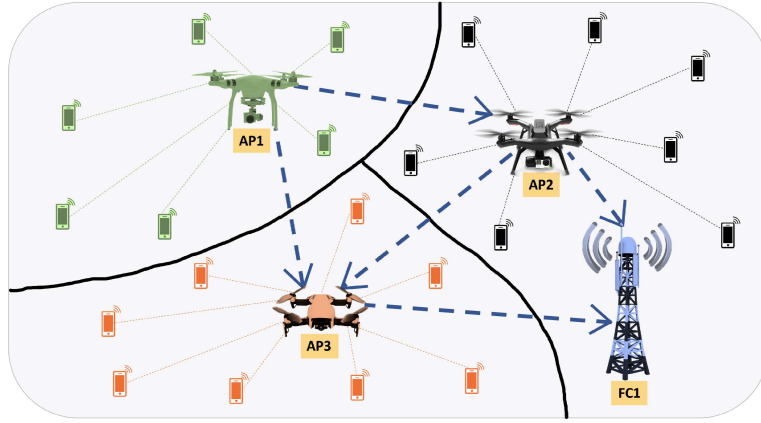}
\captionsetup{justification=justified}
\vspace{5mm}
\caption{System model.}
\label{Example1}
\end{figure}

In this section, we study the system model of heterogeneous multi-hop WSNs, as shown in Fig. \ref{Example1}, consisting of three types of nodes: homogeneous sensors, heterogeneous APs and heterogeneous FCs. Given the target region $\Omega\in\mathbb{R}^2$ which is a convex polygon including its interior, $N$ APs and $M$ FCs are deployed to collect information from densely deployed sensors. Let $\mathcal{I_A}=\{1, \cdots, N\}$ and $\mathcal{I_F}=\{N+1, \cdots, N+M\}$ denote the set of node indices for APs and FCs, respectively. If $n\in \mathcal{I_A}$, Node $n$ refers to AP $n$; however, when $n\in\mathcal{I_F}$, Node $n$ refers to FC $(n-N)$. 
The location of Node $n$ is denoted by $p_n\subset \Omega$ and collectively the node deployment is denoted by $\mathbf{P} = \left(p_1, \cdots, p_N, p_{N+1}, \cdots, p_{N+M} \right)$.
Throughout this paper, we assume that each sensor only sends data to one AP; therefore, for each $n\in \mathcal{I_A}$, AP $n$ gathers data from sensors within the region $W_n\subseteq \Omega$, and $\mathbf{W}=\left(W_1, \cdots, W_N \right)$ provides a set partitioning of the target region. The density of sensors is denoted via a continuous and differentiable function $f:\Omega\longrightarrow \mathbb{R}^+$. The total amount of data collected from sensors within the region $W_n$ in one time unit is $R_b\int_{W_n}f(\omega)d\omega$, where the bit-rate $R_b$ is a constant due to the homogeneity of sensors \cite{guo2018source}. For each $n\in\mathcal{I_A}$, the volume and centroid of the region $W_n$ is defined as $v(W_n)\triangleq\int_{W_n}f(\omega)d\omega$ and $c(W_n)\triangleq\frac{\int_{W_n}\omega f(\omega)d\omega}{\int_{W_n}f(\omega)d\omega}$, respectively. The data gathered from each sensor is forwarded to other nodes in the network until it eventually reaches to one or more FCs.

As shown in Fig. \ref{Example1}, the network can be regarded as a directed acyclic graph $\mathcal{G}(\mathcal{I}_{A}\bigcup \mathcal{I}_{F},\mathcal{E})$ where APs and FCs are source and sink nodes, respectively, and $\mathcal{E}$ is the set of directed edges $(i, j)$ such that $i\in \mathcal{I_A}$ and $j\in \mathcal{I_A}\bigcup \mathcal{I_F}$ \cite{chang2000energy}. Note that any cycle in the network's graph can be removed by reducing the flow of data along the cycle without changing the in-flow and out-flow links to that cycle. Let $\mathbf{F}=[F_{i,j}]_{N\times (N+M)}$ be the flow matrix, where $F_{i,j}$ is the amount of data transmitted through the link $(i,j)$ in one time unit. Since the in-flow to each AP, say $i$, should be equal to the out-flow, we have $\sum_{j=1}^{N}F_{j,i} + R_b\int_{W_i}f(\omega)d\omega = \sum_{j=1}^{N+M}F_{i,j}$. For $i\in\mathcal{I_A}$, we define $F_i \triangleq \sum_{j=1}^{N+M}F_{i,j}$ to be the total flow originated from AP $i$. Let $\mathbf{S}=[s_{i,j}]_{N\times (N+M)}$ be the normalized flow matrix, where $s_{i,j}\triangleq\frac{F_{i,j}}{\sum_{j=1}^{N+M}F_{i,j}}$ is the ratio of the in-flow data to AP $i$ that is transmitted to node $j$. The normalized flow matrix $\mathbf{S}$ satisfies the following properties: (a) $s_{i,j}\in[0, 1]$;\footnote{For time-invariant routing algorithms, such as Bellman-Ford Algorithm \cite{ford1956network, bellman1958routing}, the flows construct a tree-structured graph in which each node has only one successor. Under such circumstances, the normalized flow from Node $i$ to Node $j$ is either $0$ or $1$, i.e., $s_{i,j}\in\{0,1\}$. However, the time-variant routing algorithms, such as Flow Augmentation Algorithm \cite{chang2000energy},  generate different flows during different time periods. As a result, the overall normalized flow from Node $i$ to Node $j$ can be a real number between $0$ and $1$, i.e., $s_{i,j}\in[0,1]$.} (b) $\sum_{j=1}^{N+M}s_{i,j}=1$, $\forall i\in\{1, \cdots, N\}$; (c) No cycle: if there exists a path in the network's graph such as $l_0 \rightarrow l_1 \rightarrow \cdots \rightarrow l_K$, i.e., $\prod_{k=1}^{K}s_{l_{k-1},l_k}>0$, then we have $s_{l_K,l_0}=0$. In particular, we have $s_{i.i}=0$, $\forall i\in \{1, \cdots, N\}$. Since the flow matrix $\mathbf{F}$ can be uniquely determined by the set partitioning $\mathbf{W}$ and the normalized flow matrix $\mathbf{S}$, in the remaining of this paper, we use the notation $\mathbf{F}\left(\mathbf{W}, \mathbf{S}\right)$ instead of $\mathbf{F}$. The following example describes how to calculate $\mathbf{F}\left(\mathbf{W}, \mathbf{S}\right)$ in terms of $\mathbf{W}$ and $\mathbf{S}$.

\emph{Example 1.} We consider a heterogeneous multi-hop WSN with three APs and one FC, i.e. $N=3$ and $M=1$, and the bit-rate $R_b=20$. For a cell partitioning $\mathbf{W}$ with cell volumes $v(W_1)=v(W_2)=0.3$, $v(W_3)=0.4$, and the normalized flow matrix $S=\left[s_{i,j} \right]_{N\times(N+M)}$ with non-zero entries $s_{1,2} = 0.4$, $s_{1,3} = 0.6$, $s_{2,3} = 0.25$, $s_{2,4} = 0.75$ and $s_{3,4} = 1$,
the corresponding flow network is illustrated in Fig. \ref{Example1}. The amount of data generated from sensors within each cell can be calculated as: $\Gamma(W_1)\!=\!R_b v(W_1)\!=\!6$, $\Gamma(W_2)\!=\!R_b v(W_2)\!=\!6$, and $\Gamma(W_3)\!=\!R_b v(W_3)\!=\!8$. AP $1$ does not receive data from any other AP, and only transmits its collected sensed data; thus, $F_1(\mathbf{W},\mathbf{S})\!=\!\Gamma(W_1)\!=\!6$. The flows from AP $1$ are then $F_{1,2}(\mathbf{W},\mathbf{S})\!=\!s_{1,2}\!\times\!F_{1}(\mathbf{W},\mathbf{S})\!=\!2.4$ and $F_{1,3}(\mathbf{W},\mathbf{S})\!=\!s_{1,3}\!\times\!F_{1}(\mathbf{W},\mathbf{S})\!=\!3.6$, respectively. AP $2$'s flows come from $F_{1,2}(\mathbf{W},\mathbf{S})$ and the data gathered from the region $W_2$. Hence, $F_2(\mathbf{W},\mathbf{S})\!=\!\Gamma(W_2)\!+\!F_{1,2}(\mathbf{W},\mathbf{S})\!=\!8.4$. Therefore, the flows from AP $2$ are $F_{2,3}(\mathbf{W},\mathbf{S})\!=\!s_{2,3}\!\times\!F_{2}(\mathbf{W},\mathbf{S})\!=\!2.1$ and $F_{2,4}(\mathbf{W},\mathbf{S})\!=\!s_{2,4}\!\times\!F_{2}(\mathbf{W},\mathbf{S})\!=\!6.3$. Similarly, for AP $3$, we have $F_3(\mathbf{W},\mathbf{S})\!=\!\Gamma(W_3)\!+\!F_{1,3}(\mathbf{W},\mathbf{S})\!+\!F_{2,3}(\mathbf{W},\mathbf{S})\!=\!13.7$; hence, the unique flow from AP $3$ is $F_{3,4}(\mathbf{W},\mathbf{S})\!=\!s_{3,4}\times F_3(\mathbf{W},\mathbf{S})\!=\!13.7$.

In what follows, we formulate the wireless communication power consumption of the network. Also, we focus on the power consumption of sensors and APs, since FCs are usually supplied with reliable energy sources and their power consumption is not the main concern. First, we focus on the sensor's power consumption. According to \cite{guo2018source}, due to the path-loss, the instant transmission power is equal to the square of the distance between the two nodes multiplied by a constant that depends on the characteristics of both nodes, i.e., $\eta\times \|p_n - \omega\|^2$ for a sensor positioned at $\omega$ that transmits its data to AP $n$, $n\in\mathcal{I_A}$. As shown in \cite{heinzelman2000application}, the parameter $\eta$ is given by $\eta = \frac{P_{th} \left(4 \pi \right)^2}{R_b G_t G_r \lambda_c^2}$, where $P_{th}$ is the minimum receiver power threshold for successful reception, $R_b$ is the bit-rate, $G_t$ and $G_r$ are the antenna gains of the transmitter and receiver, respectively, and $\lambda_c$ is the carrier signal wavelength. In the homogeneous setting, all nodes have the same characteristics; thus, the parameter $\eta$ is the same and will not affect the optimization. However, in a heterogeneous multi-hop WSN, AP nodes can have different antenna gains and SNR thresholds; hence, the parameter $\eta$ will be a function of the node index. Therefore, the sensors' transmission power consumption can be written as
\begin{equation}\label{sensor-power}
    \overline{\mathcal{P}}^T_{\mathcal{S}}\left(\mathbf{P}, \mathbf{W} \right) = \sum_{n=1}^{N} \int_{W_n} \eta_n \|p_n - \omega \|^2 R_b f(\omega)d\omega.
\end{equation}
Similarly, the instant transmission power from Node $i$ to Node $j$ can be written as $\beta \times \|p_i - p_j \|^2$ where the parameter $\beta$ depends on the antenna gain and SNR threshold of Node $j$ and the antenna gain of Node $i$ \cite{heinzelman2000application}. Therefore, it is the same for the homogeneous setting and will not affect the optimization. However, in a heterogeneous multi-hop WSN, the heterogeneity of the nodes causes the parameter $\beta$ to be a function of the node indices. Hence, the average transmission power through link $(i,j)$ is equal to $\beta_{i,j} \|p_i - p_j \|^2 F_{i,j}(\mathbf{W}, \mathbf{S})$, and the APs' total transmission power consumption can be written as
\begin{equation}\label{access-point-power}
    \overline{\mathcal{P}}^T_{\mathcal{A}}\left(\mathbf{P}, \mathbf{W}, \mathbf{S} \right) = \sum_{i=1}^{N}\sum_{j=1}^{N+M} \beta_{i,j} \|p_i - p_j\|^2 F_{i,j}\left(\mathbf{W}, \mathbf{S} \right).
\end{equation}
According to \cite{hou2005energy}, power at the receiver of AP $n$ can be modeled as $\sum_{i=1}^{N}\rho_n F_{i,n}(\mathbf{W}, \mathbf{S}) + \rho_n R_b \int_{W_n}f(\omega)d\omega$, where $\rho_n$ is the power consumption coefficient for receiving data at AP $n$, and depends on digital coding, modulation and filtering of the signal before transmission \cite{heinzelman2000application}. Therefore, the APs' total receiver power consumption can be written as:
\begin{equation}\label{ap-receiver-power}
    \overline{\mathcal{P}}^R_{\mathcal{A}}\left(\mathbf{W}, \mathbf{S} \right) = \sum_{n=1}^{N}\rho_n \left[ \sum_{i=1}^{N} F_{i,n}\left(\mathbf{W}, \mathbf{S} \right) + R_b\int_{W_n}f(\omega)d\omega \right].
\end{equation}
Thus, the total communication power consumption of the multi-hop WSN can be written as:
\begin{align}\label{total-power-consumption}
\mathcal{D}\left(\mathbf{P}, \mathbf{W}, \mathbf{S} \right) = \overline{\mathcal{P}}^T_{\mathcal{S}}\left(\mathbf{P}, \mathbf{W} \right) + \lambda \left[\overline{\mathcal{P}}^T_{\mathcal{A}}\left(\mathbf{P}, \mathbf{W}, \mathbf{S} \right) + \overline{\mathcal{P}}^R_{\mathcal{A}}\left(\mathbf{W}, \mathbf{S} \right) \right],
\end{align}
where the Lagrangian multiplier $\lambda\geq 0$ provides a trade-off between the sensor and AP power consumption. Our main objective in this paper is to minimize the multi-hop power consumption defined in (\ref{total-power-consumption}) over the node deployment $\mathbf{P}$, cell partitioning $\mathbf{W}$, and the normalized flow matrix $\mathbf{S}$ in both static and mobile WSNs with constrained movement energy.

\section{Optimal Node Deployment in Static Heterogeneous Multi-Hop WSNs}\label{optimal_deployment-without-movement-energy-constraint}

As shown in (\ref{total-power-consumption}), the total power consumption depends on three variables $\mathbf{P}$, $\mathbf{W}$ and $\mathbf{S}$. Thus, our goal is to find the optimal AP and FC deployments, cell partitioning and normalized flow matrix, denoted by $\mathbf{P}^* = \left(p_1^*, \cdots, p_N^*, p_{N+1}^*, \cdots, p_{N+M}^* \right)$, $\mathbf{W}^* = \left(W_1^*, \cdots, W_N^* \right)$ and $\mathbf{S}^* = \left[s^*_{i,j}\right]_{N\times(N+M)}$, respectively, that minimizes the multi-hop power consumption. Note that not only the variables $\mathbf{P}$, $\mathbf{W}$ and $\mathbf{S}$ are interdependent, i.e., the optimal value for each of them depends on the value of the other two variables, but also this optimization problem is NP-hard. Our aim is to design an iterative algorithm that optimizes the value of one variable while the other two variables are held fixed. For this purpose, first we introduce a few concepts, and then we derive the necessary conditions for optimal deployment at each step.

Without loss of generality, we assume that AP $n$'s gathered data goes through $K_n$ paths in the network's graph before it reaches to one or more fusion centers. We denote these paths by $\left\{L_k^{(n)}\left(\mathbf{S}\right) \right\}_{k\in\{1, \cdots, K_n\}}$, where $L_k^{(n)}\left(\mathbf{S}\right) = l_{k,0}^{(n)}\rightarrow l_{k,1}^{(n)}\rightarrow\cdots\rightarrow l_{k,J_k^{(n)}}^{(n)}$, $l_{k,0}^{(n)} = n$, $l_{k,i}^{(n)}\in\mathcal{I_A}$ for $i\in\{0, \cdots, J_k^{(n)}-1\}$, $l_{k,J_k^{(n)}}^{(n)} \in \mathcal{I_F}$ and $J_k^{(n)}$ is the number of nodes on the $k$-th path excluding Node $n$. The portion of the total flow originated from AP $n$ that goes through the $k$-th path can then be calculated as
\begin{equation}\label{data-rate-kth-path}
    \mu_k^{(n)}\left(\mathbf{W}, \mathbf{S} \right) = F_n\left(\mathbf{W}, \mathbf{S} \right)\prod_{i=1}^{J_k^{(n)}}s_{l_{k,i-1}^{(n)}, l_{k,i}^{(n)}}.
\end{equation}
In particular, we have $\sum_{k=1}^{K_n} \mu_k^{(n)}\left(\mathbf{W}, \mathbf{S} \right) = F_n\left(\mathbf{W}, \mathbf{S} \right)$ that indicates the data from AP $n$ eventually reaches to one or more FCs. Next, for each link $(i,j)$ in the network's graph, we define the energy cost (Watt/bit) to be:
\begin{equation}\label{energy-cost}
e_{i,j}\left(\mathbf{P} \right)\triangleq 
    \begin{cases}
      \beta_{i,j}\|p_i - p_j\|^2 + \rho_j, & \text{if}\ j\in \mathcal{I_A} \\
      \beta_{i,j}\|p_i - p_j\|^2, & \text{if}\ j\in \mathcal{I_F}.
    \end{cases}
  \end{equation}
Hence, we define the path cost corresponding to the $k$-th path from AP $n$ to FCs as:
\begin{equation}\label{path-cost}
    \overline{e}_k^{(n)}\left(\mathbf{P}, \mathbf{S} \right)=\sum_{i=1}^{J_k^{(n)}}e_{l_{k,i-1}^{(n)}, l_{k,i}^{(n)}}\left(\mathbf{P} \right).
\end{equation}
Now, AP $n$'s power coefficient, denoted by $g_n\left(\mathbf{P}, \mathbf{S} \right)$ is defined to be the power consumption (Joules/bit) for transmitting $1$ bit data from AP $n$ to the FCs, i.e., we have:
\begin{align}\label{power-coefficient}
g_n\left(\mathbf{P}, \mathbf{S} \right) &= \frac{\sum_{k=1}^{K_n}\mu_k^{(n)}\left(\mathbf{W}, \mathbf{S} \right)\overline{e}_k^{(n)}\left(\mathbf{P}, \mathbf{S} \right)}{F_n\left(\mathbf{W}, \mathbf{S} \right)}\\
&=\sum_{k=1}^{K_n}\left[\prod_{i=1}^{J_k^{(n)}}s_{l_{k,i-1}^{(n)}, l_{k,i}^{(n)}}\left(\sum_{j=1}^{J_k^{(n)}}\beta_{l_{k,j-1}^{(n)}, l_{k,j}^{(n)}} \Big|\! \Big|p_{l_{k,j-1}^{(n)}} - p_{l_{k,j}^{(n)}}  \Big| \! \Big|^2 +\sum_{j=1}^{J_k^{(n)}-1}\rho_{l_{k,j}^{(n)}}   \right) \right].
\end{align}
Note that the term $F_n\left(\mathbf{W}, \mathbf{S} \right)$ is canceled in (\ref{power-coefficient}), implying that power coefficient $g_n\left(\mathbf{P}, \mathbf{S} \right)$ is independent of $\mathbf{W}$. Below we provide an example to clarify how to calculate the AP power coefficients.

\emph{Example 2.} Consider the WSN described in Example 1, and let $\mathbf{P}=((0,0), (0,1), (1,0), (1,1))$, $\beta_{i,j}\!=\!1$ and $\rho_i\!=\!1$ for all $i\in\mathcal{I_A}$ and $j\in\mathcal{I_A}\bigcup\mathcal{I_F}$. We aim to find AP $1$'s power coefficient $g_1(\mathbf{P},\mathbf{S})$. The link energy costs for this network can be calculated as $e_{1,2}(\mathbf{P})=e_{1,3}(\mathbf{P})\!=\!2$, $e_{2,3}(\mathbf{P})\!=\!3$, and $e_{2,4}(\mathbf{P})\!=\!e_{3,4}(\mathbf{P})\!=\!1$. Note that AP $1$'s data goes through the following $3$ paths: $L^{(1)}_{1}(\mathbf{S})\!=\!1\!\to\!2\!\to\!4$, $L^{(1)}_{2}(\mathbf{S})\!=\!1\!\to\!3\!\to\!4$, and 
$L^{(1)}_{3}(\mathbf{S})\!=\!1\!\to\!2\!\to\!3\!\to\!4$. The data rate through the above paths are, respectively,
$\mu_1^{(1)}(\mathbf{W},\mathbf{S})\!=\!F_1(\mathbf{W},\mathbf{S})\!\times\!s_{1,2}\!\times\!s_{2,4}\!=\!0.3F_1(\mathbf{W},\mathbf{S})$,
$\mu_2^{(1)}(\mathbf{W},\mathbf{S})\!=\!F_1(\mathbf{W},\mathbf{S})\!\times\!s_{1,3}\!\times\!s_{3,4}\!=\!0.6F_1(\mathbf{W},\mathbf{S})$, and
$\mu_3^{(1)}(\mathbf{W},\mathbf{S})\!=\!F_1(\mathbf{W},\mathbf{S})\!\times\!s_{1,2}\!\times\!s_{2,3}\!\times\!s_{3,4}\!=\!0.1F_1(\mathbf{W},\mathbf{S})$.
Moreover, we can calculate the path costs using  (\ref{path-cost}) as follows:
$\overline{e}^{(1)}_{1}(\mathbf{P})=e_{1,2}(\mathbf{P})+e_{2,4}(\mathbf{P})=3$, $\overline{e}^{(1)}_{2}(\mathbf{P})=e_{1,3}(\mathbf{P})+e_{3,4}(\mathbf{P})=3$, and 
$\overline{e}^{(1)}_{3}(\mathbf{P})=e_{1,2}(\mathbf{P})+e_{2,3}(\mathbf{P})+e_{3,4}(\mathbf{P})=6$.
Then, AP $1$'s power coefficient is $g_1(\mathbf{P},\mathbf{S})=0.3\times3+0.6\times3+0.1\times6=3.3$.

To derive the necessary condition for an optimal cell partitioning, first, we need to rewrite the objective function in (\ref{total-power-consumption}).

\begin{lemma}\label{power-coef-equality}
For the AP power coefficient defined in (\ref{power-coefficient}), we have:
\begin{equation}\label{power-coef-equality-eq}
    \sum_{n=1}^{N}g_n\left(\mathbf{P}, \mathbf{S} \right)R_b \int_{W_n}f(\omega)d\omega = \sum_{i=1}^{N}\left[\sum_{j=1}^{N+M} \beta_{i,j} \|p_i - p_j\|^2 F_{i,j}\left(\mathbf{W}, \mathbf{S} \right) + \sum_{j=1}^{N}\rho_j F_{i,j}\left(\mathbf{W}, \mathbf{S} \right) \right].
\end{equation}
\end{lemma}
The proof is provided in Appendix \ref{proof-power-coef-equality}. Using Lemma \ref{power-coef-equality}, the objective function is:
\begin{equation}\label{rewrite-objective-function}
    \mathcal{D}\left(\mathbf{P}, \mathbf{W}, \mathbf{S} \right) = \sum_{n=1}^{N} \int_{W_n} \left(\eta_n \|p_n - \omega \|^2 R_b + \lambda g_n\left(\mathbf{P}, \mathbf{S} \right)R_b + \lambda \rho_n R_b  \right) f(\omega)d\omega.
\end{equation}
Now, we study the properties of the optimal cell partitioning. For each $n\in\mathcal{I_A}$, the Voronoi cell $\mathcal{V}_n$ for a node deployment $\mathbf{P}$ and normalized flow matrix $\mathbf{S}$ is defined to be:
\begin{align}\label{voronoi-cell}
    \mathcal{V}_n\!\left(\mathbf{P}, \mathbf{S} \right) \!\triangleq\! \big \{\omega\!:\!\eta_n \|p_n \!-\! \omega \|^2 \!+\! \lambda g_n\!\left(\mathbf{P}, \mathbf{S} \right) \!+\! \lambda \rho_n  \leq  \eta_k \|p_k \!-\! \omega \|^2  \!+\! \lambda g_k\!\left(\mathbf{P}, \mathbf{S} \right) \!+\! \lambda \rho_k , \forall k\neq n   \big \}.
\end{align}
Ties are broken in the favor of the smaller index to ensure that each Voronoi cell $\mathcal{V}_n$ is a Borel set. For brevity, we write $\mathcal{V}_n$ instead of $\mathcal{V}_n\left(\mathbf{P}, \mathbf{S} \right)$ when it is clear from the context. The collection
\begin{equation}\label{generalized-voronoi-diagram}
    \mathbf{\mathcal{V}}\left(\mathbf{P}, \mathbf{S} \right) = \left(\mathcal{V}_1, \mathcal{V}_2, \cdots, \mathcal{V}_N \right)
\end{equation}
is referred to as the generalized Voronoi diagram \cite{karimi2020energy}. Note that in contrast to the regular Voronoi diagrams, the Voronoi cells defined in (\ref{voronoi-cell}) can be non-convex, not star-shaped and even disconnected. The following proposition indicates that given a node deployment $\mathbf{P}$ and normalized flow matrix $\mathbf{S}$, the generalized Voronoi diagram provides the optimal cell partitioning. 
\begin{prop}\label{optimal-cell-partitioning}
For any node deployment $\mathbf{P}$, cell partitioning $\mathbf{W}$ and normalized flow matrix $\mathbf{S}$, we have:
\begin{equation}\label{optimal-partitoning-proposition}
    \mathcal{D}\left(\mathbf{P}, \mathbf{W}, \mathbf{S} \right) \geq \mathcal{D}\left(\mathbf{P}, \mathbf{\mathcal{V}}\left(\mathbf{P}, \mathbf{S} \right), \mathbf{S} \right).
\end{equation}
\end{prop}
The proof is provided in Appendix \ref{proof-voronoi-optimality}. Now, given the link costs $\{e_{i,j}\left(\mathbf{P} \right)\}$s and generated sensing data rate from each cell partition, the total multi-hop power consumption can be minimized by Bellman-Ford Algorithm \cite{ford1956network, bellman1958routing}. For convenience, we show the functionality of Bellman-Ford Algorithm by $\mathcal{R}\left(\mathbf{P},\mathbf{W} \right)$, where $\mathbf{P}$ and $\mathbf{W}$ are inputs and $\mathbf{S}$ is the output, i.e., $\mathcal{R}\left(\mathbf{P},\mathbf{W} \right)=\argmin_{\mathbf{S}}\left[\overline{\mathcal{P}}^T_{\mathcal{A}}\left(\mathbf{P},\mathbf{W},\mathbf{S} \right) + \overline{\mathcal{P}}^R_{\mathcal{A}}\left(\mathbf{W},\mathbf{S} \right) \right]$. Since the sensors' power consumption is independent of $\mathbf{S}$, we have:
\begin{equation}\label{bellman-ford-functionality}
    \mathcal{R}\!\left(\mathbf{P},\mathbf{W} \right) = \argmin_{\mathbf{S}} \overline{\mathcal{P}}^T_{\mathcal{S}}\!\left(\mathbf{P}, \mathbf{W} \right) + \lambda \left[\overline{\mathcal{P}}^T_{\mathcal{A}}\!\left(\mathbf{P}, \mathbf{W}, \mathbf{S} \right) + \overline{\mathcal{P}}^R_{\mathcal{A}}\!\left(\mathbf{W}, \mathbf{S} \right) \right] = \argmin_{\mathbf{S}}\mathcal{D}\!\left(\mathbf{P}, \mathbf{W}, \mathbf{S} \right).
\end{equation}
Hence, the optimal flow matrix for a given $\mathbf{P}$ and $\mathbf{W}$ is $\mathbf{F}\left(\mathbf{W}, \mathcal{R}\left(\mathbf{P},\mathbf{W} \right) \right)$.
For notational brevity, we define the point $z_i\left(\mathbf{P}, \mathbf{W}, \mathbf{S} \right)$, or $z_i$ for short, to be:
\begin{align}\label{z1-AP}
    z_i = \frac{\eta_i R_b v_i c_i + \lambda\left(\sum_{j=1}^{N+M}\beta_{i,j}F_{i,j} p_j + \sum_{j=1}^{N}\beta_{j,i}F_{j,i} p_j \right)}{\eta_i R_b v_i + \lambda\left(\sum_{j=1}^{N+M}\beta_{i,j}F_{i,j} + \sum_{j=1}^{N}\beta_{j,i}F_{j,i} \right)}, \qquad \forall i\in \mathcal{I_A} \\
    z_i = \frac{\sum_{j=1}^{N}\beta_{j,i}F_{j,i} p_j}{\sum_{j=1}^{N}\beta_{j,i}F_{j,i}}.\qquad\qquad\qquad\qquad \forall i\in \mathcal{I_F} \label{z1-FC}
\end{align}
The following theorem provides the necessary conditions for the optimal deployment.
\begin{prop}\label{necessary-condition-optimal-deployment}
The necessary conditions for the optimal deployments in heterogeneous multi-hop WSNs with communication power consumption defined in (\ref{total-power-consumption}) are
\begin{align}\label{necessary-condition-optimal-deployment-APs}
    p_i^* &= z_i^*, \qquad\qquad \forall i\in \mathcal{I_A} \bigcup \mathcal{I_F} \\
    \mathbf{W}^* &= \mathcal{V}\left(\mathbf{P}^*, \mathbf{S}^* \right),\label{necessary-condition-optimal-deployment-partitioning} \\
    \mathbf{S}^* &= \mathcal{R}\left(\mathbf{P}^*, \mathbf{W}^* \right),\label{necessary-condition-optimal-deployment-S}
\end{align}
where $z^*_i=z_i\left(\mathbf{P}^*, \mathbf{W}^*, \mathbf{S}^* \right)$ is given by Eqs. (\ref{z1-AP}) and (\ref{z1-FC}).
\end{prop}
The proof of Proposition \ref{necessary-condition-optimal-deployment} is provided in Appendix \ref{proof-necessary-condition-optimal-deployment}.

Note that depending on the cell partitioning and normalized flow matrix, there may not be any flow through some links in the network's graph. Let $\mathcal{N}^P_i(\mathbf{S})\triangleq\{j|F_{j,i}(\mathbf{W},\mathbf{S})>0\}$ be the set of Node $i$'s predecessors, and $\mathcal{N}^S_i(\mathbf{S})\triangleq\{j|F_{i,j}(\mathbf{W},\mathbf{S})>0\}$ be the set of Node $i$'s successors. We can then simplify Eq. (\ref{necessary-condition-optimal-deployment-APs}) as:
\begin{equation}\label{rewrite-AP-necessary-conditions}
    p^*_i = \frac{\eta_i R_b v^*_i c^*_i + \lambda \left( \sum\limits_{j\in\mathcal{N}^S_i(\mathbf{S}^*)}\beta_{i,j}F^*_{i,j}p^*_j + \sum\limits_{j\in\mathcal{N}^P_i(\mathbf{S}^*)}\beta_{j,i}F^*_{j,i}p^*_j \right) }{\eta_i R_b v^*_i+\lambda\left(\sum\limits_{j\in\mathcal{N}^S_i(\mathbf{S}^*)}\beta_{i,j}F_{i,j}^*+\sum\limits_{j\in\mathcal{N}^P_i(\mathbf{S}^*)}\beta_{j,i}F^*_{j,i}\right)}, \qquad \forall i\in\mathcal{I_A}
\end{equation}
\begin{equation}\label{rewrite-FC-necessary-conditions}
    p^*_i=\frac{\sum\limits_{j\in\mathcal{N}^P_i(\mathbf{S}^*)}\beta_{j,i}F^*_{j,i}p^*_j}{\sum\limits_{j\in\mathcal{N}^P_i(\mathbf{S}^*)}\beta_{j,i}F^*_{j,i}}, \qquad \forall i\in\mathcal{I_F}.
\end{equation}
In other words, AP $i$'s optimal location is a linear combination of its geometric centroid, predecessors, and successors while FC $i$'s optimal location is a linear combination of its predecessors. 

In what follows, first, we quickly review the conventional Lloyd Algorithm \cite{lloyd1982least}, then we propose an algorithm to optimize the communication power consumption defined in Eq. (\ref{total-power-consumption}) for heterogeneous multi-hop WSNs. Lloyd Algorithm iterates between two steps: (i) Voronoi partitioning and (ii) Moving each node to the geometric centroid of its corresponding Voronoi region. Although the conventional Lloyd Algorithm can be used for one-tier quantizers or one-tier node deployment tasks \cite{guo2016sensor}, it cannot be applied to WSNs with multi-hop wireless communications. Based on the properties explored in this section, we design a Routing-aware Lloyd (RL) Algorithm, as outlined in Algorithm \ref{RL}, to optimize the node deployment in heterogeneous multi-hop WSNs and minimize the objective function in (\ref{total-power-consumption}). 

\begin{algorithm}[ht!]
\SetAlgoLined
\SetKwRepeat{Do}{do}{while}
\KwResult{Optimal node deployment $\mathbf{P}$, cell partitioning $\mathbf{W}$ and normalized flow matrix $\mathbf{S}$.}
Input: Convergence error threshold $\epsilon\in \mathbb{R}^+$ \;
 \Do{$\frac{\mathcal{D}_{\mathrm{old}} - \mathcal{D}_{\mathrm{new}}}{\mathcal{D}_{\mathrm{old}}} \geq \epsilon$}{ -- Calculate the objective function $\mathcal{D}_{\textrm{old}}=\mathcal{D}\left(\mathbf{P},\mathbf{W},\mathbf{S}\right)$\;
  1. Update the cell partitioning $\mathbf{W}$ according to the Eq.  (\ref{necessary-condition-optimal-deployment-partitioning})\; 
  2. Update the normalized flow matrix $\mathbf{S}$ using to the Bellman-Ford algorithm\;
  
  3. Update the node deployment $\mathbf{P}$ as follows:\
    $$p_n = \frac{\eta_n R_b v_nc_n + \lambda\left(\sum_{j=1}^{N+M}\beta_{n,j}F_{n,j} p_j + \sum_{j=1}^{N}\beta_{j,n}F_{j,n} p_j \right)}{\eta_n R_b v_n + \lambda\left(\sum_{j=1}^{N+M}\beta_{n,j}F_{n,j} + \sum_{j=1}^{N}\beta_{j,n}F_{j,n} \right)}, \qquad \forall n\in \mathcal{I_A}$$\
    $$p_n = \frac{\sum_{j=1}^{N}\beta_{j,n}F_{j,n} p_j}{\sum_{j=1}^{N}\beta_{j,n}F_{j,n}},\qquad \forall n\in \mathcal{I_F}$$\
    
  -- Calculate the objective function $\mathcal{D}_{\textrm{new}}=\mathcal{D}\left(\mathbf{P},\mathbf{W},\mathbf{S}\right)$\; }
 \caption{Routing-aware Lloyd Algorithm}
 \label{RL}
\end{algorithm}

\begin{prop}\label{RL-convergence}
RL Algorithm is an iterative improvement algorithm, i.e., the objective function is non-increasing and the algorithm converges.
\end{prop}
The proof of Proposition \ref{RL-convergence} is provided in Appendix \ref{proof-RL-convergence}.

\section{The Node Deployment with a Total Energy Constraint in Mobile WSNs}\label{Total-Energy-Constraint}

\subsection{Problem formulation}\label{Total-Energy-Constraint-problem-formulation}

In Section \ref{optimal_deployment-without-movement-energy-constraint}, we studied the scenario where nodes are directly placed at the optimal locations calculated via RL Algorithm. However, here we study mobile heterogeneous multi-hop WSNs in which each node moves from its initial position to its optimal location that minimizes the communication power consumption in (\ref{total-power-consumption}) while the total movement energy consumption of the network is constrained. More precisely, given the linear model for movement energy consumption in \cite{wang2005optimizing}, for each $n\in \mathcal{I_A}\bigcup\mathcal{I_F}$, Node $n$'s movement energy can be modeled as:
\begin{equation}\label{movement-energy}
    E_n\left(\mathbf{P}\right) = \zeta_n \|p_n - \tilde{p}_n \|,
\end{equation}
where the moving cost parameter $\zeta_n$ depends on Node $n$'s energy efficiency, $p_n$ and $\tilde{p}_n$ are its destination and initial locations, respectively. Therefore, the total movement energy consumption of the network is
\begin{equation}\label{total-movement-energy}
E\left(\mathbf{P} \right) = \sum_{n=1}^{N+M} E_n\left(\mathbf{P}\right) = \sum_{n=1}^{N+M}\zeta_n \|p_n - \tilde{p}_n \|.
\end{equation}
Our main objective in this section is to minimize the multi-hop communication power consumption in Eq. (\ref{total-power-consumption}) while the total movement energy is limited, i.e., the constrained optimization problem is defined as
\begin{align}\label{total-movement-constraint-objective}
    \minimize_{\mathbf{P}, \mathbf{W}, \mathbf{S}} \mathcal{D}\left(\mathbf{P}, \mathbf{W}, \mathbf{S} \right), \\
    \textrm{s.t.}\qquad E\left(\mathbf{P}\right) \leq \gamma \label{movement-constraint}
\end{align}
where $\gamma \geq 0$ is the maximum movement energy consumption of the network.

\subsection{The Optimal Node Deployment}\label{Optimal-Node-Deployment-Total-Move-Constraint}
Here, we aim to find the optimal node deployment $\mathbf{P}^*$, cell partitioning $\mathbf{W}^*$ and normalized flow matrix $\mathbf{S}^*$ that minimizes the total multi-hop communication power consumption while the movement energy consumption is constrained. Note that the movement energy in (\ref{movement-constraint}) is independent of the cell partitioning and normalized flow matrix; therefore, the generalized Voronoi diagram and Bellman-Ford Algorithm, represented in Eqs. (\ref{generalized-voronoi-diagram}) and (\ref{bellman-ford-functionality}), respectively, still provide the optimal cell partitioning and normalized flow matrix. Now, we discuss the optimal node deployment for the constrained optimization problem in Eqs. (\ref{total-movement-constraint-objective}) and (\ref{movement-constraint}).
\begin{lemma}\label{total-movment-constraint-between-initial-and-optimal}
Let $\mathbf{P}^*$, $\mathbf{W}^*$ and $\mathbf{S}^*$ be the optimal node deployment, cell partitioning and normalized flow matrix for the constrained optimization problem in Eqs. (\ref{total-movement-constraint-objective}) and (\ref{movement-constraint}). We have:
\begin{align}\label{total-movment-constraint-between-initial-and-optimal-eq1}
p^*_i = \delta_i \tilde{p}_i + (1-\delta_i)\times z_i^*, \qquad \forall i\in \mathcal{I_A} \bigcup \mathcal{I_F}
\end{align}
where $\delta_i\in [0, 1]$ and $\tilde{p}_i$ is the initial location of Node $i$.
\end{lemma}
The proof is provided in Appendix \ref{optimal-between-initial-z}. 

Lemma \ref{total-movment-constraint-between-initial-and-optimal} states that the optimal location for Node $i$ is on the line connecting its initial position to the point $z^*_i=z_i\left(\mathbf{P}^*, \mathbf{W}^*, \mathbf{S}^* \right)$. Note that this is in contrast to the optimal node deployment without movement energy constraint in Section \ref{optimal_deployment-without-movement-energy-constraint}, i.e., $p_i^* = z_i^*$,  as shown in Proposition \ref{necessary-condition-optimal-deployment}. The difference is because of the constraint in Eq. (\ref{movement-constraint}). Intuitively, for $\gamma = 0$ we have $\delta_i=1$ for all $i\in\mathcal{I_A}\bigcup\mathcal{I_F}$, i.e., each node will remain at its initial position since there is zero total available movement energy. However, for sufficiently large enough $\gamma$, we have $\delta_i=0$, i.e., $p_i^*=z_i^*$ for all $i\in\mathcal{I_A}\bigcup\mathcal{I_F}$. In general, nodes can be classified into two groups based on whether they have positive moving distance or they stand still. Let $\mathcal{I}_d=\left\{n \mid \|p_n-\tilde{p}_n\| > 0, \forall n\in \mathcal{I_A}\bigcup\mathcal{I_F} \right\}$ and $\mathcal{I}_s=\left\{n \mid \|p_n-\tilde{p}_n\| = 0, \forall n\in \mathcal{I_A}\bigcup\mathcal{I_F} \right\}$ be the set of dynamic and static nodes, respectively. The following theorem provides the necessary condition for the optimal node deployment in multi-hop WSNs with total movement energy constraint:
\begin{prop}\label{necessary-condition-total-move-constraint}
Let $\mathbf{P}^*, \mathbf{W}^*$ and $\mathbf{S}^*$ be the optimal node deployment, cell partitioning and normalized flow matrix for the constrained optimization problem in Eqs. (\ref{total-movement-constraint-objective}) and (\ref{movement-constraint}). Then:
\begin{align}\label{necessary-condition-total-move-constraint-eq}
    \chi^*_n &= \chi^*_m\geq \chi^*_k, \qquad\qquad\qquad\qquad\qquad\qquad\qquad\qquad \forall n,m\in\mathcal{I}_d , k\in\mathcal{I}_s \\
    p_n^* &= \tilde{p}_n + \Gamma_n^*\times\left[1-\frac{\max\left(0, \sum_{i\in\mathcal{I}_d}\zeta_i \|\Gamma_i^*\| - \gamma \right)}{\|\Gamma_n^*\| \times \frac{\psi_n^*}{\zeta_n}\times\sum_{i\in\mathcal{I}_d}\frac{\zeta_i^2}{\psi_i^*}} \right], \qquad \forall n\in\mathcal{I}_d \label{necessary-condition-total-move-constraint-eq2}
\end{align}
where $\Gamma_n^* = z_n^* - \tilde{p}_n$ and $\psi^*_n$ is defined to be
\begin{equation}\label{psi}
    \psi_n^*\triangleq 
    \begin{cases}
      \eta_n R_b v^*_n + \lambda\left[\sum_{k=1}^{N+M} \beta_{n,k}F_{n,k}^* + \sum_{k=1}^N \beta_{k,n}F_{k,n}^* \right], & \text{if}\ n\in \mathcal{I_A} \\
      \lambda\sum_{k=1}^N \beta_{k,n}F_{k,n}^*, & \text{if}\ n\in \mathcal{I_F}
    \end{cases}
\end{equation}
and the moving efficiency $\chi^*_n$ is defined as
\begin{equation}\label{chi}
    \chi^*_n = \frac{\psi^*_n \|p^*_n - z^*_n\|^2}{\zeta_n\|p^*_n - z^*_n\|}= \frac{\psi^*_n}{\zeta_n}\|p^*_n - z^*_n\|, \qquad\qquad\qquad \forall n\in \mathcal{I_A}\bigcup\mathcal{I_F}
\end{equation}
to reflect Node $n$'s ability to reduce the communication power consumption by movement.
\end{prop}
The proof is provided in Appendix \ref{proof-necessary-condition-total-constraint}. Proposition \ref{necessary-condition-total-move-constraint} captures the intuition in Lemma \ref{total-movment-constraint-between-initial-and-optimal} that in an optimal deployment, Node $n$ is located on the line connecting its initial position $\tilde{p}_n$ to the point $z^*_n$, for all $n\in\mathcal{I_A}\bigcup\mathcal{I_F}$. Furthermore, for a sufficiently large enough available movement energy $\gamma$, say $\gamma\geq \sum_{i\in\mathcal{I}_d}\zeta_i \|\Gamma_i^*\|$, we have $p^*_n=z^*_n$ for all $n\in\mathcal{I}_d$. Based on the necessary conditions in Proposition \ref{necessary-condition-total-move-constraint}, we propose a Movement-Efficient Routing-aware Lloyd (MERL) Algorithm, as outlined in Algorithm \ref{MERL}, to optimize the node deployment in heterogeneous multi-hop WSNs with constrained movement energy, and minimize the objective function in Eqs. (\ref{total-movement-constraint-objective}) and (\ref{movement-constraint}).


\begin{algorithm}[ht!]
\SetAlgoLined
\SetKwRepeat{Do}{do}{while}
\KwResult{Optimal node deployment $\mathbf{P}$, cell partitioning $\mathbf{W}$ and normalized flow matrix $\mathbf{S}$.}
Input: Initial node deployment $\tilde{\mathbf{P}}$, convergence error threshold $\epsilon\in \mathbb{R}^+$ \;
 \Do{$\frac{\mathcal{D}_{\mathrm{old}} - \mathcal{D}_{\mathrm{new}}}{\mathcal{D}_{\mathrm{old}}} \geq \epsilon$}{    -- Calculate the objective function $\mathcal{D}_{\textrm{old}}=\mathcal{D}\left(\mathbf{P},\mathbf{W},\mathbf{S}\right)$\;
  1. Update the cell partitioning $\mathbf{W}$ according to the Eq.  (\ref{necessary-condition-optimal-deployment-partitioning})\; 
  2. Update the normalized flow matrix $\mathbf{S}$ using to the Bellman-Ford algorithm\;
  3. Set $\mathcal{I}_d = \{1,\cdots,N+M\}$ and calculate $r_n \triangleq \left[1-\frac{\max\left(0, \sum_{i\in\mathcal{I}_d}\zeta_i \|\Gamma_i\| - \gamma \right)}{\|\Gamma_n\| \times \frac{\psi_n}{\zeta_n}\times\sum_{i\in\mathcal{I}_d}\frac{\zeta_i^2}{\psi_i}} \right]$, $\forall n\in \mathcal{I}_d$\;
  4. \While{$\exists n\in \mathcal{I}_d$ \textnormal{such that} $r_n\leq 0$}{
  4.1. Update $\mathcal{I}_d = \mathcal{I}_d - \bigcup_{r_n\leq 0} n$\;
  4.2. Update $\{r_n\}_{n\in\mathcal{I}_d}$;
  }
  5. $p_n = \tilde{p}_n + \Gamma_n\times\left[1-\frac{\max\left(0, \sum_{i\in\mathcal{I}_d}\zeta_i \|\Gamma_i\| - \gamma \right)}{\|\Gamma_n\| \times \frac{\psi_n}{\zeta_n}\times\sum_{i\in\mathcal{I}_d}\frac{\zeta_i^2}{\psi_i}} \right]\times \mathbf{1}_{\mathcal{I}_d}(n), \qquad\qquad\qquad \forall n\in\mathcal{I_A}\bigcup\mathcal{I_F}$\;
    -- Calculate the objective function $\mathcal{D}_{\textrm{new}}=\mathcal{D}\left(\mathbf{P},\mathbf{W},\mathbf{S}\right)$\; }
 \caption{Movement-Efficient Routing-aware Lloyd Algorithm}
 \label{MERL}
\end{algorithm}

\begin{prop}\label{MERL-convergence}
MERL Algorithm is an iterative improvement algorithm, i.e., the objective function is non-increasing and the algorithm converges.
\end{prop}
The proof of Proposition \ref{MERL-convergence} is provided in Appendix \ref{proof-MERL-convergence}.

\section{The Node Deployment with a Network Lifetime Constraint in Mobile WSNs}\label{Network-Lifetime-Constraint}

\subsection{Problem formulation}\label{Network-Lifetime-Constraint-problem-formulation}

In Section \ref{Total-Energy-Constraint}, we studied the node deployment with a total movement energy constraint, which can be seen as a resource allocation problem. This is because we can calculate how much movement energy each node requires once an optimal deployment is obtained. In this section, we focus on minimizing the communication power consumption given a constraint on the network lifetime. Let $\nu_n$ be the residual movement energy on Node $n$, and $\alpha_n$ be the power consumption for Node $n$ after relocation. To ensure a network lifetime of $T$, the following condition
\begin{equation}\label{network-lifetime-first-condition}
    \nu_n - E_n\left(\mathbf{P}\right) \geq \alpha_n T, \qquad \forall n\in\mathcal{I_A}\bigcup\mathcal{I_F}
\end{equation}
has to be satisfied. Hence, the network lifetime of $T$ can be achieved by setting a maximum individual movement energy consumption for each node. Here, our main objective is to find the optimal node deployment for the following constrained optimization problem:
\begin{align}\label{network-lifetime-objective}
    \minimize_{\mathbf{P}, \mathbf{W}, \mathbf{S}} \mathcal{D}\left(\mathbf{P}, \mathbf{W}, \mathbf{S} \right)& \\
    \textrm{s.t.}\qquad E_n\left(\mathbf{P}\right) \leq \gamma_n,& \qquad \forall n\in\mathcal{I_A}\bigcup\mathcal{I_F} \label{lifetime-constraint}
\end{align}
where $\gamma_n = \nu_n - \alpha_n T$ is the maximum individual movement energy consumption of Node $n$.

\subsection{The Optimal Node Deployment}

Here, our goal is to find the optimal node deployment $\mathbf{P}^*$, cell partitioning $\mathbf{W}^*$ and normalized flow matrix $\mathbf{S}^*$ that minimizes the multi-hop communication power consumption while each individual movement energy consumption is constrained. The following theorem provides the necessary condition for optimal node deployment in the constrained optimization problem in Eqs. (\ref{network-lifetime-objective}) and (\ref{lifetime-constraint}).
\begin{prop}\label{necessary-condition-network-lifetime}
Let $\mathbf{P}^*$, $\mathbf{W}^*$ and $\mathbf{S}^*$ be the optimal node deployment, cell partitioning and normalized flow matrix for the constrained optimization problem in Eqs. (\ref{network-lifetime-objective}) and (\ref{lifetime-constraint}). Then,
\begin{equation}\label{necessary-condition-network-lifetime-eq}
    p^*_n = \tilde{p}_n + \Gamma^*_n \times \min\left(1, \frac{\gamma_n}{\zeta_n\|\Gamma^*_n\|} \right), \qquad \forall n\in\mathcal{I_A}\bigcup\mathcal{I_F}
\end{equation}
where $\Gamma_n^* = z_n^* - \tilde{p}_n$.
\end{prop}
The proof of Proposition \ref{necessary-condition-network-lifetime} is provided in Appendix \ref{proof-necessary-condition-network-lifetime}. Based on the optimal condition in Proposition \ref{necessary-condition-network-lifetime}, we design the Lifetime-Optimized Routing-aware Lloyd (LORL) Algorithm, as outlined in Algorithm \ref{LORL}, to optimize the node deployment in heterogeneous multi-hop WSNs with network lifetime constraint, and minimize the objective function in Eqs. (\ref{network-lifetime-objective}) and (\ref{lifetime-constraint}). 

\begin{algorithm}[ht!]
\SetAlgoLined
\SetKwRepeat{Do}{do}{while}
\KwResult{Optimal node deployment $\mathbf{P}$, cell partitioning $\mathbf{W}$ and normalized flow matrix $\mathbf{S}$.}
Input: Initial node deployment $\tilde{\mathbf{P}}$, convergence error threshold $\epsilon\in \mathbb{R}^+$ \;
 \Do{$\frac{\mathcal{D}_{\mathrm{old}} - \mathcal{D}_{\mathrm{new}}}{\mathcal{D}_{\mathrm{old}}} \geq \epsilon$}{    -- Calculate the objective function $\mathcal{D}_{\textrm{old}}=\mathcal{D}\left(\mathbf{P},\mathbf{W},\mathbf{S}\right)$\;
  1. Update the cell partitioning $\mathbf{W}$ according to the Eq.  (\ref{necessary-condition-optimal-deployment-partitioning})\; 
  2. Update the normalized flow matrix $\mathbf{S}$ using to the Bellman-Ford algorithm\;
  3. $    p_n = \tilde{p}_n + \Gamma_n \times \min\left(1, \frac{\gamma_n}{\zeta_n\|\Gamma_n\|} \right), \qquad\qquad\qquad\qquad\qquad\qquad\qquad \forall n\in\mathcal{I_A}\bigcup\mathcal{I_F}$\;
    -- Calculate the objective function $\mathcal{D}_{\textrm{new}}=\mathcal{D}\left(\mathbf{P},\mathbf{W},\mathbf{S}\right)$\; }
 \caption{Lifetime-Optimized Routing-aware Lloyd Algorithm}
 \label{LORL}
\end{algorithm}

\begin{prop}\label{LORL-convergence}
LORL Algorithm is an iterative improvement algorithm, i.e., the objective function is non-increasing and the algorithm converges.
\end{prop}
The proof of Proposition \ref{LORL-convergence} is provided in Appendix \ref{proof-LORL-convergence}.

\section{Experiments}\label{Experiments}

Simulations are carried out for a heterogeneous wireless sensor network consisting of $30$ APs and $3$ FCs. We consider a square field of size $10$km $\times$ $10$km, i.e., $\Omega = \left[0, 10000\right]^2$. Simulations are performed for two different sensor density functions, i.e., a uniform distribution $f\left(\omega\right) = \frac{1}{\int_{\Omega}d\omega} = 10^{-8}$, and a mixture of Gaussian where sensors are distributed according to:
\begin{align}\label{Mixture_of_Gaussian_Distribution}
    f(\omega) &= \frac{1}{2} \! \times  \mathcal{N}\! \left(\! \begin{bmatrix}
    3e\!+\!3 \\
    3e\!+\!3 
\end{bmatrix} \! , \! \begin{bmatrix}
    1.5e\!+\!6 & 0 \\
    0 & 1.5e\!+\!6 
\end{bmatrix} \! \right)\! +  \frac{1}{4} \! \times  \mathcal{N}\! \left(\! \begin{bmatrix}
    6e\!+\!3 \\
    7e\!+\!3 
\end{bmatrix} \! , \! \begin{bmatrix}
    2e\!+\!6 & 0 \\
    0 & 2e\!+\!6 
\end{bmatrix}\! \right) \! \nonumber\\&+  \frac{1}{4} \! \times  \mathcal{N}\! \left(\! \begin{bmatrix}
    7.5e\!+\!3 \\
    2.5e\!+\!3 
\end{bmatrix} \! , \! \begin{bmatrix}
    1e\!+\!6 & 0 \\
    0 & 1e\!+\!6
\end{bmatrix} \! \right).
\end{align}
All homogeneous densely deployed sensors share the transmitter antenna gain of $G_{t_{\textrm{sensor}}} = 1$. We consider a radio bit-rate of $R_b = 1$Mbps, and assume that the wavelength of the carrier signal is $\lambda_c = 0.3$m. In order for APs and FCs to receive the signal without error, the received power at each node $n \in \mathcal{I_A}\bigcup\mathcal{I_F}$ should be greater than some threshold $P_{th_n}$. Moreover, the transceiver electronics in each AP $n$ consumes $\rho_n$ Joules per bit for digital coding, modulation and filtering before signal transmission. Table \ref{P_threshold_rho} summarizes the values of $P_{th_n}$ and $\rho_n$ for all nodes \cite{heinzelman2000application}.

\begin{table}[!bth]
\centering
  \caption{Simulation parameters}
  \vspace{-3mm}
\begin{tabular}{ccccccccc}
 \toprule
                   \multicolumn{4}{c}{{\bf minimum received power (nW)}}                           && \multicolumn{4}{c}{{\bf electronics energy dissipation (nJ/bit)}} \\\cline{1-4}\cline{6-9}
$\mathbf{P_{th_{1:15}}}$&$\mathbf{P_{th_{16:30}}}$&$\mathbf{P_{th_{31}}}$&$\mathbf{P_{th_{32:33}}}$&&$\mathbf{\rho_{1:7}}$ &&$\mathbf{\rho_{8:16}}$&$\mathbf{\rho_{17:30}}$\\
          $10$          &           $6$           &        $6$           &         $10$            &&         $40$         &&           $50$       &         $60$          \\ 
 \bottomrule 
\end{tabular}
  \label{P_threshold_rho}
\end{table}

For each AP $n$, we denote its transmitter antenna gain via $G_{t_n}$. In addition, for each node $n\in\mathcal{I_A}\bigcup\mathcal{I_F}$, let $G_{r_n}$ be its receiver antenna gain. Table \ref{antenna_gains} summarizes the values of the transmitter and receiver antenna gains for all nodes \cite{heinzelman2000application}.

\begin{table}[!bth]
\centering
  \caption{Transmitter and receiver antenna gains}
  \vspace{-3mm}
\begin{tabular}{ccccc}
 \toprule
\multicolumn{2}{c}{{\bf transmitter antenna gain}}&& \multicolumn{2}{c}{{\bf receiver antenna gain}} \\\cline{1-2}\cline{4-5}
$\mathbf{G_{t_{1:7,15:22}}}$&$\mathbf{G_{t_{8:14,23:30}}}$&&$\mathbf{G_{r_{1:3,8:11,15:18,23:26,31:32}}}$ &$\mathbf{G_{r_{4:7,12:14,19:22,27:30,33}}}$\\
             $1$            &           $2$               &&                     $1$                      &                        $2$                \\ 
 \bottomrule 
\end{tabular}
  \label{antenna_gains}
\end{table}

Note that parameters $\eta_i$ and $\beta_{i,j}$, for all $i\in\mathcal{I_A}$ and $j\in\mathcal{I_A}\bigcup\mathcal{I_F}$, can be calculated using the explained experimental setup. For example, we have $\eta_7 = \frac{P_{th_7} \times \left(4\pi\right)^2}{R_b G_{t_{\textrm{sensor}}} G_{r_7} \lambda_c^2} = \frac{10^{-8}\times \left(4\pi\right)^2}{10^6 \times 1 \times 2 \times (0.3)^2}=8.77 \textrm{ pJ/bit/}\textrm{m}^2$ and $\beta_{10, 20} = \frac{P_{th_{20}} \times \left(4\pi\right)^2}{R_b G_{t_{10}} G_{r_{20}} \lambda_c^2} = \frac{6\times 10^{-9}\times \left(4\pi\right)^2}{10^6 \times 2 \times 2 \times (0.3)^2}=2.63 \textrm{ pJ/bit/}\textrm{m}^2$. For performance evaluation, $10$ initial AP and FC deployments are generated randomly on $\Omega$, i.e., the location of each node is generated according to a uniform distribution on $\Omega$. The maximum number of iterations for all algorithms is set to $200$ and the Lagrangian multiplier is set to $\lambda = 0.25$.

\subsection{Static Heterogeneous Multi-Hop WSNs}\label{Static_Heterogeneous_Simulations}

We compare the total weighted communication power consumption of our proposed RL Algorithm with Cluster-Formation Algorithm \cite{chatterjee2015multiple}, Global Algorithm \cite{vincze2007deploying}, HTTL Algorithm \cite{karimi2020energy}, PSO Algorithm \cite{dandekar2013energy}, and SHMS Algorithm \cite{jain2015lifetime}. To reduce the number of hops that data packets have to travel to reach the fusion centers, the Cluster-Formation algorithm employs a graph theoretic approach to optimize both the number of clusters and their corresponding diameters. To reduce the communication distance between the nodes, the Global algorithm deploys nodes such that the average Euclidean distance between access points and their corresponding fusion centers is minimized. For a two-tier hierarchy of APs and FCs, the HTTL algorithm iteratively updates the node deployment, cell partitioning, and connections between APs and FCs while the flow of data from each sensor to its corresponding FC is mediated by exactly one access point. PSO is a population based iterative algorithm for finding the optimal node deployment and minimizing the non-linear objective function. For a given node deployment, the SHMS algorithm determines the connections between APs and FCs such that the maximum energy consumed by each node is minimized. 

The weighted power consumption of Cluster-Formation, Global, HTTL, PSO, SHMS and RL algorithms for the uniform sensor density function are summarized in Table \ref{weighted_power_static_uniform}. The RL algorithm outperforms other algorithms, and achieves a lower weighted communication power consumption. Note that although the HTTL algorithm proposed in \cite{karimi2020energy} deploys nodes based on the necessary conditions of optimality, the network architecture is restricted to a two-tier hierarchy while the RL algorithm simultaneously optimizes over the node deployment and data routing. As a result, the  node deployment based on the RL algorithm results in a WSN that saves about $21\%$ of the energy consumed by the node deployment based on HTTL Algorithm.

\begin{table}[!bth]
\centering
  \caption{Weighted power comparison for the uniform sensor density function}
  \vspace{-3mm}
\begin{tabular}{cccccc}
 \toprule
{\bf Cluster-Formation}&{\bf Global}&{\bf HTTL}&{\bf PSO}&{\bf SHMS}&    {\bf RL}    \\
        $15.49$        &   $14.98$  & $12.80$  & $19.98$ & $22.39$  &$\mathbf{10.12}$\\ 
 \bottomrule 
\end{tabular}
  \label{weighted_power_static_uniform}
\end{table}

\begin{figure}[!htb]
\centering
\subfloat[]{\includegraphics[width=50.5mm]{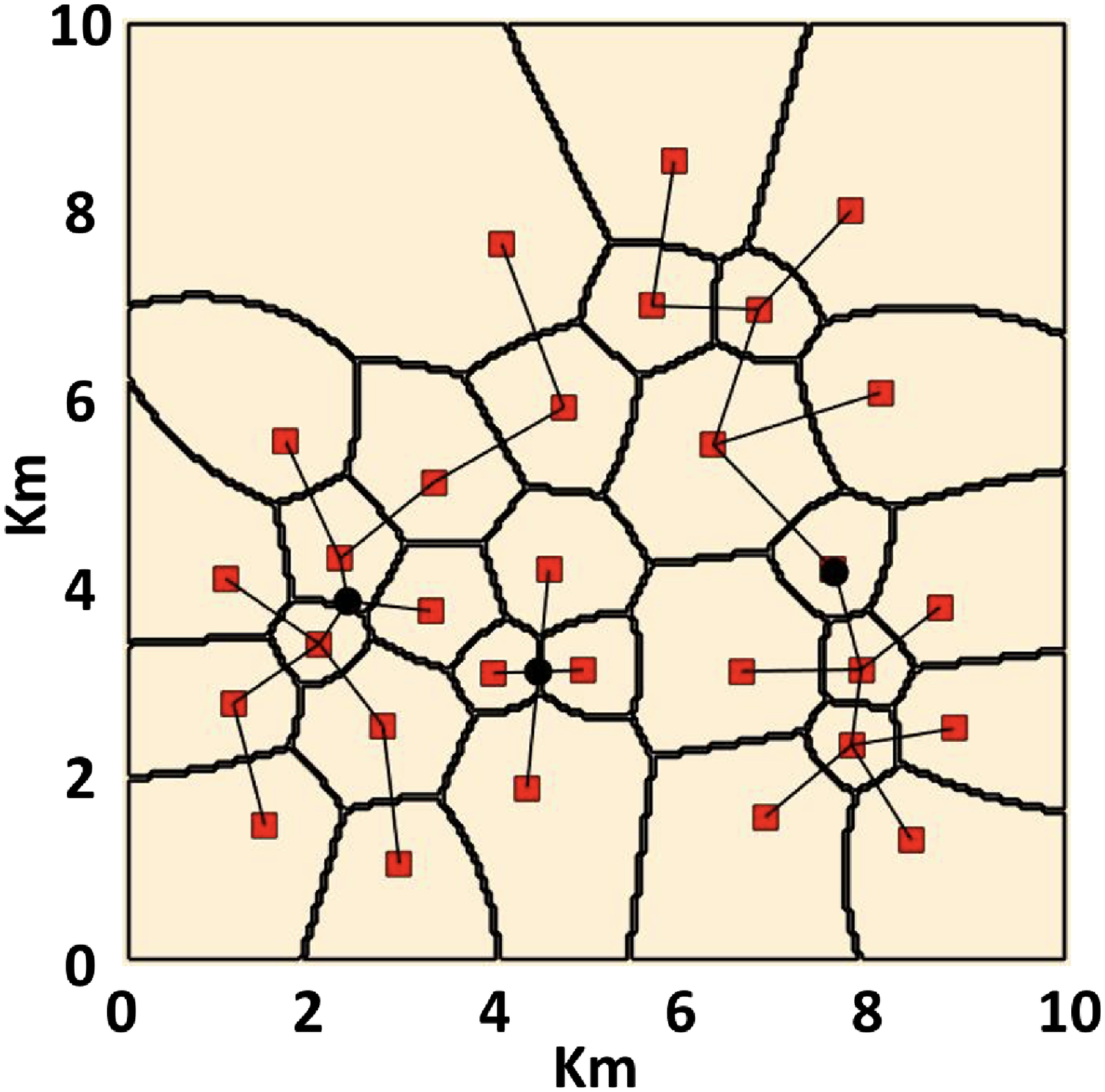}}
\label{Gaussian_static_sample_deployment_cluster}
\hspace{3.0mm}
\subfloat[]{\includegraphics[width=50.5mm]{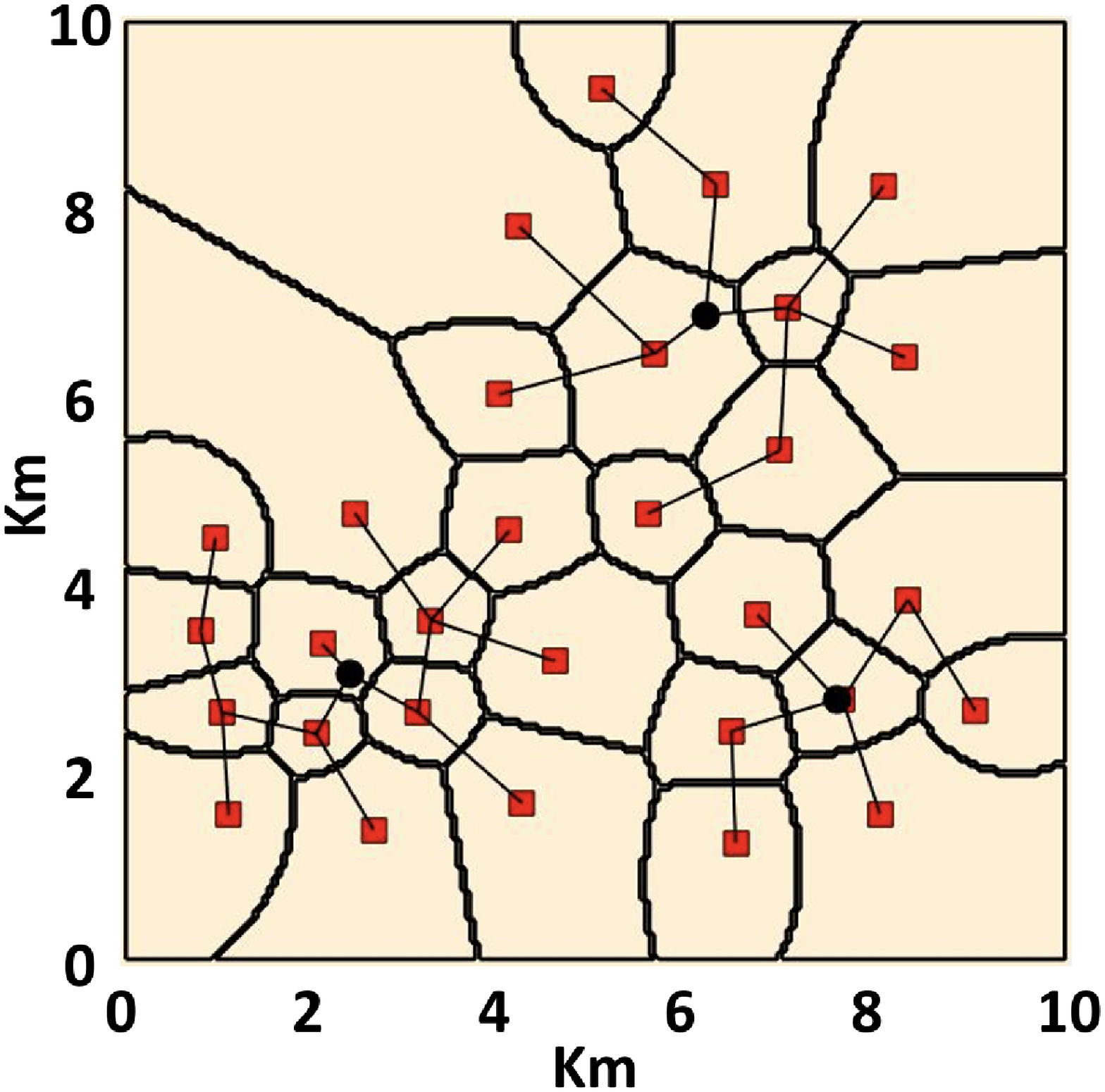}}
\label{Gaussian_static_sample_deployment_global}
\hspace{3.0mm}
\subfloat[]{\includegraphics[width=50.5mm]{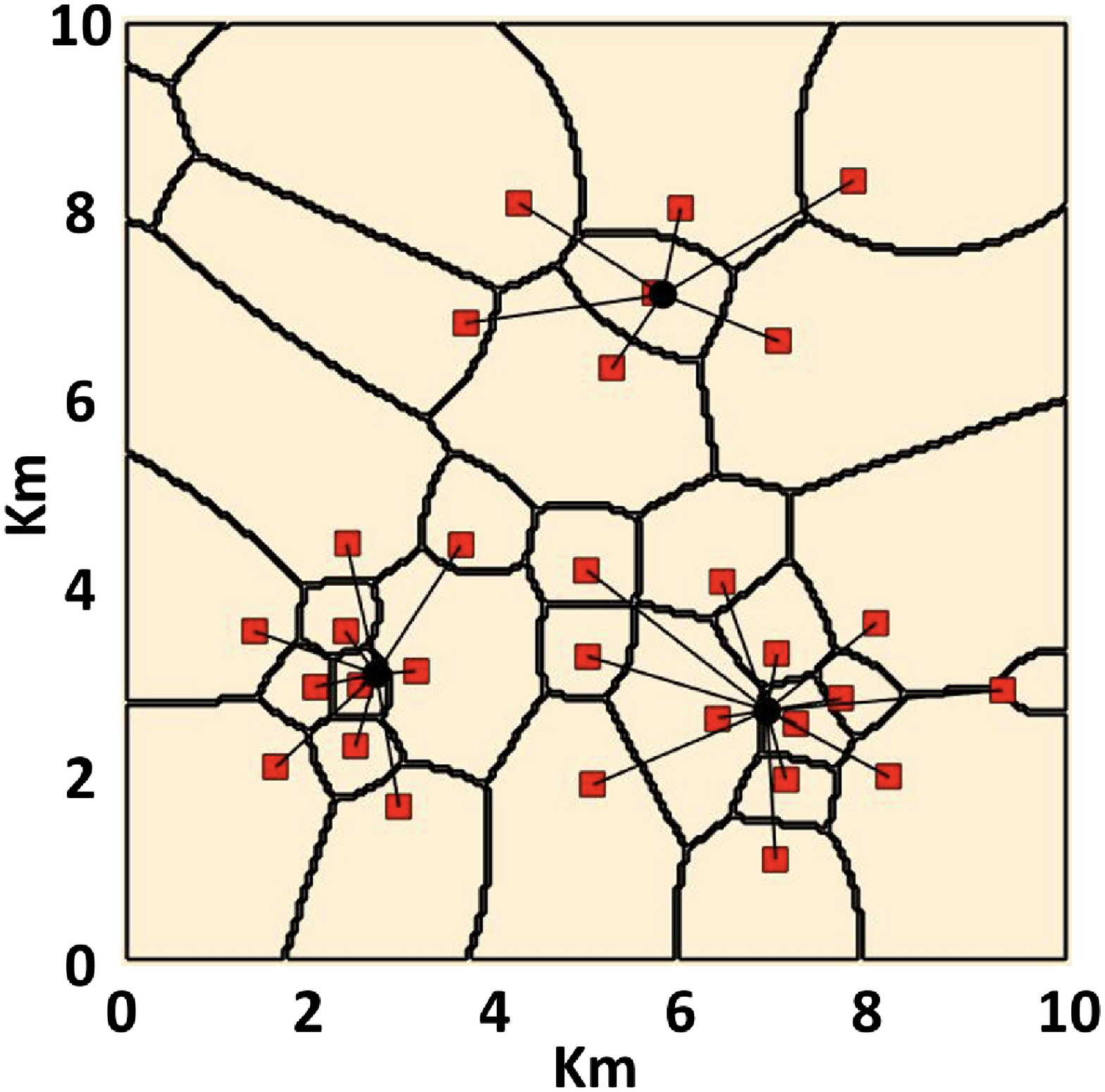}}
\label{Gaussian_static_sample_deployment_httl}
\hspace{3.0mm}
\vspace{1mm}
\subfloat[]{\includegraphics[width=50.5mm]{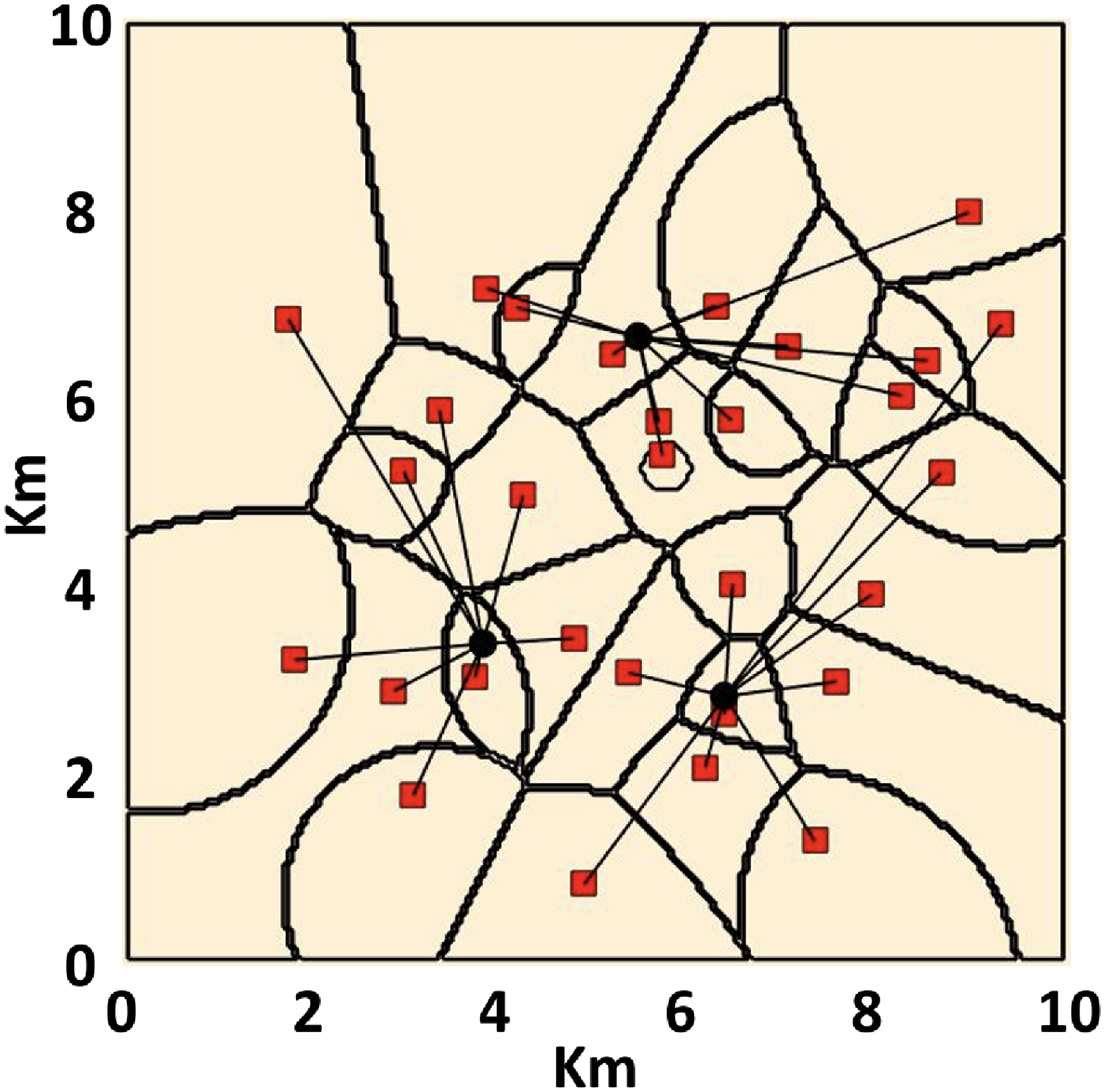}}
\label{Gaussian_static_sample_deployment_pso}
\hspace{3.0mm}
\subfloat[]{\includegraphics[width=50.5mm]{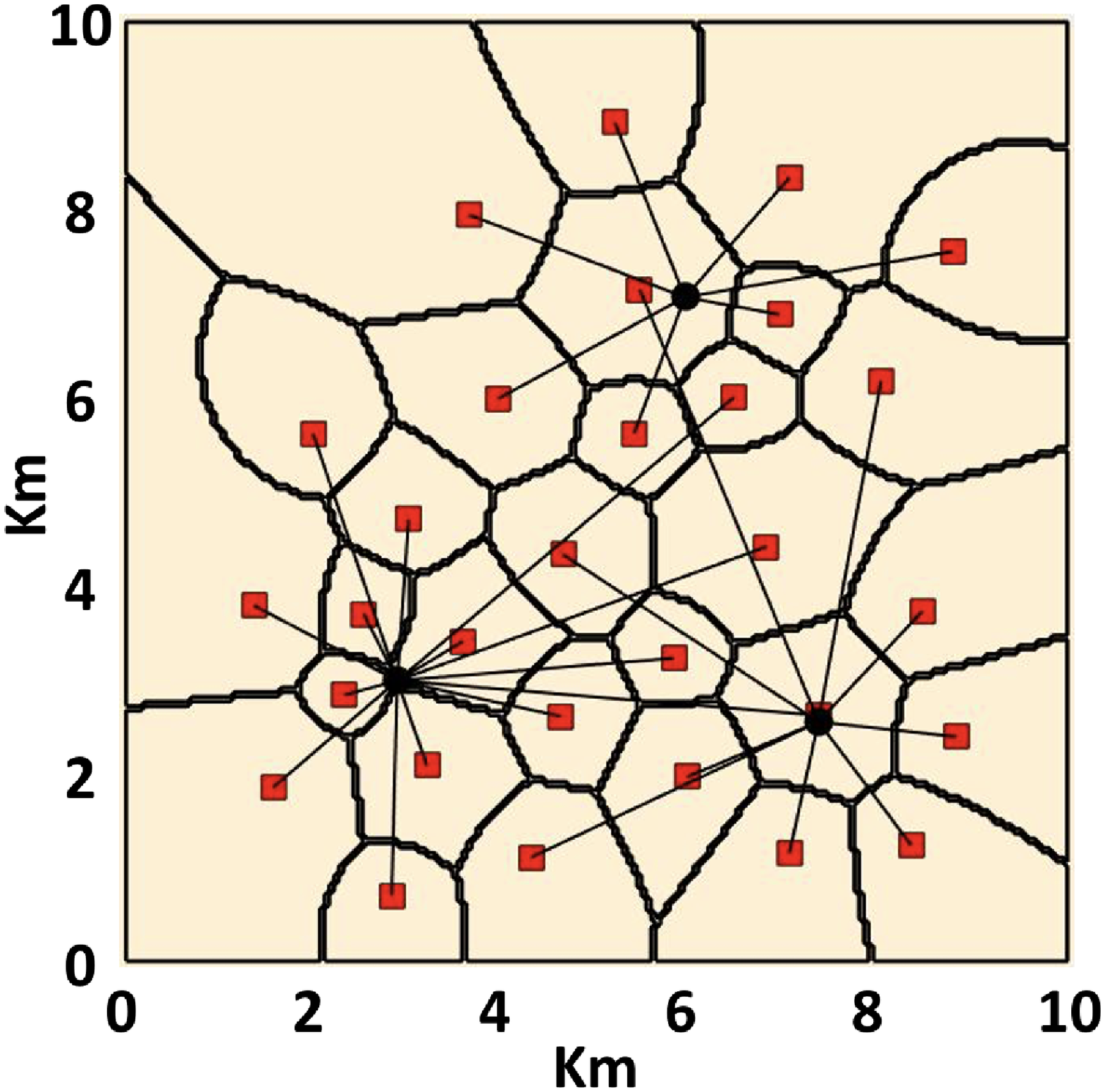}}
\label{Gaussian_static_sample_deployment_shms}
\hspace{3.0mm}
\subfloat[]{\includegraphics[width=50.5mm]{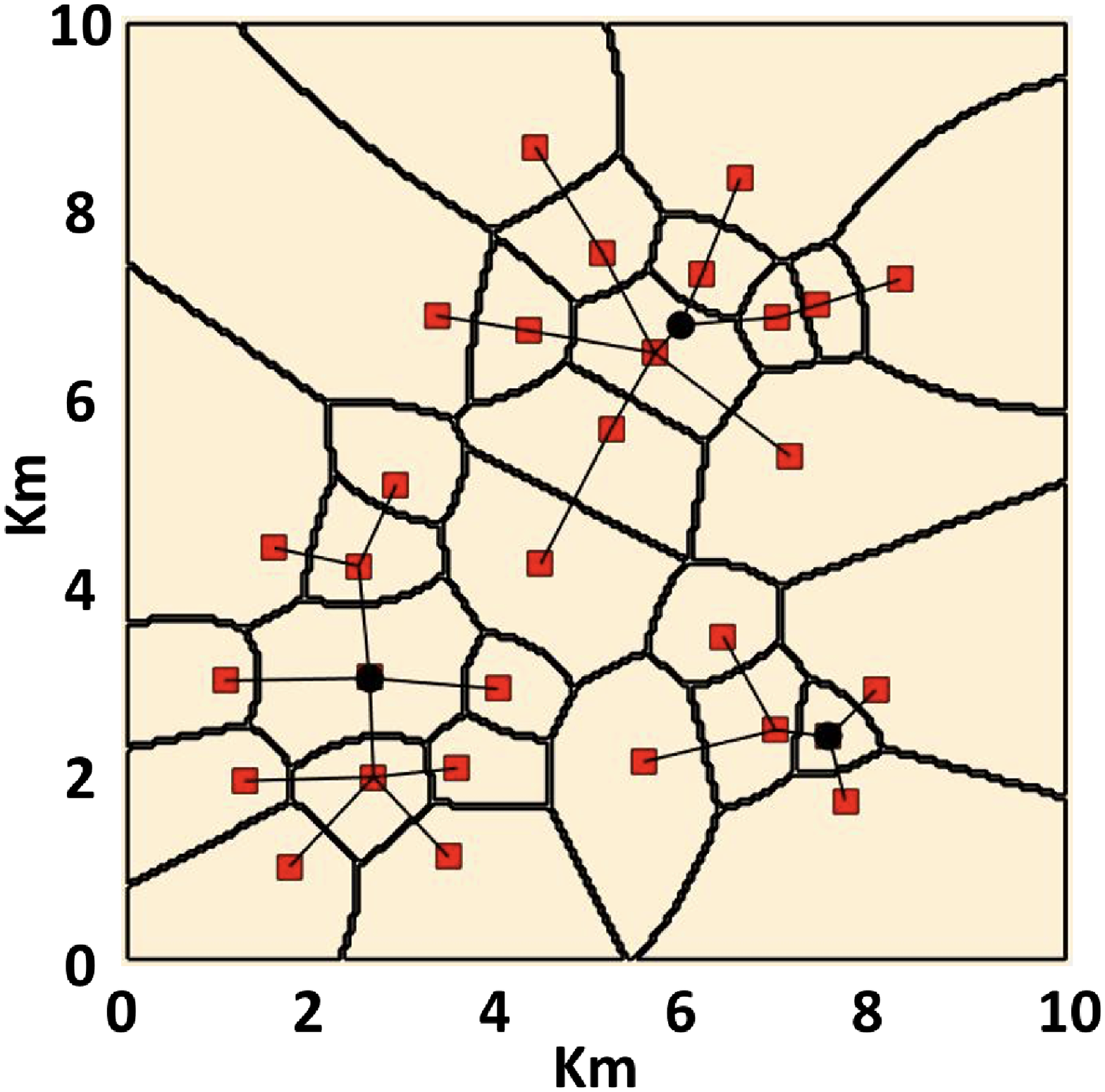}}
\label{Gaussian_static_sample_deployment_rl}

\captionsetup{justification=justified}
\caption{\small{Node deployment for different algorithms and the mixture of Gaussian sensor density function. (a) Cluster-Formation (b) Global (c) HTTL (d) PSO (e) SHMS (f) RL.}}
\label{Gaussian_static_sample_deployment}
\end{figure}

Table \ref{weighted_power_static_gaussian} summarizes the weighted communication power consumption of Cluster-Formation, Global, HTTL, PSO, SHMS and RL algorithms for the mixture of Gaussian sensor density function in Eq. (\ref{Mixture_of_Gaussian_Distribution}). The RL algorithm results in a power consumption of $5.58$ Watts, and outperforms other methods. Furthermore, the RL algorithm leads to a network architecture that exhaust its available communication energy in a time period that is longer by about $10\%$ of that of HTTL Algorithm. Figure \ref{Gaussian_static_sample_deployment} shows the optimal node deployment for different algorithms where APs and FCs are denoted by red squares and black circles, respectively.

\begin{table}[!bth]
\centering
  \caption{Weighted power comparison for the mixture of Gaussian sensor density function}
  \vspace{-3mm}
\begin{tabular}{cccccc}
 \toprule
{\bf Cluster-Formation}&{\bf Global}&{\bf HTTL}&{\bf PSO}&{\bf SHMS}&    {\bf RL}    \\
        $7.07$         &   $6.81$   & $6.23$   & $9.97$  & $16.62$  &$\mathbf{5.58}$ \\ 
 \bottomrule 
\end{tabular}
  \label{weighted_power_static_gaussian}
\end{table}

\subsection{Mobile Heterogeneous Multi-Hop WSNs with a Total Movement Energy Constraint}\label{Mobile_Total_Movement_Energy_Simulations}

The underlying assumption in all deployment strategies studied in Section \ref{Static_Heterogeneous_Simulations} is that the optimal node locations are calculated offline, then each node is placed at its corresponding position. However, in many applications, e.g. when the target region is a hostile environment, static deployment is not feasible. Instead, nodes are initially deployed in the target region, e.g. by airdropping them using small drones or manual placement in an accessible sub-region of the field, then each node moves to its optimal location based on the initial deployment and available movement energy. When the total available movement energy is constrained, the optimization problem is translated into a resource allocation problem where the optimal energy supply for each node is determined such that the resulting total communication power consumption after optimal deployment is minimized. The performance evaluation under this scenario is the focus of this section. In Section \ref{Mobile_Lifetime_Simulations}, we study the performance evaluation when the available movement energy for each node is predetermined, and the optimization problem is translated to that of enhancing the network lifetime.

The same experimental setup described at the beginning of Section \ref{Experiments} and in Tables \ref{P_threshold_rho} and \ref{antenna_gains} is used for the simulations. Furthermore, Table \ref{Moving_cost_parameters} provides the moving cost parameters $\zeta_n$ for each node $n\in\mathcal{I_A}\bigcup\mathcal{I_F}$  \cite{shigei2012battery, el2011mobile}. We consider a total available movement energy of $\gamma = 40000$ Joules for the constrained objective function in Eqs. (\ref{total-movement-constraint-objective}) and (\ref{movement-constraint}). 

\begin{table}[!bth]
\centering
  \caption{Moving cost parameters (J/m)}
  \vspace{-3mm}
\begin{tabular}{cccccc}
 \toprule
$\mathbf{\zeta_{1:8}}$&$\mathbf{\zeta_{9:22}}$&$\mathbf{\zeta_{23:30}}$&$\mathbf{\zeta_{31}}$&$\mathbf{\zeta_{32}}$&$\mathbf{\zeta_{33}}$\\
          $2$         &          $4$          &           $6$          &          $4$        &          $5$        &          $6$        \\ 
 \bottomrule 
\end{tabular}
  \label{Moving_cost_parameters}
\end{table}

We compare the total weighted communication power consumption of our proposed MERL Algorithm with Lloyd-$\alpha$ Algorithm \cite{song2013distributed}, OMF Algorithm \cite{chellappan2007deploying}, and VFA Algorithm \cite{zou2004sensor}. The Lloyd-$\alpha$ algorithm applies a penalty term to the Lloyd algorithm to reduce the movement steps and save traveling energy while guaranteeing the convergence property. The OMF algorithm optimizes the movement plan for nodes such that each region in the network has a minimum number of nodes to relay the data to fusion centers while the sum of nodes' traveling distances is minimized. The VFA algorithm uses attractive and repulsive virtual forces on nodes such that not only every two nodes in the final deployment maintain a minimum distance from each other, but also the communication distances are minimized by avoiding nodes to be located very far from each other. For a fair comparison, the same initial deployment is used for all algorithms. 

The weighted communication power consumption of Lloyd-$\alpha$, OMF, VFA, and MERL algorithms for the uniform sensor density function are summarized in Table \ref{weighted_power_mobile_total_uniform}. All algorithms exhausted the available movement energy $\gamma$ to move the AP and FC nodes from their initial deployment to their designated optimal locations. The MERL algorithm leads to a deployment that consumes communication energy in a rate that is almost half of other algorithms. The superior performance of the MERL algorithm is due to the optimal energy allocation among nodes, as it is implicit in Eq. (\ref{necessary-condition-total-move-constraint-eq2}). Note that if the total movement energy $\gamma$ is large enough, e.g. $\gamma \geq \sum_{i=1}^{N+M}\zeta_i \|\tilde{p}_i - z^*_i \|$, then the performance of the MERL algorithm will converge to that of the RL algorithm. However, since the value of $\gamma$ in our experiments is not large enough, nodes will run out of their allocated movement energy, and MERL algorithm leads to a communication power consumption that is larger than that of the RL algorithm in Section \ref{Static_Heterogeneous_Simulations}.

\begin{table}[!bth]
\centering
  \caption{Weighted power comparison for the uniform sensor density function}
  \vspace{-3mm}
\begin{tabular}{cccc}
 \toprule
{\bf Lloyd-$\alpha$}&{\bf OMF}&{\bf VFA} &  {\bf MERL}    \\
        $29.12$     & $27.35$ & $27.85$  &$\mathbf{14.49}$\\ 
 \bottomrule 
\end{tabular}
  \label{weighted_power_mobile_total_uniform}
\end{table}

Table \ref{weighted_power_mobile_total_gaussian} also summarizes the weighted communication power consumption of Lloyd-$\alpha$, OMF, VFA, and MERL algorithms for the mixture of Gaussian sensor density function in Eq. (\ref{Mixture_of_Gaussian_Distribution}). The MERL algorithm significantly outperforms other methods and leads to a communication power consumption that is less than half of what other algorithms achieve. This is because the MERL algorithm can optimally adapt to any underlying sensor density function $f(\omega)$ and deploy nodes accordingly, as we studied in Section \ref{Total-Energy-Constraint}.

\begin{table}[!bth]
\centering
  \caption{Weighted power comparison for the mixture of Gaussian sensor density function}
  \vspace{-3mm}
\begin{tabular}{cccc}
 \toprule
{\bf Lloyd-$\alpha$}&{\bf OMF}&{\bf VFA} &  {\bf MERL}    \\
        $17.38$     & $17.29$ & $18.76$  &$\mathbf{7.64}$ \\ 
 \bottomrule 
\end{tabular}
  \label{weighted_power_mobile_total_gaussian}
\end{table}

\begin{figure}[!htb]
\centering
\subfloat[]{\includegraphics[width=37.5mm]{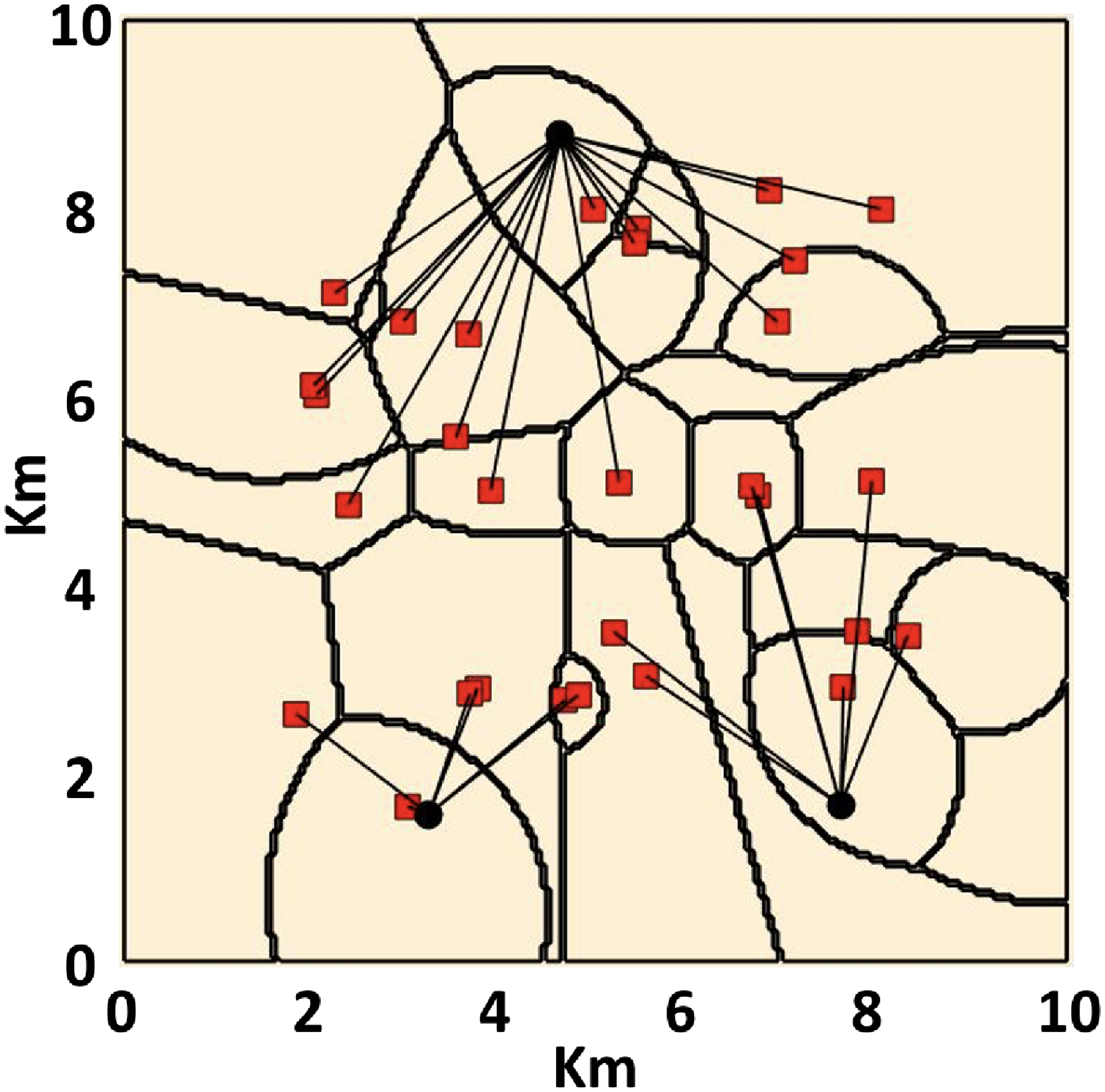}}
\label{Gaussian_mobile_total_sample_deployment_lloyd_alpha}
\hspace{2mm}
\subfloat[]{\includegraphics[width=37.5mm]{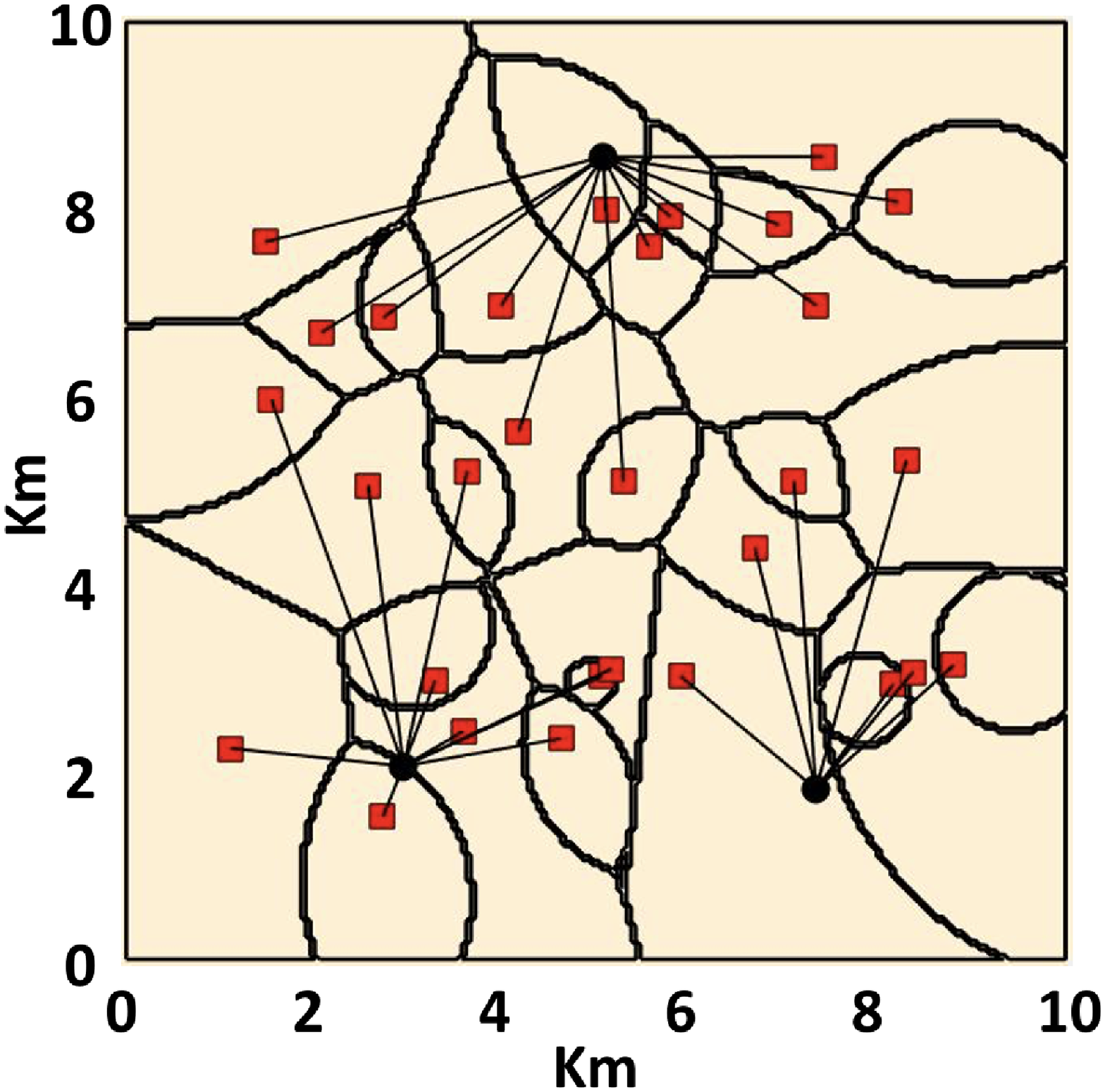}}
\label{Gaussian_mobile_total_sample_deployment_OMF}
\hspace{2mm}
\subfloat[]{\includegraphics[width=37.5mm]{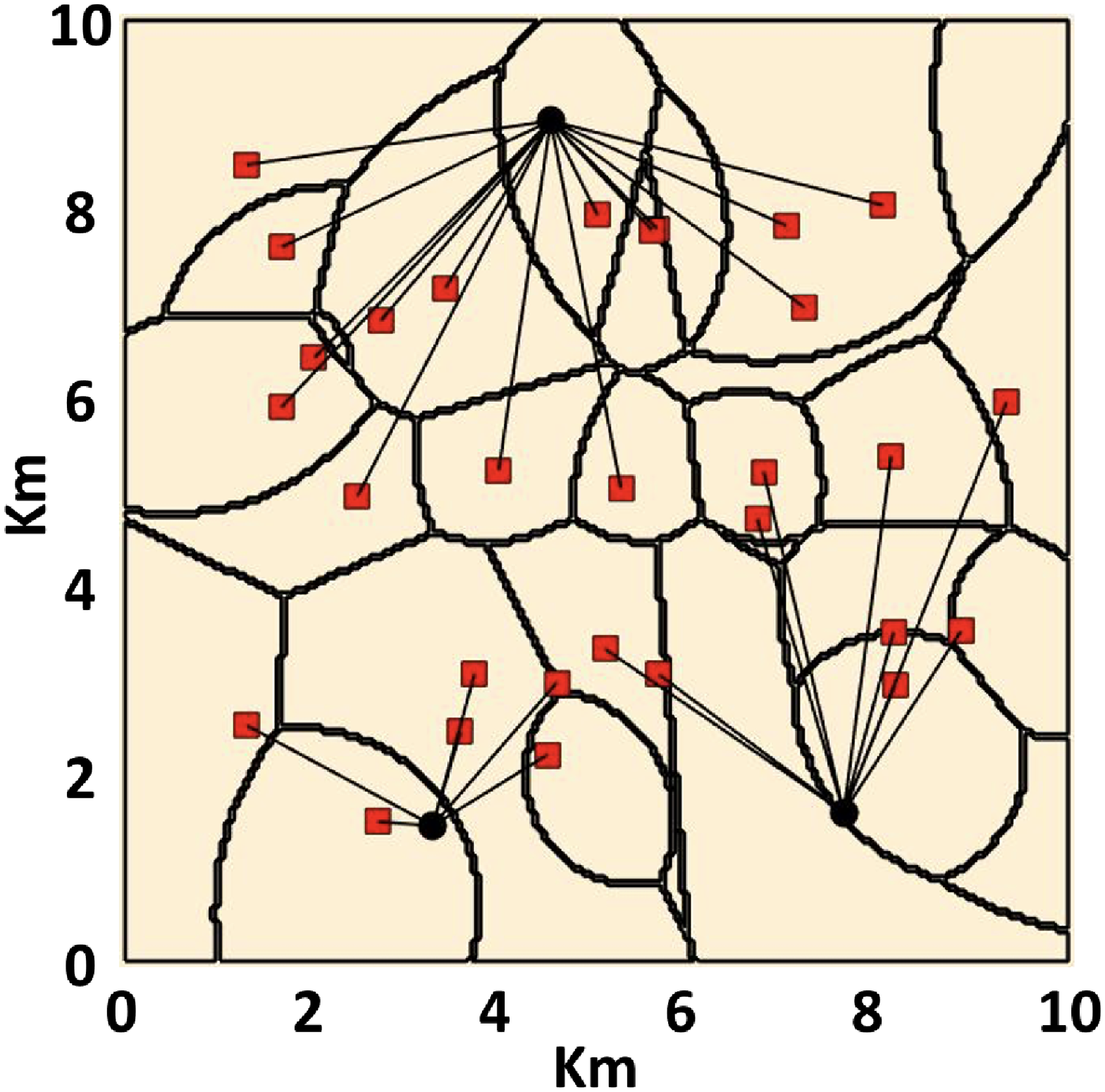}}
\label{Gaussian_mobile_total_sample_deployment_VFA}
\hspace{2mm}
\subfloat[]{\includegraphics[width=37.5mm]{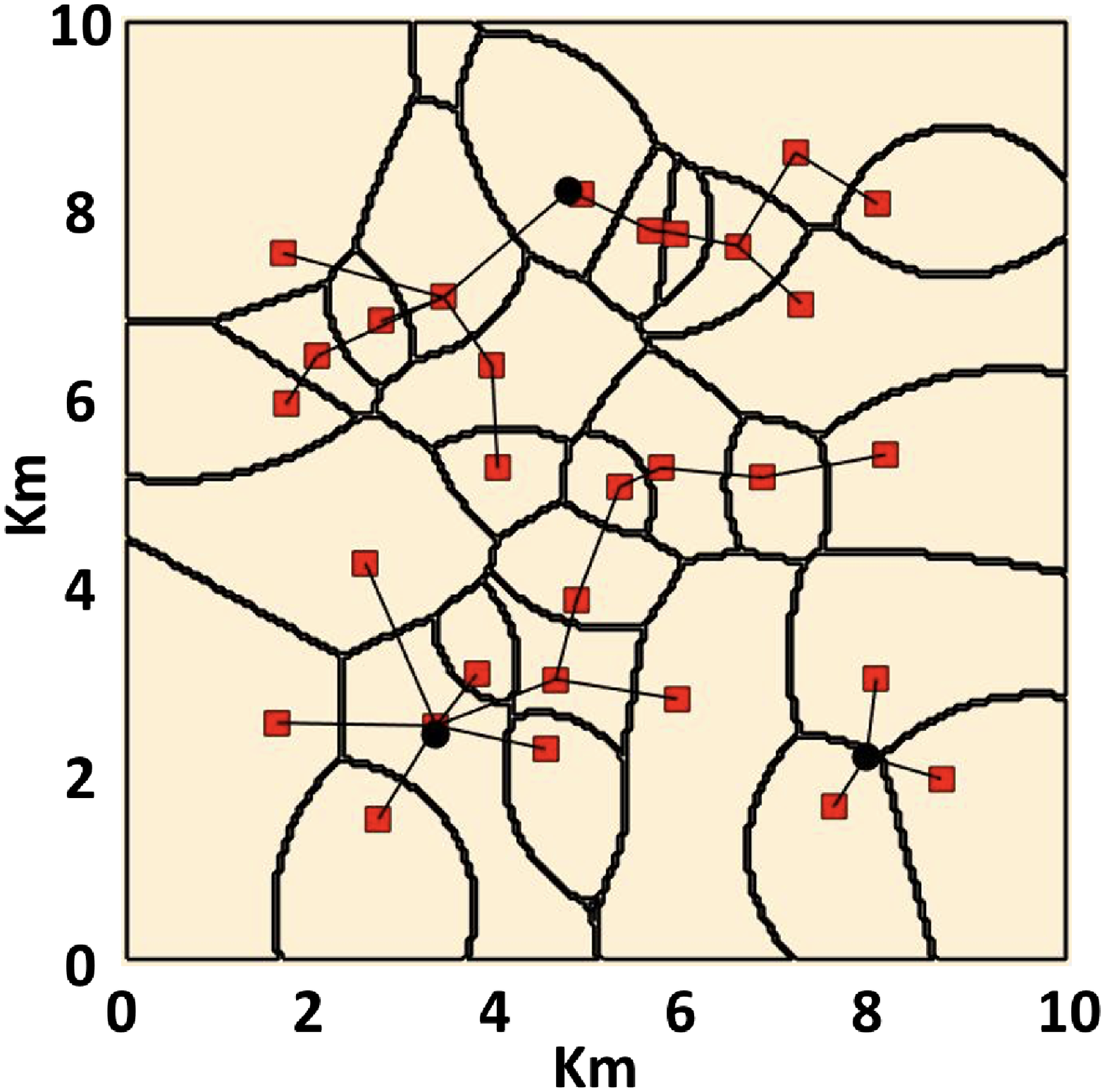}}
\label{Gaussian_mobile_total_sample_deployment_MERL}

\vspace{-2mm}
\captionsetup{justification=justified}
\caption{\small{Node deployment for different algorithms and the mixture of Gaussian sensor density function. (a) Lloyd-$\alpha$ (b) OMF (c) VFA (d) MERL.}}
\label{Gaussian_mobile_total_sample_deployment}
\end{figure}

Figure \ref{Gaussian_mobile_total_sample_deployment} shows the final deployment for different algorithms where APs and FCs are denoted by red squares and black circles, respectively.

\subsection{Mobile Heterogeneous Multi-Hop WSNs with a Network Lifetime Constraint}\label{Mobile_Lifetime_Simulations}

While in Section \ref{Mobile_Total_Movement_Energy_Simulations} we studied the performance evaluation of mobile WSNs under a total movement energy constraint, here we focus on enhancing the network lifetime, which necessitates nodes to have individual movement energy constraints, as formulated in Eqs. (\ref{network-lifetime-objective}) and (\ref{lifetime-constraint}). We use the same experimental setup and node characterization as described at the beginning of Section \ref{Experiments} and in Tables \ref{P_threshold_rho}, \ref{antenna_gains} and \ref{Moving_cost_parameters} for performance evaluation. In addition, Table \ref{gamma_ns} provides the maximum individual movement energy consumption $\gamma_n$ for all nodes $n\in\mathcal{I_A}\bigcup\mathcal{I_F}$.

\begin{table}[!bth]
\centering
  \caption{Movement energy constraints (J)}
  \vspace{-3mm}
\begin{tabular}{cccccc}
 \toprule
$\mathbf{\gamma_{1:8}}$&$\mathbf{\gamma_{9:22}}$&$\mathbf{\gamma_{23:30}}$&$\mathbf{\gamma_{31}}$&$\mathbf{\gamma_{32}}$&$\mathbf{\gamma_{33}}$\\
         $800$         &          $1100$        &          $1400$         &         $2000$       &         $2400$       &         $2600$       \\ 
 \bottomrule 
\end{tabular}
  \label{gamma_ns}
\end{table}

We compare the weighted communication power consumption of our proposed LORL Algorithm with those of Lloyd-$\alpha$ Algorithm, OMF Algorithm, and VFA Algorithm described in Section \ref{Mobile_Total_Movement_Energy_Simulations}. For a fair comparison, the same initial deployment as in Section \ref{Mobile_Total_Movement_Energy_Simulations} is used for all algorithms.

The weighted communication power consumption of Lloyd-$\alpha$, OMF, VFA, and LORL algorithms for the uniform sensor density function are provided in Table \ref{weighted_power_mobile_individual_uniform}. The LORL algorithm outperforms other methods, and achieves a significantly lower power consumption. For instance, the LORL algorithm leads to a node deployment in which the network consumes its residual energy with a rate that is less than $70\%$ of that of the VFA algorithm. This in turn prolongs the network lifetime, which is a prominent factor in wireless sensor networks.

\begin{table}[!bth]
\centering
  \caption{Weighted power comparison for the uniform sensor density function}
  \vspace{-3mm}
\begin{tabular}{cccc}
 \toprule
{\bf Lloyd-$\alpha$}&{\bf OMF}&{\bf VFA} &  {\bf LORL}    \\
        $27.64$     & $30.12$ & $25.24$  &$\mathbf{17.33}$\\ 
 \bottomrule 
\end{tabular}
  \label{weighted_power_mobile_individual_uniform}
\end{table}

\begin{figure}[!htb]
\centering
\subfloat[]{\includegraphics[width=37.5mm]{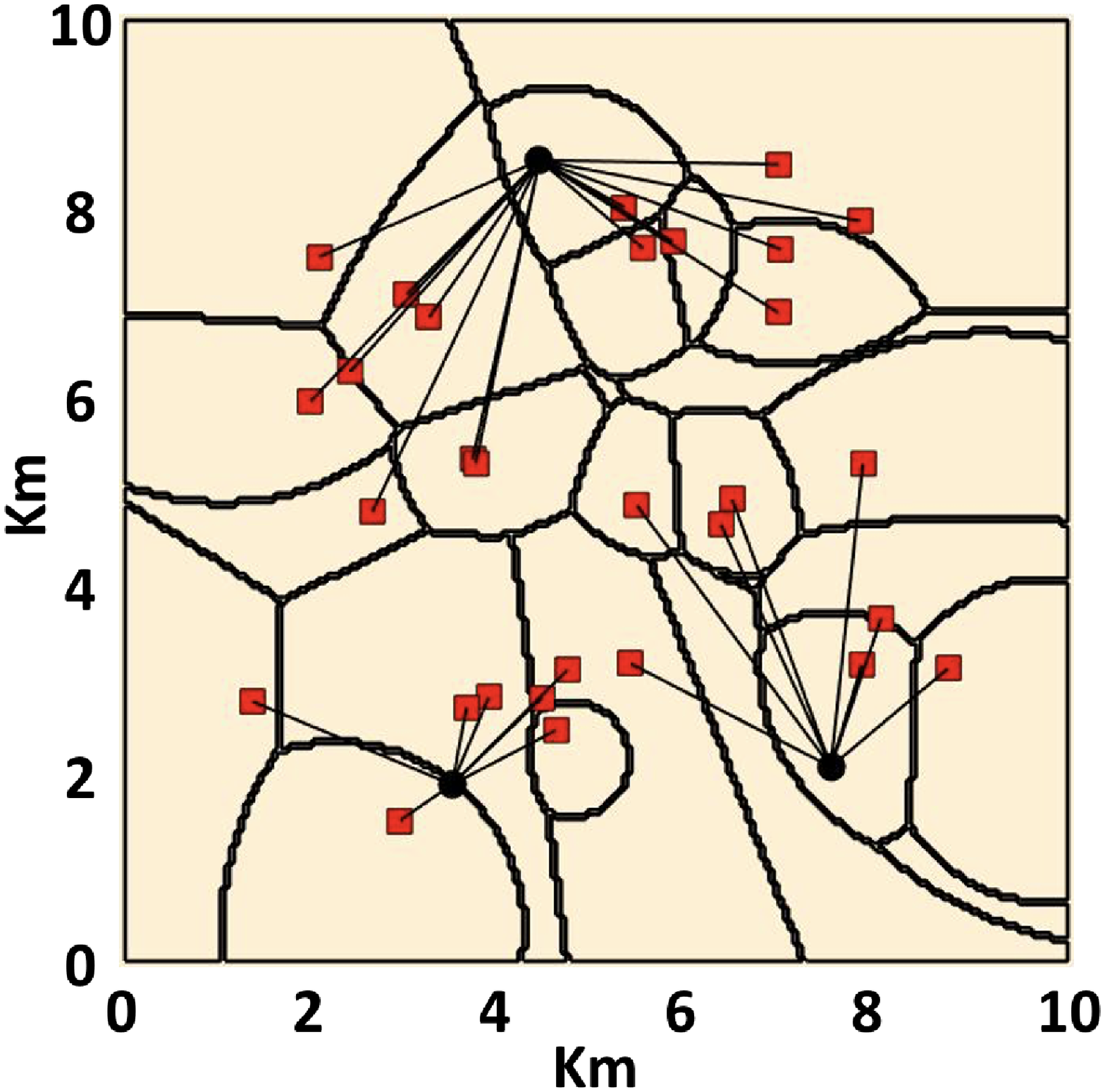}}
\label{Gaussian_mobile_individual_sample_deployment_lloyd_alpha}
\hspace{2mm}
\subfloat[]{\includegraphics[width=37.5mm]{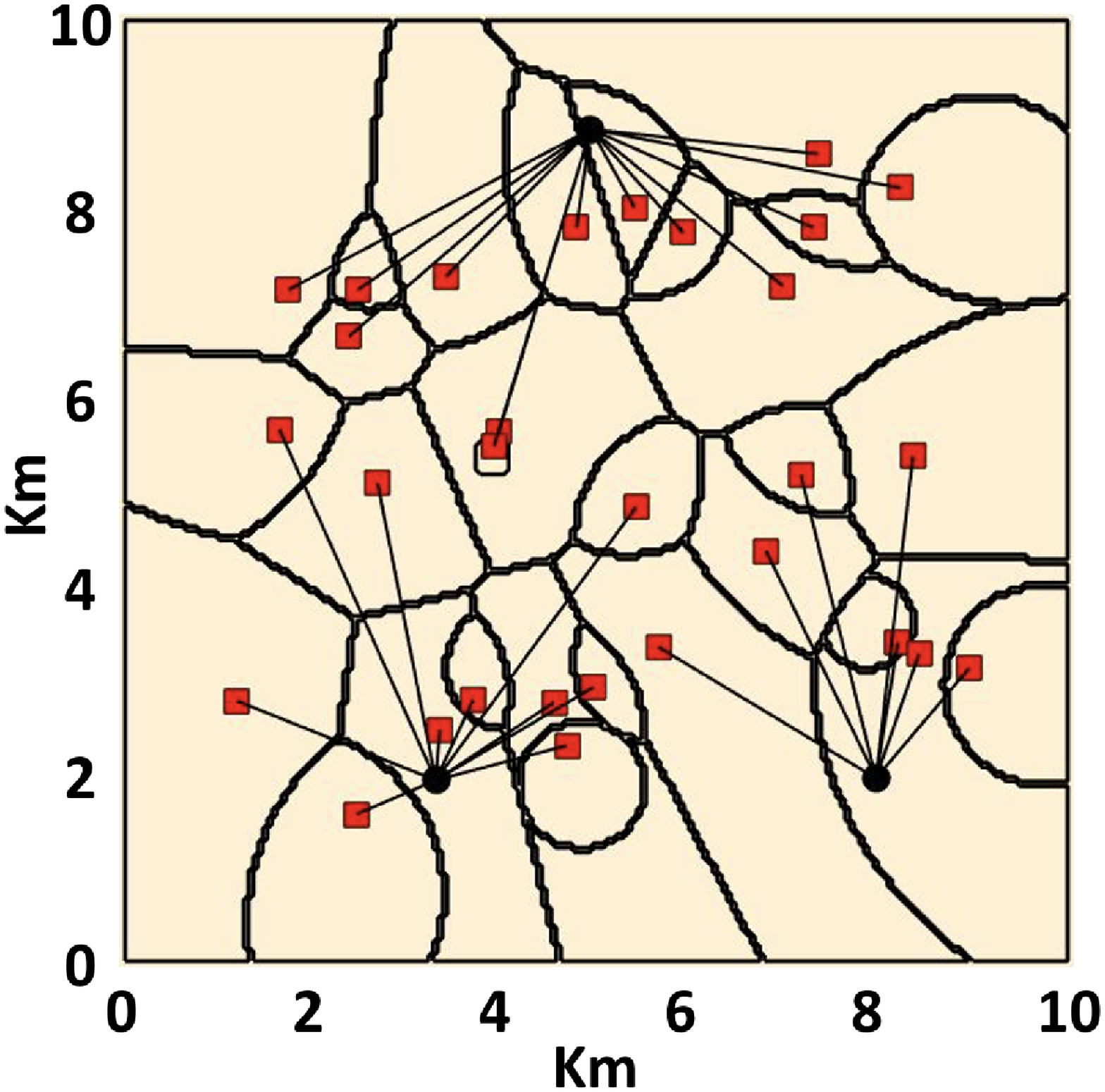}}
\label{Gaussian_mobile_individual_sample_deployment_OMF}
\hspace{2mm}
\subfloat[]{\includegraphics[width=37.5mm]{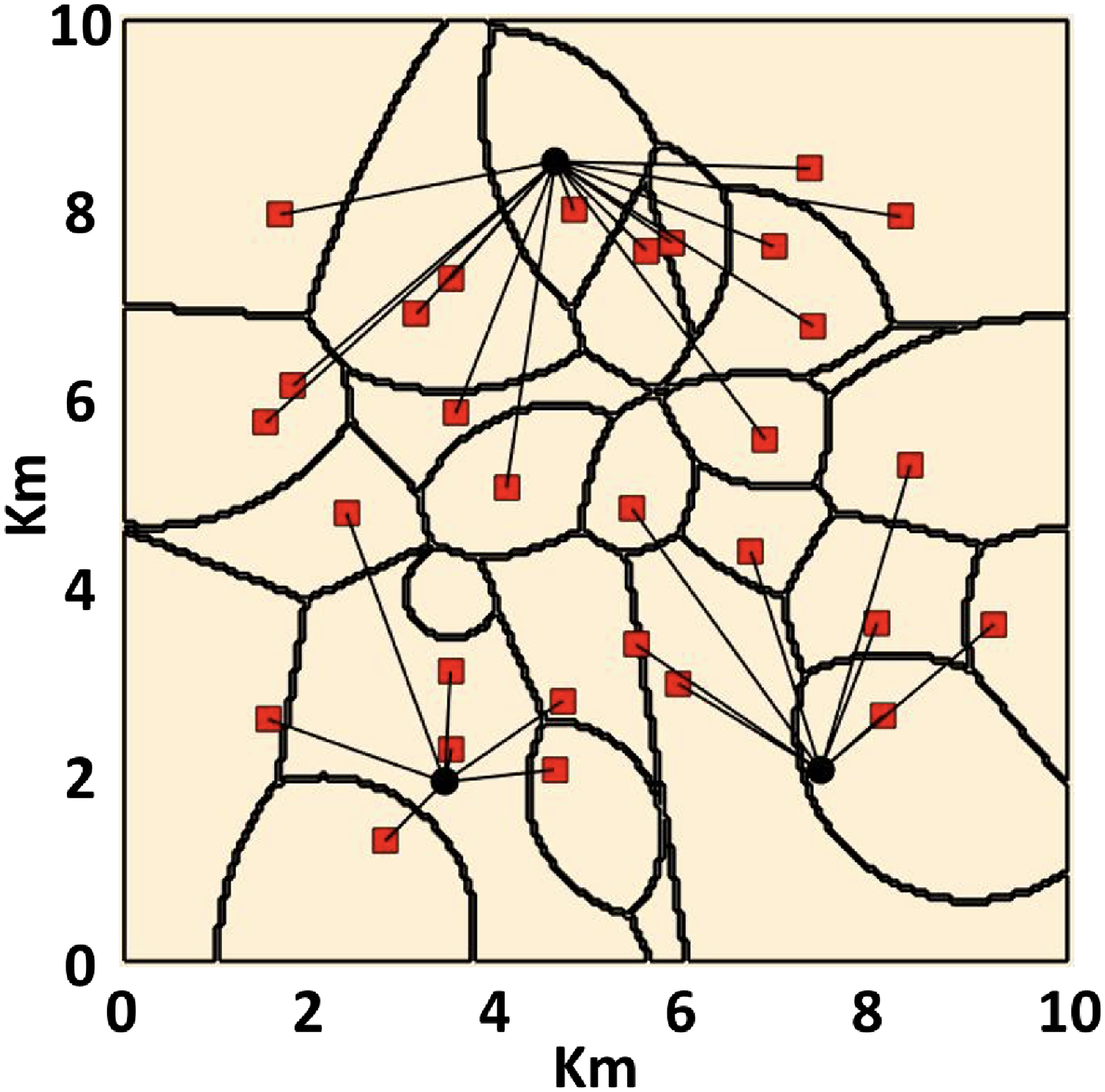}}
\label{Gaussian_mobile_individual_sample_deployment_VFA}
\hspace{2mm}
\subfloat[]{\includegraphics[width=37.5mm]{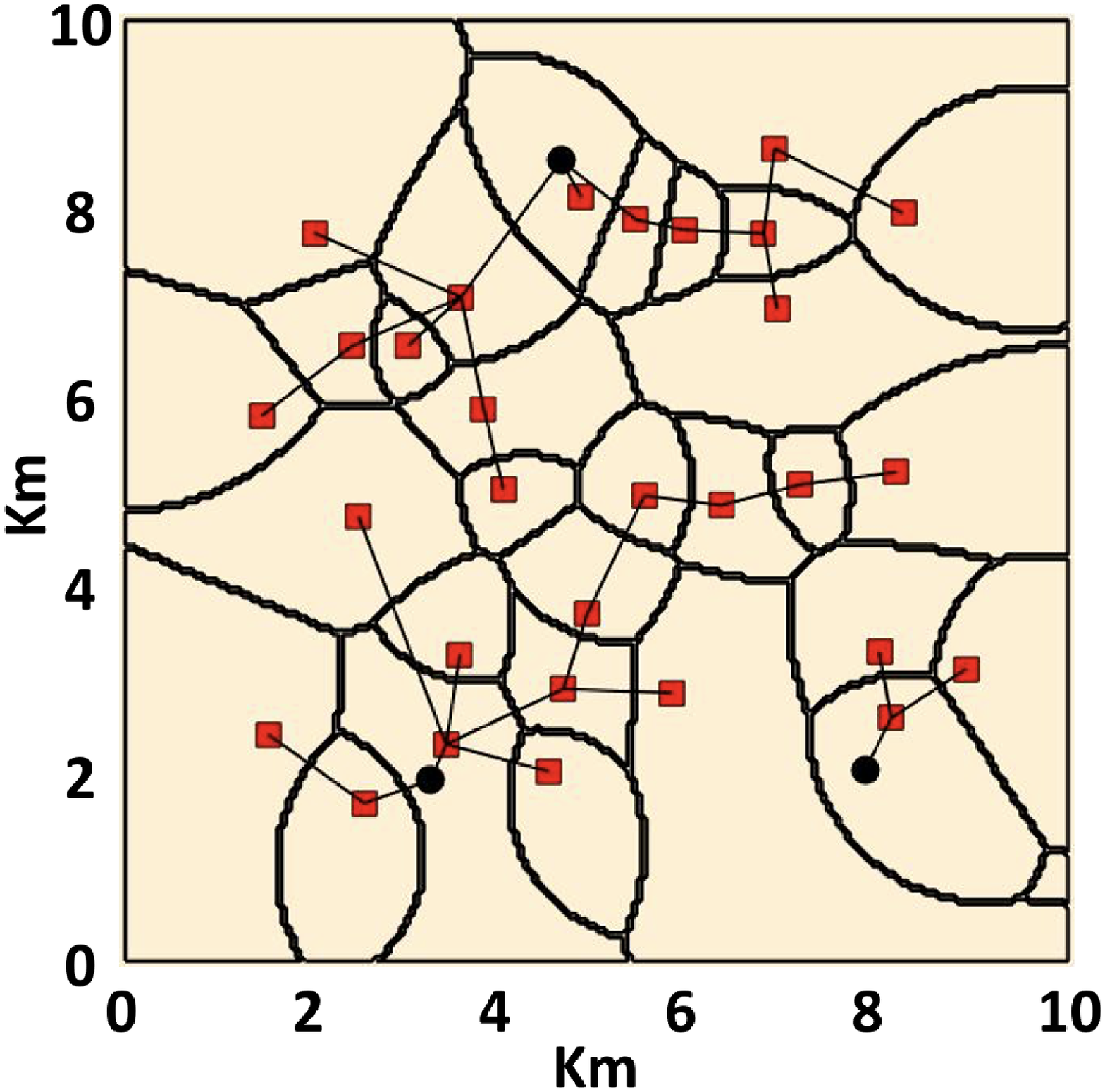}}
\label{Gaussian_mobile_individual_sample_deployment_LORL}

\vspace{-2mm}
\captionsetup{justification=justified}
\caption{\small{Node deployment for different algorithms and the mixture of Gaussian sensor density function. (a) Lloyd-$\alpha$ (b) OMF (c) VFA (d) LORL.}}
\label{Gaussian_mobile_individual_sample_deployment}
\end{figure}

Table \ref{weighted_power_mobile_individual_gaussian} also summarizes the weighted power consumption of different algorithms for the mixture of Gaussian sensor density function given in Eq. (\ref{Mixture_of_Gaussian_Distribution}). The LORL algorithm achieves a power consumption of $9.59$ Watts and outperforms other methods. Figure \ref{Gaussian_mobile_individual_sample_deployment} shows the final node deployment for different algorithms.

\begin{table}[!bth]
\centering
  \caption{Weighted power comparison for the mixture of Gaussian sensor density function}
  \vspace{-3mm}
\begin{tabular}{cccc}
 \toprule
{\bf Lloyd-$\alpha$}&{\bf OMF}&{\bf VFA} &  {\bf LORL}    \\
        $17.24$     & $20.12$ & $14.60$  &$\mathbf{9.59}$ \\ 
 \bottomrule 
\end{tabular}
  \label{weighted_power_mobile_individual_gaussian}
\end{table}

The sum of individual movement energies in Table \ref{gamma_ns}, i.e. $\sum_{i=1}^{N+M}\gamma_i$, is equal to the value of $\gamma$ in Section \ref{Mobile_Total_Movement_Energy_Simulations}. In other words, Table \ref{gamma_ns} represents one exemplary distribution of the total movement energy $\gamma$ among the AP and FC nodes; however, it is different from the optimal energy allocation provided by the MERL algorithm in Section \ref{Mobile_Total_Movement_Energy_Simulations}. 
The results in Tables \ref{weighted_power_mobile_total_uniform}, \ref{weighted_power_mobile_total_gaussian}, \ref{weighted_power_mobile_individual_uniform} and \ref{weighted_power_mobile_individual_gaussian}
verify that the MERL algorithm achieves a lower total power consumption compared to the LORL algorithm although it does not guarantee any individual power constraint.

\section{Conclusion}\label{Conclusion}

In this work, a heterogeneous multi-hop wireless sensor network is discussed where data is collected from densely deployed sensors and transferred to heterogeneous fusion centers using heterogeneous access points as relay nodes. We modeled the minimum communication power consumption of such networks as an optimization problem, and studied the necessary conditions of optimal deployment under both static and mobile network settings. A novel generalized Voronoi diagram is proposed to provide the best cell partition for the heterogeneous multi-hop network. When manual deployment is feasible, the necessary conditions of optimal deployment are explored under the static network setup, and accordingly a Routing-aware Lloyd 
algorithm is proposed  to deploy nodes. However, when static placement is not doable, the necessary conditions of the optimal deployment are studied under a mobile network setting where nodes move from their initial locations to their optimal positions. We consider both total and individual movement energy constraints and formulate them as resource allocation and lifetime optimizations, respectively. Based on the derived necessary conditions, we propose Movement-Efficient Routing-aware Lloyd 
and Lifetime-Optimized Routing-aware Lloyd 
algorithms to deploy nodes under total and individual energy constraints, respectively. Simulation results show that our proposed RL, MERL, and LORL algorithms significantly save communication power in such networks and provide superior results compared to other methods in the literature.


%

\appendices

\section{}\label{proof-power-coef-equality}

Proof of Lemma \ref{power-coef-equality}: 
The AP power coefficient $g_n\left(\mathbf{P}, \mathbf{S}\right)$ defined in Eq. (\ref{power-coefficient}) is the power consumption for transmitting $1$ bit data from AP $n$ to the FCs. This includes both the transmission power at each node, including AP $n$, on the paths connecting AP $n$ to the FCs, and the receiver power at each node, excluding AP $n$, on the paths connecting AP $n$ to the FCs. Since $R_b\int_{W_n}f(\omega)d\omega$ is the total amount of data collected by AP $n$ from sensors within the region $W_n$ in a unit time, the term $g_n\left(\mathbf{P}, \mathbf{S}\right)R_b\int_{W_n}f(\omega)d\omega$ is the required communication power for transmitting the sensory data collected within the region $W_n$ from AP $n$ to the FCs. Hence, the left-hand-side of Eq. (\ref{power-coef-equality-eq}) is the required communication power for transmitting the sensory data collected within the target region from APs to FCs. This can be decomposed into the APs' total transmission power in addition to the required receiver power for the data to reach FCs from AP nodes. This proves  Eq. (\ref{power-coef-equality-eq}) since the right-hand-side of  Eq. (\ref{power-coef-equality-eq}) can be rewritten as $\overline{\mathcal{P}}^T_{\mathcal{A}} + \sum_{i=1}^{N}\sum_{j=1}^{N}\rho_j F_{i,j}\left(\mathbf{W},\mathbf{S}\right)$, i.e. the sum of APs' total transmission power and the receiver power for all links $(i,j)$ connecting AP $i$ and AP $j$. $\hfill\blacksquare$

\section{}\label{proof-voronoi-optimality}

Proof of Proposition \ref{optimal-cell-partitioning}:
Using Eq. (\ref{rewrite-objective-function}), we have:
\begin{align}\label{proof-voronoi-optimality-equations}
    \mathcal{D}\left(\mathbf{P}, \mathbf{W}, \mathbf{S} \right) &= \sum_{n=1}^{N} \int_{W_n} \left(\eta_n \|p_n - \omega \|^2 R_b + \lambda g_n\left(\mathbf{P}, \mathbf{S} \right)R_b + \lambda \rho_n R_b  \right) f(\omega)d\omega \nonumber \\
    & \geq \sum_{n=1}^{N} \int_{W_n} \min_j \left(\eta_j \|p_j - \omega \|^2 R_b + \lambda g_j\left(\mathbf{P}, \mathbf{S} \right)R_b + \lambda \rho_j R_b  \right)  f(\omega)d\omega \nonumber \\
    &= \int_{\Omega} \min_j \left(\eta_j \|p_j - \omega \|^2 R_b + \lambda g_j\left(\mathbf{P}, \mathbf{S} \right)R_b + \lambda \rho_j R_b  \right)  f(\omega)d\omega \nonumber \\
    &= \sum_{n=1}^{N} \int_{\mathcal{V}_n} \min_j \left(\eta_j \|p_j - \omega \|^2 R_b + \lambda g_j\left(\mathbf{P}, \mathbf{S} \right)R_b + \lambda \rho_j R_b  \right)  f(\omega)d\omega \nonumber \\
    &= \sum_{n=1}^{N} \int_{\mathcal{V}_n} \left(\eta_n \|p_n - \omega \|^2 R_b + \lambda g_n\left(\mathbf{P}, \mathbf{S} \right)R_b + \lambda \rho_n R_b  \right)  f(\omega)d\omega \nonumber \nonumber \\
    &= \mathcal{D}\left(\mathbf{P},\mathcal{V}\left(\mathbf{P},\mathbf{S} \right),\mathbf{S} \right). 
\end{align}
Hence, the generalized Voronoi diagram provides the optimal cell partitioning for any given node deployment $\mathbf{P}$ and normalized flow matrix $\mathbf{S}$. $\hfill \blacksquare$

\section{}\label{proof-necessary-condition-optimal-deployment}

Proof of Proposition \ref{necessary-condition-optimal-deployment}:
Eq. (\ref{necessary-condition-optimal-deployment-partitioning}) is a direct implication of Proposition \ref{optimal-cell-partitioning}. Eq. (\ref{necessary-condition-optimal-deployment-S}) is directly followed from Eq. (\ref{bellman-ford-functionality}). Here, we prove Eq. (\ref{necessary-condition-optimal-deployment-APs})  for the optimal locations of APs and FCs.
First, we study the shape of the Voronoi regions in (\ref{voronoi-cell}). Let $\mathcal{B}(c,r)=\{\omega|\|\omega-c\|\leq r\}$ be a disk centered at $c$ with radius $r$ in two-dimensional space. In particular, $\mathcal{B}(c,r)=\varnothing$ when $r < 0$. Let $\mathcal{HS}(a, b) = \left\{\omega|a\cdot\omega + b\leq 0\right\}$ be a half space, where $a\in\mathbb{R}^2$ is a vector and $b\in\mathbb{R}$ is a constant. For $i,j \in \mathcal{I_A}$, we define 
\begin{equation}
    \mathcal{V}_{ij}(\mathbf{P}, \mathbf{S})\triangleq\{\omega|\eta_i \|p_i - \omega \|^2  + \lambda g_i\left(\mathbf{P}, \mathbf{S} \right) + \lambda \rho_i  \leq  \eta_j \|p_j - \omega \|^2  + \lambda g_j\left(\mathbf{P}, \mathbf{S} \right) + \lambda \rho_j  \}
\label{Vij}
\end{equation}
to be the pairwise Voronoi region of AP $i$ where only APs $i$ and $j$ are considered. Then, AP $i$'s Voronoi region can be represented as $\mathcal{V}_i(\mathbf{P}, \mathbf{S}) = \left[\bigcap_{j\neq i}\mathcal{V}_{ij}(\mathbf{P},\mathbf{S})\right]\bigcap\Omega$. By expanding (\ref{Vij}) and straightforward algebraic calculations, the pairwise Voronoi region $\mathcal{V}_{ij}$ is derived as:
\begin{align}
    V_{ij}=\Omega\cap\begin{cases}
         \mathcal{HS}\left(a_{ij}, b_{ij}\right)& \quad ,  \eta_i=\eta_j  \\
        \mathcal{B}\left(c_{ij}, r_{ij}\right) &\quad , \eta_i>\eta_j , L_{ij}\geq 0\\
        \varnothing & \quad ,  \eta_i>\eta_j , L_{ij}< 0\\
        \mathcal{B}^c\left(c_{ij}, r_{ij}\right) &\quad , \eta_i<\eta_j , L_{ij} \geq 0 \\
        \mathbb{R}^2 & \quad , \eta_i<\eta_j , L_{ij}< 0
        \end{cases},
\end{align}
where $a_{ij} = \eta_j p_j - \eta_i p_i$, $b_{ij}=\frac{\left(\eta_i\|p_i\|^2-\eta_j\|p_j\|^2+\lambda g_i\left(\mathbf{P}, \mathbf{S} \right) + \lambda \rho_i - \lambda g_j\left(\mathbf{P}, \mathbf{S} \right) - \lambda \rho_j\right)}{2}$, $c_{ij}=\frac{\eta_ip_i-\eta_jp_j}{\eta_i-\eta_j}$, $L_{ij} = \frac{\eta_i\eta_j\|p_i-p_j\|^2}{\left(\eta_i-\eta_j\right)^2} -\lambda \times \frac{g_i\left(\mathbf{P}, \mathbf{S} \right) + \rho_i - g_j\left(\mathbf{P}, \mathbf{S} \right) - \rho_j}{(\eta_i-\eta_j)}$, $r_{ij} = \sqrt{\max\left(L_{ij}, 0\right)}$, and $\mathcal{B}^c(c_{ij},r_{ij})$ is the complementary of  $\mathcal{B}(c_{ij},r_{ij})$. Note that for two distinct indices such as $i,j\in \mathcal{I_A}$, if $\eta_i>\eta_j$ and $L_{ij}<0$, then two regions $\Omega\cap\mathcal{B}(c_{ij},r_{ij})$ and $\varnothing$ differ only in the point $c_{ij}$. Similarly, for $\eta_i<\eta_j$ and $L_{ij}<0$, two regions $\Omega\cap\mathcal{B}^c(c_{ij},r_{ij})$ and $\Omega$ differ only in the point $c_{ij}$. If we define:
\begin{equation}\label{voronoi_v_n}
    \overline{V}_k = \left [ \bigcap_{i:\eta_k>\eta_i} \mathcal{B}(c_{ki},r_{ki})  \right] \bigcap \left[\bigcap_{i:\eta_k=\eta_i} \mathcal{HS}(a_{ki},b_{ki}) \right] \bigcap \left[ \bigcap_{i:\eta_k<\eta_i} \mathcal{B}^c(c_{ki},r_{ki})  \right] \bigcap \Omega,
\end{equation}
then two regions $\overline{V}_k$ and $V_k$ differ only in finite number of points. As a result, integrals over both $\overline{V}_k$ and $V_k$ have the same value since the density function $f$ is continuous and differentiable, and removing finite number of points from the integral region does not change the integral value. Note that if $V_k$ is empty, the Proposition 1 in \cite{guo2016sensor} holds since the integral over an empty region is zero. If $V_k$ is not empty, the same arguments as in Appendix A of \cite{guo2016sensor} can be replicated since $\overline{V}_k$ in (\ref{voronoi_v_n}) is similar to (31) in \cite{guo2016sensor}.

Using parallel axis theorem \cite{paul1979kinematics}, the heterogeneous multi-hop communication power consumption can be written as:
\begin{align}
& \mathcal{D}\left(\mathbf{P}, \mathbf{W}, \mathbf{S} \right) =    \sum_{n=1}^{N} \int_{W_n} \eta_n \|c_n - \omega \|^2 R_b f(\omega)d\omega + \sum_{n=1}^{N}\eta_n\|p_n - c_n\|^2R_b v_n \nonumber \\ & + \lambda\sum_{i=1}^{N}\sum_{j=1}^{N+M} \beta_{i,j} \|p_i - p_j\|^2 F_{i,j}\left(\mathbf{W}, \mathbf{S} \right)  + \lambda\sum_{n=1}^{N}\rho_n \left[ \sum_{i=1}^{N} F_{i,n}\left(\mathbf{W}, \mathbf{S} \right) + R_b\int_{W_n}f(\omega)d\omega \right], \label{objective-after-parallel-axis-theorem}
\end{align}
where $v_n=v\left(W_n\right)$ and $c_n$ are the volume and centroid of the region $W_n$, respectively.
Using Proposition 1 in \cite{guo2016sensor}, since the optimal deployment $\mathbf{P}^*$ should have a zero gradient, we take the partial derivatives of (\ref{objective-after-parallel-axis-theorem}) with respect to node locations. For each $i\in \mathcal{I}_{A}$, we have
\begin{equation}\label{pd1}
    \frac{\partial \mathcal{D}}{\partial p^*_i} =
    2\eta_i(p^*_i-c^*_i)R_bv^*_i +2\lambda\sum\limits_{j=1}^{N+M}\beta_{i,j}(p^*_i-p^*_j)F^*_{i,j}
    +2\lambda\sum\limits_{j=1}^{N}\beta_{j,i}(p^*_i-p^*_j)F^*_{j,i} = 0,
\end{equation}
and for each $i\in\mathcal{I_F}$, we have
\begin{equation}\label{pd2}
    \frac{\partial \mathcal{D}}{\partial p^*_i} =
    2\lambda\sum\limits_{j=1}^{N}\beta_{j,i}(p^*_i-p^*_j)F^*_{j,i} = 0.
\end{equation}
By solving Eqs. (\ref{pd1}) and (\ref{pd2}), we obtain Eq. (\ref{necessary-condition-optimal-deployment-APs}) and the proof is complete. $\hfill\blacksquare$

\section{}\label{proof-RL-convergence}

Proof of Proposition \ref{RL-convergence}:
Note that RL Algorithm iterates between three steps. In what follows, we show that none of these steps will increase the objective function $\mathcal{D}\left(\mathbf{P}, \mathbf{W}, \mathbf{S}\right)$. For a fixed node deployment $\mathbf{P}$ and normalized flow matrix $\mathbf{S}$, the cell partitioning $\mathbf{W}$ is updated according to Eq. (\ref{necessary-condition-optimal-deployment-partitioning}) which was shown to be optimal for a given $\mathbf{P}$ and $\mathbf{S}$ in Proposition \ref{optimal-cell-partitioning}. Therefore, the first step of RL Algorithm does not increase the objective function. Next, since $\mathcal{R}\left(\mathbf{P},\mathbf{W}\right)$ is the optimal normalized flow matrix for a given node deployment $\mathbf{P}$ and cell partitioning $\mathbf{W}$, the second step of RL Algorithm does not increase the objective function either. Finally, note that when $\mathbf{W}$, $\mathbf{S}$ and $\left\{p_j \right\}_{j\neq i}$ are fixed, the objective function $\mathcal{D}\left(\mathbf{P}, \mathbf{W}, \mathbf{S}\right)$ in Eq. (\ref{total-power-consumption}) is a convex function of the node position $p_i$; hence, by solving the zero-gradient equations and updating the node locations according to the Eq. (\ref{necessary-condition-optimal-deployment-APs}), the objective function does not increase. Therefore, the objective function of RL Algorithm is nonincreasing. In addition, the objective function is lower bounded by $0$, i.e., $\mathcal{D}\left(\mathbf{P},\mathbf{W},\mathbf{S}\right)\geq 0$. As a result, RL Algorithm is an iterative improvement algorithm and it converges. $\hfill\blacksquare$

\section{}\label{optimal-between-initial-z}

Proof of Lemma \ref{total-movment-constraint-between-initial-and-optimal}:
Before going through the proof, we state the following lemma: 

\begin{lemma}\label{geometric-locus-weighted-sum-square-distances}
Given a set of points $q_i\in\mathbb{R}^2$ and non-negative scalar weights $a_i$ for $i\in\{1,\cdots,K\}$, and a scalar $m$, the geometric locus of the point $p\in\mathbb{R}^2$ such that the equality
\begin{equation}\label{geometric-locus-weighted-sum-square-distances-eq}
    \sum_{i=1}^K a_i \|p - q_i\|^2 = m
\end{equation}
holds, is either an empty set, a single point, or a circle centered at the point $c = \frac{\sum_{i=1}^K a_i q_i}{\sum_{i=1}^K a_i}$.
\end{lemma}
Proof: Let $p=\left(p_x, p_y\right)$ and $q_i = \left( q_{i,x}, q_{i,y}\right)$. Then, we can rewrite Eq. (\ref{geometric-locus-weighted-sum-square-distances-eq}) as
\begin{align}\label{expand-equation}
    \left(\sum_{i=1}^K a_i \right) \left( p_x^2 + p_y^2 \right) - 2\left(\sum_{i=1}^K a_i q_{i,x} \right) p_x - 2\left(\sum_{i=1}^K a_i q_{i,y} \right) p_y = m - \sum_{i=1}^K a_i \|q_i\|^2 .
\end{align}
By manipulating both sides, we can rewrite  Eq. (\ref{expand-equation}) as follows:
\begin{align}\label{manipulated-expanded-equation}
    \left[p_x - \frac{\sum\limits_{i=1}^K a_i q_{i,x}}{\sum\limits_{i=1}^K a_i} \right]^2 \!+\! \left[p_y - \frac{\sum\limits_{i=1}^K a_i q_{i,y}}{\sum\limits_{i=1}^K a_i} \right]^2 \!=\! \frac{m - \sum\limits_{i=1}^K a_i \|q_i\|^2}{\sum\limits_{i=1}^K a_i}  + \frac{\left(\sum\limits_{i=1}^K a_i q_{i,x} \right)^2 \!+\!  \left(\sum\limits_{i=1}^K a_i q_{i,y} \right)^2}{\left(\sum\limits_{i=1}^K a_i\right)^2}.
\end{align}
Hence, the geometric locus of the point $p=\left(p_x, p_y\right)$ is an empty set or a single point if the right-hand-side of Eq. (\ref{manipulated-expanded-equation}) is negative or zero, respectively; otherwise, the geometric locus is a circle centered at the point $c = \frac{\sum_{i=1}^K a_i q_i}{\sum_{i=1}^K a_i}$ with the radius $r = \sqrt{\frac{m - \sum_{i=1}^K a_i \|q_i\|^2}{\sum_{i=1}^K a_i}  + \frac{\left(\sum_{i=1}^K a_i q_{i,x} \right)^2 +  \left(\sum_{i=1}^K a_i q_{i,y} \right)^2}{\left(\sum_{i=1}^K a_i\right)^2}}$, and Lemma \ref{geometric-locus-weighted-sum-square-distances} is proved.

\begin{corollary}\label{point_inside_circle}
If the geometric locus in Lemma \ref{geometric-locus-weighted-sum-square-distances} is a circle centered at $c$ with radius $r$, then for any point $p$ within this circle we have $\sum_{i=1}^{K}a_i \|p - q_i\|^2 < m$, i.e. moving the point $p$ inside this circle  reduces the weighted squared sum in Eq. (\ref{geometric-locus-weighted-sum-square-distances-eq}).
\end{corollary}

Now, assume that there exists at least one node, say $n$, for which Eq. (\ref{total-movment-constraint-between-initial-and-optimal-eq1}) in Lemma \ref{total-movment-constraint-between-initial-and-optimal} does not hold for an optimal node deployment $\mathbf{P}^*$, cell partitioning $\mathbf{W}^*$ and normalized flow matrix $\mathbf{S}^*$, i.e. $p^*_n$ does not lie on the segment $\overline{z^*_n\tilde{p}_n}$. We aim to find another deployment such as $\mathbf{P}'$, $\mathbf{W}'$ and $\mathbf{S}'$ so that $E\left(\mathbf{P}' \right)\leq \gamma$ and $\mathcal{D}\left(\mathbf{P}',\mathbf{W}',\mathbf{S}'\right) < \mathcal{D}\left(\mathbf{P}^*,\mathbf{W}^*, \mathbf{S}^* \right)$; hence, contradicting the optimality assumption of $\mathbf{P}^*$, $\mathbf{W}^*$ and $\mathbf{S}^*$, and concluding that Eq. (\ref{total-movment-constraint-between-initial-and-optimal-eq1}) holds for all nodes. For this purpose, let $\mathbf{W}' = \mathbf{W}^*$, $\mathbf{S}' = \mathbf{S}^*$ and $p'_i = p^*_i$ for all $i\in\mathcal{I_A}\bigcup\mathcal{I_F}\backslash \{n\}$. We aim to determine the node location $p'_n$ accordingly.
Using the parallel axis theorem \cite{paul1979kinematics}, we can rewrite  $\mathcal{D}\left(\mathbf{P}^*,\mathbf{W}^*,\mathbf{S}^* \right)$ as:
\begin{align}\label{proof-lemma2-parallel-axis}
    \mathcal{D}\left(\mathbf{P}^*,\mathbf{W}^*,\mathbf{S}^*\right) & = \sum_{i=1}^{N}\int_{W^*_i}\eta_i \|c^*_i - \omega\|^2 R_b f(\omega)d\omega + \sum_{i=1}^{N}\eta_i R_b v^*_i \|p^*_i - c^*_i\|^2 \nonumber \\& + \lambda \sum_{i=1}^{N}\sum_{j=1}^{N+M} \beta_{i,j} \|p^*_i - p^*_j\|^2 F_{i,j}\left(\mathbf{W}^*, \mathbf{S}^* \right) + \lambda\overline{\mathcal{P}}^R_{\mathcal{A}}\left(\mathbf{W}^*,\mathbf{S}^*\right),
\end{align}
where $v^*_i$ and $c^*_i$ are the volume and centroid of the region $W^*_i$, respectively. In what follows, we assume that $n\in\mathcal{I_A}$, i.e. node $n$ is an AP. Similar proof can be carried out for $n\in\mathcal{I_F}$. Note that Eq. (\ref{proof-lemma2-parallel-axis}) can be split
as $\mathcal{D}\left(\mathbf{P}^*,\mathbf{W}^*,\mathbf{S}^*\right) = \mathcal{D}_1\left(\mathbf{P}^*,\mathbf{W}^*,\mathbf{S}^*\right) + \mathcal{D}_2\left(\mathbf{P}^*,\mathbf{W}^*,\mathbf{S}^*\right)$, where 
\begin{align}\label{D1}
    \mathcal{D}_1\!\left(\mathbf{P}^*,\mathbf{W}^*,\mathbf{S}^*\right) \!=\! \eta_n R_b v^*_n \|p^*_n \!-\! c^*_n \|^2 +\! \sum_{j=1}^{N+M}\lambda\beta_{n,j}F^*_{n,j} \|p^*_n \!-\! p^*_j \|^2 +\! \sum_{j=1}^{N}\lambda\beta_{j,n}F^*_{j,n} \|p^*_n \!-\! p^*_j \|^2,
\end{align}
i.e. $\mathcal{D}_1$ includes those terms in Eq. (\ref{proof-lemma2-parallel-axis}) that involve $p^*_n$. In particular, regardless of the node $n$'s position, we have $\mathcal{D}_2\left(\mathbf{P}^*,\mathbf{W}^*,\mathbf{S}^* \right) = \mathcal{D}_2\left(\mathbf{P}',\mathbf{W}',\mathbf{S}' \right)$. According to Lemma \ref{geometric-locus-weighted-sum-square-distances}, the geometric locus of points such as $p^*_n$ for which the value of $\mathcal{D}_1\left(\mathbf{P}^*,\mathbf{W}^*,\mathbf{S}^* \right)$ in Eq. (\ref{D1}) remains the same is a circle $\Phi^*_n$ centered at the point $z^*_n = z_n\left(\mathbf{P}^*,\mathbf{W}^*,\mathbf{S}^* \right)$ defined in Eq. (\ref{z1-AP}), with radius $r^*_n = \|z^*_n - p^*_n \|$. Note that if $\|z^*_n - \tilde{p}_n\| < \|z^*_n - p^*_n\|$, then setting $p'_n=\tilde{p}_n$ not only leads to the movement energy $E\left(\mathbf{P}' \right)<E\left(\mathbf{P}^*\right)$, but also results in $\mathcal{D}_1\left(\mathbf{P}',\mathbf{W}',\mathbf{S}' \right) < \mathcal{D}_1\left(\mathbf{P}^*,\mathbf{W}^*,\mathbf{S}^* \right)$ since $p'_n$ lies inside $\Phi^*_n$. Therefore, we have $\mathcal{D}\left(\mathbf{P}',\mathbf{W}',\mathbf{S}' \right) < \mathcal{D}\left(\mathbf{P}^*,\mathbf{W}^*,\mathbf{S}^* \right)$ which is in contradiction with the optimality of $\mathbf{P}^*$, $\mathbf{W}^*$ and $\mathbf{S}^*$; hence, we have $\|z^*_n - \tilde{p}_n\| \geq \|z^*_n - p^*_n \|$. Let $\hat{p}_n$ be the intersection point of the circle $\Phi^*_n$ and segment $\overline{z^*_n \tilde{p}_n}$. Since $\|\tilde{p}_n - \hat{p}_n\| < \|\tilde{p}_n - p^*_n \|$, there exists an $\epsilon_n\in\mathbb{R}^+$ such that $\|\tilde{p}_n - \hat{p}_n\| + \epsilon_n < \|\tilde{p}_n - p^*_n \|$. If $p'_n = \hat{p}_n + \epsilon_n\times\frac{z^*_n - \hat{p}_n}{\|z^*_n - \hat{p}_n\|}$, then not only we have $E\left(\mathbf{P}'\right) < E\left(\mathbf{P}^*\right)$ since $E\left(\mathbf{P}^*\right) - E\left(\mathbf{P}'\right) > \zeta_n \epsilon_n > 0 $, but also $\mathcal{D}_1\left(\mathbf{P}',\mathbf{W}',\mathbf{S}' \right) < \mathcal{D}_1\left(\mathbf{P}^*,\mathbf{W}^*,\mathbf{S}^* \right)$ since $p'_n$ lies inside the circle $\Phi^*_n$. Therefore, we have $\mathcal{D}\left(\mathbf{P}',\mathbf{W}',\mathbf{S}' \right) < \mathcal{D}\left(\mathbf{P}^*,\mathbf{W}^*,\mathbf{S}^* \right)$ which contradicts the optimality of $\mathbf{P}^*$, $\mathbf{W}^*$ and $\mathbf{S}^*$ and concludes the proof. $\hfill\blacksquare$

\section{}\label{proof-necessary-condition-total-constraint}

Proof of Proposition \ref{necessary-condition-total-move-constraint}:
If $p^*_i = z^*_i$ for all $i\in\mathcal{I}_d$, then Eq. (\ref{movement-constraint}) implies that $E\left(\mathbf{P}^*\right) = \sum_{i\in\mathcal{I}_d}\zeta_i \|\Gamma^*_i\| \leq \gamma$; hence, Eq. (\ref{necessary-condition-total-move-constraint-eq2}) reduces to the trivial statement $p^*_n = \tilde{p}_n + \Gamma^*_n$ and the proof is complete. Therefore, we assume that there exists at least one node, say $n$, for which $p^*_n \neq z^*_n$. Note that if any residual movement energy is left in the optimal deployment, i.e. $E\left(\mathbf{P}^*\right)<\gamma$, then there exists an $\epsilon\in\mathbb{R}^+$ such that $E\left(\mathbf{P}^*\right) + \epsilon < \gamma$ and $\overline{p}_n=p^*_n + \epsilon\times\frac{z^*_n-p^*_n}{\|z^*_n-p^*_n\|}$ lies inside the circle centered at $z^*_n$ and radius $\|z^*_n - p^*_n \|$. Then, according to Lemma \ref{geometric-locus-weighted-sum-square-distances} and Corollary \ref{point_inside_circle}, by fixing the cell partitioning, normalized flow matrix and the location of all nodes except Node $n$, and placing Node $n$ at $\overline{p}_n$ we can achieve a lower total multi-hop communication power without exhausting the available movement energy, which contradicts the optimality of $\mathbf{P}^*$, $\mathbf{W}^*$ and $\mathbf{S}^*$. Therefore, $p^*_n\neq z^*_n$ implies that $E\left(\mathbf{P}^*\right)=\gamma$. Now, given the optimal node deployment $\mathbf{P}^*$, $\mathbf{W}^*$ and $\mathbf{S}^*$, we construct the node deployment $\mathbf{P}'$, $\mathbf{W}'$ and $\mathbf{S}'$ as follows. Let $\mathbf{W}' = \mathbf{W}^*$, $\mathbf{S}' = \mathbf{S}^*$ and $p'_i = p^*_i$ for all $i\in\mathcal{I_A}\bigcup\mathcal{I_F}\backslash\{m, n\}$. Let $\epsilon_m, \epsilon_n \in \mathbb{R}^+$ be small values and define
\begin{equation}\label{proof-eq1}
    p'_m = p^*_m - \epsilon_m\times \frac{z^*_m - \tilde{p}_m}{\|z^*_m - \tilde{p}_m\|} \qquad , \qquad
    p'_n = p^*_n + \epsilon_n\times \frac{z^*_n - \tilde{p}_n}{\|z^*_n - \tilde{p}_n\|}.
\end{equation}
To satisfy the equality $E\left(\mathbf{P}'\right) = \gamma$, we have $\zeta_n \epsilon_n = \zeta_m \epsilon_m$.
Now, we calculate the change in the multi-hop communication power, i.e. $\mathcal{D}\left(\mathbf{P}',\mathbf{W}',\mathbf{S}'\right) - \mathcal{D}\left(\mathbf{P}^*,\mathbf{W}^*,\mathbf{S}^*\right)$. Assume that Node $m$ is fixed at $p^*_m$ and we move Node $n$ from $p^*_n$ to $p'_n$. Note that this movement only changes the term $\mathcal{D}_1$ defined in Eq. (\ref{D1}); thus, according to Lemma \ref{geometric-locus-weighted-sum-square-distances} and Eq. (\ref{manipulated-expanded-equation}), this change is proportional to the difference between the squared radii, i.e.
\begin{equation}\label{delta_1}
    \Delta_1 = \left[\|p'_n - z^*_n \|^2 - \|p^*_n - z^*_n \|^2 \right]\times \psi^*_n,
\end{equation}
where $\psi^*_n$ is defined in Eq. (\ref{psi}). Now, with Node $n$ placed at $p'_n$, we move Node $m$ from $p^*_m$ to $p'_m$. Similar to the above argument, the term $\Delta_2$ defined as
\begin{equation}\label{delta_2}
    \Delta_2 = \left[\|p'_m - z^*_m \|^2 - \|p^*_m - z^*_m \|^2 \right]\times \psi^*_m
\end{equation}
captures the change in $\mathcal{D}$ with the assumption that Node $n$ was located at $p^*_n$. Now, we take into account that Node $n$ was located at $p'_n$ instead of $p^*_n$ during Node $m$'s movement. 
\begin{align}\label{delta_3}
    \Delta_3 &= \lambda \beta_{n,m} F^*_{n,m}\times\left[\left(\|p'_n - p'_m \|^2 - \|p'_n - p^*_m \|^2 \right) - \left(\|p^*_n - p'_m \|^2 - \|p^*_n - p^*_m\|^2 \right) \right] \\
    &= \lambda \beta_{n,m} F^*_{n,m}\times\big[\left(\|p'_n - p^*_m \|^2 + \epsilon_m^2 - 2\epsilon_m\|p'_n - p^*_m \| \cos{\measuredangle p'_n p^*_m p'_m} - \|p'_n - p^*_m \|^2 \right) \nonumber \\ &- \left(\|p^*_n - p'_m\|^2 - \|p^*_n - p'_m \|^2 - \epsilon_m^2 - 2\epsilon_m \|p^*_n - p'_m\| \cos{\measuredangle p^*_n p'_m \tilde{p}_m} \right)   \big] \label{delta_3_eq_2} \\
    &= \lambda \beta_{n,m} F^*_{n,m}\times\left[2\epsilon_m^2 - 2\epsilon_m\left(\|p'_n - p^*_m \| \cos{\measuredangle p'_n p^*_m p'_m} - \|p^*_n - p'_m\| \cos{\measuredangle p^*_n p'_m \tilde{p}_m} \right) \right] \label{delta_3_eq_3} \\
    &= \lambda \beta_{n,m} F^*_{n,m}\times\left[2\epsilon_m^2 - 2\epsilon_m \left(\epsilon_m - \epsilon_n \cos{\theta} \right) \right] \label{delta_3_eq_4} \\
    &= \lambda \beta_{n,m} F^*_{n,m}\times\left[2\frac{\zeta_m}{\zeta_n}\epsilon_m^2 \cos{\theta} \right], \label{delta_3_eq_5}
\end{align}
where $s$ and $\theta = \measuredangle z^*_n s z^*_m$ are the intersection point and the angle between the lines $\overline{z^*_n\tilde{p}_n}$ and $\overline{z^*_m\tilde{p}_m}$, respectively. Note that in Eq. (\ref{delta_3}), without any loss of generality, we have assumed that the direction of the flow of data, if any, is from Node $n$ to Node $m$. Moreover, Eq. (\ref{delta_3_eq_2}) follows from the law of cosines and Eq. (\ref{delta_3_eq_5}) follows from the equation $\zeta_n\epsilon_n = \zeta_m\epsilon_m$. Hence, we have:
\begin{align}\label{d_deltas}
    &\mathcal{D}\left(\mathbf{P}',\mathbf{W}',\mathbf{S}'\right) - \mathcal{D}\left(\mathbf{P}^*,\mathbf{W}^*,\mathbf{S}^*\right) = \Delta_1 + \Delta_2 + \Delta_3 \\ &= \left[\frac{\zeta_m^2}{\zeta_n^2}\epsilon_m^2 - 2\frac{\zeta_m}{\zeta_n}\epsilon_m \|p^*_n - z^*_n \| \right]\times \psi^*_n  + \left[\epsilon_m^2 + 2\epsilon_m \|p^*_m - z^*_m \| \right] \times \psi^*_m + 2\lambda\beta_{n,m}F^*_{n,m}\frac{\zeta_m}{\zeta_n}\epsilon_m^2 \cos{\theta}. \nonumber
\end{align}
Due to the optimality of $\mathbf{P}^*$, $\mathbf{W}^*$ and $\mathbf{S}^*$, Eq. (\ref{d_deltas}) should be non-negative, or equivalently:
\begin{equation}\label{epsilon_very_small}
    \epsilon_m \left(\frac{\zeta_m^2}{\zeta_n^2}\psi^*_n + \psi^*_m + 2\lambda\beta_{n,m}F^*_{n,m}\frac{\zeta_m}{\zeta_n}\cos{\theta} \right) \geq 2\left(\frac{\zeta_m}{\zeta_n}\psi^*_n \|p^*_n - z^*_n\| - \psi^*_m \|p^*_m - z^*_m \| \right).
\end{equation}
According to Eq. (\ref{psi}), the term $\lambda\beta_{n,m}F^*_{n,m}$ is included in both $\psi^*_n$ and $\psi^*_m$, i.e. $\psi^*_n \geq \lambda\beta_{n,m}F^*_{n,m}$ and $\psi^*_m \geq  \lambda\beta_{n,m}F^*_{n,m}$; therefore, we have:
\begin{align}\label{always_positive}
    \frac{\zeta_m^2}{\zeta_n^2}\psi^*_n + \psi^*_m + 2\lambda\beta_{n,m}F^*_{n,m}\frac{\zeta_m}{\zeta_n}\cos{\theta} &\geq \frac{\zeta_m^2}{\zeta_n^2}\lambda\beta_{n,m}F^*_{n,m} + \lambda\beta_{n,m}F^*_{n,m} + 2\lambda\beta_{n,m}F^*_{n,m}\frac{\zeta_m}{\zeta_n}\cos{\theta} \\&\geq \lambda\beta_{n,m}F^*_{n,m} \left(\frac{\zeta_m}{\zeta_n} - 1 \right)^2 \geq 0,
\end{align}
thus, the term inside the parentheses on the left hand side of Eq. (\ref{epsilon_very_small}) is always non-negative. Note that if the right hand side of Eq. (\ref{epsilon_very_small}) is strictly positive, then we can choose a small enough $\epsilon_m$ such that the inequality in Eq. (\ref{epsilon_very_small}) is contradicted. Hence, we have:
\begin{equation}\label{first_inequality}
    \zeta_m\psi^*_n \|p^*_n - z^*_n\| \leq \zeta_n \psi^*_m \|p^*_m - z^*_m \|.
\end{equation}
By swapping the indices $m$ and $n$ in Eq. (\ref{proof-eq1}) and repeating the same argument, we have:
\begin{equation}\label{second_inequality}
    \zeta_m\psi^*_n \|p^*_n - z^*_n\| \geq \zeta_n \psi^*_m \|p^*_m - z^*_m \|.
\end{equation}
Eqs. (\ref{first_inequality}) and (\ref{second_inequality}) imply that:
\begin{equation}\label{third_equality}
    \zeta_m\psi^*_n \|p^*_n - z^*_n\| = \zeta_n \psi^*_m \|p^*_m - z^*_m \|.
\end{equation}
Note that Eq. (\ref{proof-eq1}) indicates that Eq. (\ref{first_inequality}) holds for any $n$ but only for a dynamic index $m\in\mathcal{I}_d$, and similarly Eq. (\ref{second_inequality}) holds for any $m$ but only for a dynamic index $n\in\mathcal{I}_d$. Hence, Eqs. (\ref{first_inequality}) and (\ref{third_equality}) imply that $\chi^*_m \geq \chi^*_n$ if $n\in\mathcal{I}_s, m\in\mathcal{I}_d$ and $\chi^*_m = \chi^*_n$ if $n,m\in\mathcal{I}_d$, and Eq. (\ref{necessary-condition-total-move-constraint-eq}) is proved.
Now, by using Eq. (\ref{third_equality}) and the equality $E\left(\mathbf{P}^*\right)=\gamma$, we can write:
\begin{align}\label{solve_for_unknown}
    \sum_{i\in\mathcal{I}_d}\zeta_i \|\Gamma^*_i\| - \gamma &= \sum_{i\in\mathcal{I}_d}\zeta_i \|p^*_i - z^*_i\| =
    \sum_{i\in\mathcal{I}_d}\frac{\zeta_i^2\psi^*_n}{\zeta_n\psi^*_i} \|p^*_n - z^*_n\| =
    \frac{\psi^*_n}{\zeta_n}\|p^*_n - z^*_n\| \sum_{i\in\mathcal{I}_d}\frac{\zeta_i^2}{\psi^*_i},
\end{align}
or equivalently:
\begin{equation}\label{solved_unknown}
    \|p^*_n - z^*_n\| = \frac{\sum_{i\in\mathcal{I}_d}\zeta_i\|\Gamma^*_i\| - \gamma}{\frac{\psi^*_n}{\zeta_n}\sum_{i\in\mathcal{I}_d}\frac{\zeta_i^2}{\psi^*_i}}.
\end{equation}
Hence, we have:
\begin{equation}\label{final_formula}
    p^*_n = \tilde{p}_n + \frac{\Gamma^*_n}{\|\Gamma^*_n\|}\left(\|\Gamma^*_n\| - \|p^*_n - z^*_n\| \right) = \tilde{p}_n + \Gamma^*_n\left(1 - \frac{\sum_{i\in\mathcal{I}_d}\zeta_i\|\Gamma^*_i\| - \gamma}{\|\Gamma^*_n\|\times \frac{\psi^*_n}{\zeta_n}\times\sum_{i\in\mathcal{I}_d}\frac{\zeta_i^2}{\psi^*_i}} \right),
\end{equation}
and the proof is complete. $\hfill\blacksquare$

\section{}\label{proof-MERL-convergence}

Proof of Proposition \ref{MERL-convergence}:
We show that none of the steps in MERL Algorithm increases the multi-hop communication power $\mathcal{D}\left(\mathbf{P},\mathbf{W},\mathbf{S}\right)$. Since the movement energy constraint in Eq. (\ref{movement-constraint}) does not depend on the cell partitioning and normalized flow matrix, same reasoning as in Appendix \ref{proof-RL-convergence} shows that updating $\mathbf{W}$ and $\mathbf{S}$ according to the generalized Voronoi diagram and Bellman-Ford Algorithm, respectively, does not increase $\mathcal{D}\left(\mathbf{P},\mathbf{W},\mathbf{S}\right)$. In what follows, we show that updating the node deployment according to steps 4 and 5 in Algorithm \ref{MERL} will not increase the objective function as well. To show this, we first need the following concepts:

Let $\mathbf{P}^k = \left(p_1^k, \cdots ,p_N^k, p_{N+1}^k, \cdots, p_{N+M}^k \right)$ denote the node deployment after the $k$-th iteration. In particular, $\mathbf{P}^0 = \tilde{\mathbf{P}}$ is the initial deployment. We define the energy allocation after the $k$-th iteration as $\mathbf{E}^k = \left(e_1^k, \cdots, e_N^k, e_{N+1}^k, \cdots, e_{N+M}^k \right)$ where $e_n^k = \zeta_n \|p_n^k - \tilde{p}_n \|$ is node $n$'s movement energy consumption. Note that after the cell partitioning using the generalized Voronoi diagram, the partitions are fixed as $\mathcal{V}\left(\mathbf{P}^{k-1}, \mathbf{S}^{k-1} \right)$. Moreover, let $v_n^k$ and $c_n^k$ denote the volume and centroid of $\mathcal{V}_n\left(\mathbf{P}^{k}, \mathbf{S}^{k} \right)$, respectively, and define $\Gamma_n^k = z_n^k - \tilde{p}_n$ where $z_n^k$ is expressed as in Eqs. (\ref{z1-AP}) and (\ref{z1-FC}). We denote the energy consumed by moving node $n$ from its initial location to $z^k_n$ by $\tau_n^k = \zeta_n \|\Gamma_n^k \|$, and define $\kappa_n^k = \kappa_n\left(\mathbf{P}^k, \mathbf{S}^k \right) = \frac{\zeta_n^2}{\psi_n^k}$ where $\psi_n^k$ is given by Eq. (\ref{psi}). Finally, we define an auxiliary function $\hat{\chi}_n^k: \mathbb{R}^{N+M}\longrightarrow \mathbb{R}$ to be $\hat{\chi}_n^k\left(\mathbf{E} \right) = \frac{\tau_n^k - e_n}{\kappa_n^k}$. Note that $\hat{\chi}_n^k$ differs from $\chi_n$ defined in Eq. (\ref{chi}) in the sense that it depends on the energy allocation $\mathbf{E}$ rather than the node deployment and data routing.

\begin{lemma}\label{aux_lemma_1}
Let $\mathcal{I}_d^k$ and $\mathcal{I}_s^k$ denote the set of dynamic and static nodes after the $k$-th iteration of the MERL algorithm, respectively. Then, we have:
\begin{align}\label{aux_lemma_1_eq}
    \hat{\chi}_i^{k-1}\left(\mathbf{E}^k \right) &= \hat{\chi}_j^{k-1}\left(\mathbf{E}^k \right), \qquad\qquad \forall i,j\in \mathcal{I}_d^k \\
    \hat{\chi}_i^{k-1}\left(\mathbf{E}^k \right) &\geq \hat{\chi}_j^{k-1}\left(\mathbf{E}^k \right), \qquad\qquad \forall i\in \mathcal{I}_d^k, j\in \mathcal{I}_s^k \label{aux_lemma_1_eq_1}
\end{align}

Proof: At the end of the deployment step, dynamic node $n$'s location in the $k$-th iteration is:
\begin{equation}\label{app_G_eq1}
   p_n^k = \tilde{p}_n + \Gamma^{k-1}_n\left(1 - \frac{\sum_{i\in\mathcal{I}^k_d}\zeta_i\|\Gamma^{k-1}_i\| - \gamma}{\|\Gamma^{k-1}_n\|\times \frac{\psi^{k-1}_n}{\zeta_n}\times\sum_{i\in\mathcal{I}^k_d}\frac{\zeta_i^2}{\psi^{k-1}_i}} \right),
\end{equation}
thus, its movement energy consumption is:
\begin{align}\label{app_G_eq2}
    e_n^k &= \zeta_n \|p_n^k - \tilde{p}_n \| = \zeta_n \|\Gamma_n^{k-1}\| \times \left |\!\left|1 - \frac{\sum_{i\in\mathcal{I}^k_d}\zeta_i\|\Gamma^{k-1}_i\| - \gamma}{\|\Gamma^{k-1}_n\|\times \frac{\psi^{k-1}_n}{\zeta_n}\times\sum_{i\in\mathcal{I}^k_d}\frac{\zeta_i^2}{\psi^{k-1}_i}} \right |\! \right| \\
    &= \left |\! \left | \tau_n^{k-1} - \frac{\kappa_n^{k-1}\left(\sum_{i\in\mathcal{I}_d^k}\tau_i^{k-1} - \gamma \right)}{\sum_{i\in\mathcal{I}_d^k}\kappa_i^{k-1}}  \right |\! \right |, \qquad\qquad\qquad\qquad \forall n \in \mathcal{I}_d^k \label{app_G_eq2_1}
\end{align}
where $\mathcal{I}_d^k$ is the set of dynamic nodes in the $k$-th iteration, determined by the inner loop in steps 3 and 4 of the MERL algorithm. According to this inner loop, the term inside the vertical bars in Eq. (\ref{app_G_eq2_1}) is positive; hence, we have:
\begin{equation}\label{app_G_eq3}
    e_n^k =  \tau_n^{k-1} - \frac{\kappa_n^{k-1}\left(\sum_{i\in\mathcal{I}_d^k}\tau_i^{k-1} - \gamma \right)}{\sum_{i\in\mathcal{I}_d^k}\kappa_i^{k-1}} , \qquad\qquad \forall n \in \mathcal{I}_d^k.
\end{equation}
Now, by substituting Eq. (\ref{app_G_eq3}) into the definition of $\hat{\chi}_n^k$, we have:
\begin{equation}\label{app_G_eq4}
    \hat{\chi}_n^{k-1}\left(\mathbf{E}^k\right) = \frac{\tau_n^{k-1} - e_n^k}{\kappa_n^{k-1}} = \frac{\left[\sum_{i\in\mathcal{I}_d^k}\tau_i^{k-1} \right] - \gamma}{\sum_{i\in\mathcal{I}_d^k}\kappa_i^{k-1}}, \qquad\qquad \forall n\in\mathcal{I}_d^k.
\end{equation}
Therefore, all $\hat{\chi}_n^{k-1}\left(\mathbf{E}^k\right)$ for dynamic nodes are the same and Eq. (\ref{aux_lemma_1_eq}) is proved.

In order to prove Eq. (\ref{aux_lemma_1_eq_1}), we assume that $L_k$ inner iterations are performed in steps 3 and 4 of the MERL algorithm to determine the dynamic node set in the $k$-th iteration of the algorithm. For $l\in\{1, \cdots, L_k\}$, let $\mathcal{J}_l^k$ be the dynamic node set after the $l$-th inner iteration, where $k$ is the iteration index of the MERL algorithm. In particular, we have $\mathcal{J}_0^k = \mathcal{I_A}\bigcup\mathcal{I_F}$ and:
\begin{equation}\label{app_G_eq5}
    \mathcal{I}_d^k = \mathcal{J}_{L_k}^k \subsetneq \mathcal{J}_{L_k-1}^k \subsetneq \cdots \subsetneq \mathcal{J}_0^k
\end{equation}
In other words, in the $l$-th inner iteration, nodes within the set $\mathcal{J}_{l-1}^k - \mathcal{J}_{l}^k$ are removed from $\mathcal{J}_{l-1}^k$ due to their non-positive energy allocation, i.e., we have:
\begin{equation}\label{app_G_eq6}
e_j^k =  \tau_j^{k-1} - \frac{\kappa_j^{k-1}\left(\sum_{i\in\mathcal{J}_{l-1}^k}\tau_i^{k-1} - \gamma \right)}{\sum_{i\in\mathcal{J}_{l-1}^k}\kappa_i^{k-1}} \leq 0 , \qquad\qquad \forall j \in \mathcal{J}_{l-1}^k - \mathcal{J}_{l}^k
\end{equation}
hence, by rearranging the terms in Eq. (\ref{app_G_eq6}), and summation over all $j \in \mathcal{J}_{l-1}^k - \mathcal{J}_{l}^k$, we have:
\begin{equation}\label{app_G_eq7}
\left(\sum_{j\in \mathcal{J}_{l-1}^k - \mathcal{J}_{l}^k} \tau_j^{k-1}\right) \left(\sum_{i\in\mathcal{J}_{l-1}^k}\kappa_i^{k-1} \right)   \leq \left(\sum_{j\in \mathcal{J}_{l-1}^k - \mathcal{J}_{l}^k} \kappa_j^{k-1} \right) \left(\sum_{i\in\mathcal{J}_{l-1}^k}\tau_i^{k-1} - \gamma \right).
\end{equation}
Let the auxiliary function $\tilde{\chi}^k\left(\mathcal{J}\right) = \frac{\left(\sum_{i\in\mathcal{J}}\tau_i^k\right) - \gamma}{\sum_{i\in\mathcal{J}}\kappa_i^k}$ be a mapping from the node set $\mathcal{J}$ to the real numbers. For an inner iteration index $l\in\{1,\cdots,L_k\}$, we have:
\begin{align}\label{app_G_eq8}
    &\tilde{\chi}^{k-1}\left(\mathcal{J}_l^k \right) - \tilde{\chi}^{k-1}\left(\mathcal{J}_{l-1}^k \right) \\&= \frac{\left(\sum_{i\in\mathcal{J}_l^k}\tau_i^{k-1}\right) - \gamma}{\sum_{i\in\mathcal{J}_l^k}\kappa_i^{k-1}} - \frac{\left(\sum_{i\in\mathcal{J}_{l-1}^k}\tau_i^{k-1}\right) - \gamma}{\sum_{i\in\mathcal{J}_{l-1}^k}\kappa_i^{k-1}}\\
    &= \frac{\left(\sum_{i\in\mathcal{J}_{l-1}^k}\kappa_i^{k-1}\right)\left[\left(\sum_{i\in\mathcal{J}_l^k}\tau_i^{k-1}\right) - \gamma\right] -\left(\sum_{i\in\mathcal{J}_l^k}\kappa_i^{k-1}\right)\left[\left(\sum_{i\in\mathcal{J}_{l-1}^k}\tau_i^{k-1}\right) - \gamma \right]   }{\left(\sum_{i\in\mathcal{J}_l^k}\kappa_i^{k-1}\right)\left
    (\sum_{i\in\mathcal{J}_{l-1}^k}\kappa_i^{k-1}\right)} \\&=
    \frac{\left(\sum_{i\in\mathcal{J}_{l-1}^k}\kappa_i^{k-1}\right) \left[\left(\sum_{i\in\mathcal{J}_{l-1}^k}\tau_i^{k-1}\right) - \left(\sum_{i\in\mathcal{J}_{l-1}^k-\mathcal{J}_l^k}\tau_i^{k-1}\right) - \gamma\right]}{\left(\sum_{i\in\mathcal{J}_l^k}\kappa_i^{k-1}\right)\left
    (\sum_{i\in\mathcal{J}_{l-1}^k}\kappa_i^{k-1}\right)} \\ &- \frac{\left[\left(\sum_{i\in\mathcal{J}_{l-1}^k}\kappa_i^{k-1}\right) - \left(\sum_{i\in\mathcal{J}_{l-1}^k-\mathcal{J}_l^k}\kappa_i^{k-1}\right) \right]\left[\left(\sum_{i\in\mathcal{J}_{l-1}^k}\tau_i^{k-1}\right) - \gamma \right]}{\left(\sum_{i\in\mathcal{J}_l^k}\kappa_i^{k-1}\right)\left
    (\sum_{i\in\mathcal{J}_{l-1}^k}\kappa_i^{k-1}\right)} \\ &=
    \frac{\left(\sum\limits_{i\in\mathcal{J}_{l-1}^k - \mathcal{J}_{l}^k}\kappa_i^{k-1}\right)\left[\left(\sum\limits_{i\in\mathcal{J}_{l-1}^k}\tau_i^{k-1}\right) - \gamma\right] - \left(\sum\limits_{i\in\mathcal{J}_{l-1}^k- \mathcal{J}_{l}^k}\tau_i^{k-1}\right)\left(\sum\limits_{i\in\mathcal{J}_{l-1}^k}\kappa_i^{k-1}\right) }{\left(\sum_{i\in\mathcal{J}_l^k}\kappa_i^{k-1}\right)\left
    (\sum_{i\in\mathcal{J}_{l-1}^k}\kappa_i^{k-1}\right)} \geq 0,
\end{align}
where the last inequality follows from Eq. (\ref{app_G_eq7}). Thus, we have the following ordered sequence:
\begin{equation}\label{app_G_eq9}
    \tilde{\chi}^{k-1}\left(\mathcal{J}_{0}^k \right) \leq \tilde{\chi}^{k-1}\left(\mathcal{J}_{1}^k \right) \leq \cdots \leq \tilde{\chi}^{k-1}\left(\mathcal{J}_{L_k}^k \right) = \tilde{\chi}^{k-1}\left(\mathcal{I}_{d}^k \right) = \hat{\chi}_n^{k-1}\left(\mathbf{E}^k \right),\qquad \forall n\in \mathcal{I}_d^k.
\end{equation}
Let the tentative energy allocation in the $l$-th inner iteration be $\tilde{\mathbf{E}}^k(l) = \left(\tilde{e}_1^k(l), \cdots, \tilde{e}_{N+M}^k(l) \right)$. The tentative movement energy consumption of node $n$ in the $l$-th inner iteration is given by:
\begin{align}\label{app_G_eq10}
    \tilde{e}_n^k(l) = \tau_n^{k-1} - \frac{\kappa_n^{k-1}\left[\left(\sum_{i\in\mathcal{J}_l^k}\tau_i^{k-1}\right) - \gamma\right]}{\sum_{i\in\mathcal{J}_l^k}\kappa_i^{k-1}}, \qquad\qquad \forall n \in \mathcal{J}_l^k
\end{align}
hence, we can rewrite $\tilde{\chi}^{k-1}\left(\mathcal{J}_l^k \right)$ as:
\begin{equation}\label{app_G_eq11}
    \tilde{\chi}^{k-1}\left(\mathcal{J}_l^k \right) = \frac{\left[\left(\sum_{i\in\mathcal{J}_l^k}\tau_i^{k-1}\right) - \gamma\right]}{\sum_{i\in\mathcal{J}_l^k}\kappa_i^{k-1}} = \frac{\tau_n^{k-1} - \tilde{e}_n^k(l)}{\kappa_n^{k-1}}, \qquad\qquad \forall n \in \mathcal{J}_l^k.
\end{equation}
Note that each node $j\in \mathcal{J}_{l-1}^k - \mathcal{J}_l^k$ is removed from the dynamic node set in the $l$-th inner iteration of the MERL algorithm due to its non-positive tentative energy $\tilde{e}_j^k(l) \leq 0$; therefore, we have $j\in\mathcal{I}_s^k$ and its allocated movement energy consumption is $e_j^k = 0$. Then, we have:
\begin{equation}\label{app_G_eq12}
    \hat{\chi}_j^{k-1}\left(\mathbf{E}^k\right) = \frac{\tau_j^{k-1} - e_j^k}{\kappa_j^{k-1}} = \frac{\tau_j^{k-1}}{\kappa_j^{k-1}} \leq \frac{\tau_j^{k-1} - \tilde{e}_j^k(l)}{\kappa_j^{k-1}} = \tilde{\chi}^{k-1}\left(\mathcal{J}_l^k\right), \qquad \forall j \in \mathcal{J}_{l-1}^k - \mathcal{J}_l^k.
\end{equation}
Using Eqs. (\ref{app_G_eq9}) and (\ref{app_G_eq12}), we have:
\begin{equation}\label{app_G_eq13}
    \hat{\chi}_j^{k-1}\left(\mathbf{E}^k\right) \leq \hat{\chi}_i^{k-1}\left(\mathbf{E}^k\right), \qquad\qquad \forall i\in\mathcal{I}_d^k, j\in \mathcal{J}_{l-1}^k - \mathcal{J}_l^k, l\in\{1, \cdots, L_k\}.
\end{equation}
Note that the static node set $\mathcal{I}_s^k$ consists of all nodes that are removed in the inner loop, i.e. $\mathcal{I}_s = \bigcup_{l\in\{1,\cdots,L_k\}}\left(\mathcal{J}_{l-1}^k - \mathcal{J}_l^k \right)$; hence, Eq. (\ref{aux_lemma_1_eq_1}) follows from Eq. (\ref{app_G_eq13}) and the proof is finished.
\end{lemma}

\begin{lemma}\label{aux_lemma_2}
For a fixed cell partitioning 
and normalized flow matrix, 
the node deployment $\mathbf{P}^k$ given by the $k$-th iteration of MERL Algorithm is the unique minimizer to the objective function in Eqs. (\ref{total-movement-constraint-objective}) and (\ref{movement-constraint}).

Proof: Using parallel axis theorem \cite{paul1979kinematics}, the objective function in the $k$-th iteration is:
\begin{align}\label{app_g_eq1}
    \mathcal{D} &= \sum_{i=1}^{N}\int_{\mathcal{V}_i^{k-1}}\eta_i \|c_i^{k-1} - \omega\|^2 R_b f(\omega)d\omega + \sum_{i}^{N}\eta_i R_b v_i^{k-1} \|p_i - c_i^{k-1} \|^2  \nonumber \\& + 
    \lambda \sum_{i=1}^{N}\sum_{j=1}^{N+M} \beta_{i,j} \|p_i - p_j\|^2 F_{i,j}^{k-1} + 
    \lambda \overline{\mathcal{P}}^R_{\mathcal{A}}\left(\mathbf{W}^{k-1}, \mathbf{S}^{k-1} \right).
\end{align}
For a fixed partitioning and routing, a similar reasoning as in the proof of Lemma \ref{total-movment-constraint-between-initial-and-optimal} shows that node $n$'s optimal location at the end of $k$-th iteration should be placed on the segment connecting its initial location $\tilde{p}_n$ to the point $z_n^{k-1}$ given in Eqs. (\ref{z1-AP}) and (\ref{z1-FC}), 
 i.e., if we denote the node $n$'s movement energy by $e_n$, we have:
\begin{align}\label{app_g_eq2}
    p_n(e_n) = \tilde{p}_n + \frac{e_n}{\zeta_n} \times \frac{\Gamma_n^{k-1}}{\|\Gamma_n^{k-1}\|}, \qquad\qquad \forall n\in\mathcal{I_A}\bigcup\mathcal{I_F}.
\end{align}
By substituting the Eq. (\ref{app_g_eq2}) into Eq. (\ref{app_g_eq1}), we can rewrite the objective function as:
\begin{align}\label{app_g_eq3}
    &\qquad\qquad\qquad\qquad\qquad \minimize_{\mathbf{E}}\mathcal{D}(\mathbf{E}) \nonumber  \\ &\qquad \textrm{s.t.}\qquad \left(\sum_{n=1}^{N+M}e_n \right) \leq \gamma, \qquad 0\leq e_n \leq \zeta_n\|\Gamma_n^{k-1}\|,\quad \forall n\in \mathcal{I_A}\bigcup\mathcal{I_F}.
\end{align}
where:
\begin{align}\label{app_g_eq4}
    &\mathcal{D}(\mathbf{E}) = \sum_{i=1}^{N}\int_{\mathcal{V}_i^{k-1}}\eta_i \|c_i^{k-1} - \omega\|^2 R_b f(\omega)d\omega + \sum_{i}^{N}\eta_i R_b v_i^{k-1} \bigg|\! \bigg|\tilde{p}_i + \frac{e_i}{\zeta_i} \times \frac{\Gamma_i^{k-1}}{\|\Gamma_i^{k-1}\|} - c_i^{k-1} \bigg |\! \bigg |^2  \nonumber \\& + 
    \lambda \sum_{i=1}^{N}\sum_{j=1}^{N+M} \beta_{i,j} \bigg|\! \bigg|\tilde{p}_i + \frac{e_i}{\zeta_i} \times \frac{\Gamma_i^{k-1}}{\|\Gamma_i^{k-1}\|} - \tilde{p}_j - \frac{e_j}{\zeta_j} \times \frac{\Gamma_j^{k-1}}{\|\Gamma_j^{k-1}\|}\bigg|\!\bigg|^2 F_{i,j}^{k-1} + 
    \lambda \overline{\mathcal{P}}^R_{\mathcal{A}}\left(\mathbf{W}^{k-1}, \mathbf{S}^{k-1} \right),
\end{align}
Note that the objective function in Eq. (\ref{app_g_eq3}) and its constraints are convex; hence, it has a unique minimizer for a fixed partitioning and routing. If $\left(\sum_{n=1}^{N+M}\zeta_n\| \Gamma_n^{k-1}\|  \right) \leq \gamma$, then the MERL algorithm moves each node $n$ to $z_n^{k-1}$ without violating the total energy constraint, indicating an optimal deployment. On the other hand, if $\left(\sum_{n=1}^{N+M}\zeta_n\| \Gamma_n^{k-1}\|  \right) > \gamma$, then nodes will run out of movement energy before they can reach to their corresponding $z_n^{k-1}$, and the same reasoning as in Appendix \ref{proof-necessary-condition-total-constraint} shows that $\left(\sum_{n=1}^{N+M}e_n \right) = \gamma$. For the fixed partitioning and routing, let 
$\mathbf{E}^* = \left(e_1^*, \cdots, e_{N+M}^* \right)$ be the optimal 
energy allocation for the constrained objective function in Eq. (\ref{app_g_eq3}), and let $\mathbf{P}^* = \left(p_1^*, \cdots, p_{N+M}^* \right)$ be the corresponding optimal deployment. Assume that the movement energy allocation $\mathbf{E}^k$ in the $k$-th iteration is different from the optimal one, i.e., $\mathbf{E}^* \neq \mathbf{E}^k$. Since $\left(\sum_{n=1}^{N+M}e^*_n \right) = \left(\sum_{n=1}^{N+M}e^k_n \right) = \gamma$, there exist two distinct indices $i$ and $j$ such that $ 0 \leq e_i^k < e^*_i$ and $0 \leq e_j^* < e_j^k$. Note that $e_j^k > 0$ indicates that $j\in\mathcal{I}_d^k$, i.e., node $j$ is a dynamic node in the $k$-th iteration. Therefore, using Lemma \ref{aux_lemma_1} we have:
\begin{align}\label{app_g_eq5}
    \frac{\zeta_i \|\Gamma_i^{k-1}\| - e_i^*}{\frac{\zeta_i^2}{\psi_i^{k-1}}} < \frac{\zeta_i \|\Gamma_i^{k-1}\| - e_i^k}{\frac{\zeta_i^2}{\psi_i^{k-1}}} \leq \frac{\zeta_j \|\Gamma_j^{k-1}\| - e_j^k}{\frac{\zeta_j^2}{\psi_j^{k-1}}} < \frac{\zeta_j \|\Gamma_j^{k-1}\| - e_j^*}{\frac{\zeta_j^2}{\psi_j^{k-1}}}.
\end{align}
Now, we consider a new energy allocation $\overline{\mathbf{E}} = \left(\overline{e}_1, \cdots, \overline{e}_{N+M} \right)$, where $\overline{e}_i = e_i^* - \epsilon$, $\overline{e}_j = e_j^* + \epsilon$ and $\overline{e}_t = e_t^*$ for all $t\in\mathcal{I_A}\bigcup\mathcal{I_F}\backslash \{i, j\}$. Note that $\left(\sum_{n=1}^{N+M}\overline{e}_n\right) = \gamma$, and for a sufficiently small positive value of $\epsilon$, we have $0 \leq e_i^* - \epsilon = \overline{e}_i < e_i^*\leq \zeta_i\|\Gamma_i^{k-1}\|$ and $0\leq e_j^* < \overline{e}_j = e_j^* + \epsilon \leq e_j^k \leq \zeta_j \|\Gamma_j^{k-1}\|$, i.e., $\overline{\mathbf{E}}$ satisfies the constraints in Eq. (\ref{app_g_eq3}) and it is a valid energy allocation. Similar argument as in Appendix \ref{proof-necessary-condition-total-constraint}, that led to the Eq. (\ref{first_inequality}), shows that in order for the energy allocation $\overline{\mathbf{E}}$ not to achieve a lower objective function value in Eq. (\ref{app_g_eq3}) than $\mathcal{D}\left(\mathbf{E}^*\right)$, which contradicts the optimality of the movement energy allocation $\mathbf{E}^*$, we should have:
\begin{equation}\label{app_g_eq6}
    \zeta_i \psi_j^{k-1}\|p_j^* - z_j^{k-1} \| \leq \zeta_j \psi_i^{k-1} \|p_i^* - z_i^{k-1} \|,
\end{equation}
or equivalently:
\begin{equation}\label{app_g_eq7}
    \frac{\zeta_j \|p_j^* - z_j^{k-1} \|}{\frac{\zeta_j^2}{\psi_j^{k-1}}} \leq \frac{\zeta_i \|p_i^* - z_i^{k-1} \|}{\frac{\zeta_i^2}{\psi_i^{k-1}}}.
\end{equation}
According to Eq. (\ref{app_g_eq2}), each node $n\in\mathcal{I_A}\bigcup\mathcal{I_F}$ is located on the segment connecting $\tilde{p}_n$ to $z_n^{k-1}$; hence: we can rewrite the Eq. (\ref{app_g_eq7}) as:
\begin{equation}\label{app_g_eq8}
    \frac{\zeta_j \|\Gamma_j^{k-1} \| - e_j^*}{\frac{\zeta_j^2}{\psi_j^{k-1}}} \leq \frac{\zeta_i \|\Gamma_i^{k-1} \| - e_i^*}{\frac{\zeta_i^2}{\psi_i^{k-1}}}.
\end{equation}
But Eq. (\ref{app_g_eq8}) is in contradiction with Eq. (\ref{app_g_eq5}); thus, the assumption $\mathbf{E}^* \neq \mathbf{E}^k$ is wrong and we have $\mathbf{E}^* = \mathbf{E}^k$, i.e. the deployment given by the MERL algorithm is the unique minimizer of the constrained objective function and the proof is complete.
\end{lemma}
Now, we have enough materials to prove the convergence of the MERL algorithm. As mentioned in the beginning of the Appendix \ref{proof-MERL-convergence}, updating the partitioning and normalized flow matrix using the generalized Voronoi diagram and Bellman-Ford Algorithm, respectively, does not increase the objective function. Now, for a fixed partitioning and routing, Lemma \ref{aux_lemma_2} indicates that the deployment given by the MERL algorithm is the unique minimizer of the constrained objective function, i.e., the deployment step in the MERL algorithm does not increase the objective function either. Hence, the MERL algorithm generates a sequence of positive non-increasing values for the objective function $\mathcal{D}$; thus, it converges. $\hfill\blacksquare$


\section{}\label{proof-necessary-condition-network-lifetime}
Proof of Proposition \ref{necessary-condition-network-lifetime}:
If $p^*_n = z^*_n$ is an optimal deployment $\mathbf{P}^*$, $\mathbf{W}^*$ and $\mathbf{S}^*$, then Eq. (\ref{lifetime-constraint}) implies that $E_n\left(\mathbf{P}^*\right) = \zeta_n \|\Gamma^*_n\| \leq \gamma_n$. Therefore, Eq. (\ref{necessary-condition-network-lifetime-eq}) reduces to the trivial statement $p^*_n = \tilde{p}_n + \Gamma^*_n$ and the proof is complete. Hence, we assume that $p^*_n \neq z^*_n$. Now, if any residual movement energy is left in Node $n$, i.e. if $E_n\left(\mathbf{P}^*\right) < \gamma_n$, then there exists an $\epsilon_n\in\mathbb{R}^+$ such that $E_n\left(\mathbf{P}^*\right) + \epsilon_n < \gamma_n$ and the point $\overline{p}_n = p^*_n + \epsilon_n \times \frac{z^*_n - p^*_n}{\|z^*_n - p^*_n\|}$ lies inside the circle centered at $z^*_n$ with radius $\|z^*_n - p^*_n\|$. Then, according to Lemma \ref{geometric-locus-weighted-sum-square-distances}, by fixing the cell partitioning, normalized flow matrix and the location of all nodes except Node $n$, and placing Node $n$ at $\overline{p}_n$, we can achieve a lower total multi-hop communication power without exhausting the available movement energy in Node $n$, which contradicts the optimality of $\mathbf{P}^*$, $\mathbf{W}^*$ and $\mathbf{S}^*$. Therefore, $p^*_n\neq z^*_n$ implies that $E_n\left(\mathbf{P}^*\right) = \gamma_n$, that is 
\begin{equation}\label{appendix-H-eq1}
    \zeta_n\|p^*_n - \tilde{p}_n \| = \gamma_n.
\end{equation}
According to Lemma \ref{total-movment-constraint-between-initial-and-optimal}, we have 
\begin{equation}\label{appendix-H-eq2}
    p^*_n = \delta_n \tilde{p}_n + \left(1-\delta_n\right)z^*_n,
\end{equation}
where $\delta_n\in\left[0, 1\right]$, which indicates that
\begin{equation}\label{appendix-H-eq3}
    \|p^*_n - \tilde{p}_n\| = \left(1-\delta_n \right) \|z^*_n - \tilde{p}_n\|.
\end{equation}
Eqs. (\ref{appendix-H-eq1}) and (\ref{appendix-H-eq3}) imply that $\delta_n = 1 - \frac{\gamma_n}{\zeta_n \|z^*_n - \tilde{p}_n\|}$. Therefore, Eq. (\ref{appendix-H-eq2}) can be written as:
\begin{align}\label{appendix-H-eq4}
    p^*_n &= \left(1 - \frac{\gamma_n}{\zeta_n \|z^*_n - \tilde{p}_n\|}\right)\tilde{p}_n + \left(\frac{\gamma_n}{\zeta_n \|z^*_n - \tilde{p}_n\|}\right) z^*_n \\
    &= \tilde{p_n} + \left(\frac{\gamma_n}{\zeta_n \|z^*_n - \tilde{p}_n\|}\right) \left(z^*_n - \tilde{p}_n \right) \\
    &= \tilde{p}_n + \frac{\gamma_n}{\zeta_n\|\Gamma^*_n\|}\Gamma^*_n. \label{appendix-H-eq5}
\end{align}
Eqs. (\ref{appendix-H-eq1}) and (\ref{appendix-H-eq2}) imply that $\gamma_n = \zeta_n \|p^*_n-\tilde{p}_n\| \leq \zeta_n \|z^*_n-\tilde{p}_n\| = \zeta_n\|\Gamma^*_n\|$, i.e. $\frac{\gamma_n}{\zeta_n\|\Gamma^*_n\|} \leq 1$. Thus, Eq. (\ref{appendix-H-eq5}) can be rewritten as $p^*_n = \tilde{p}_n + \min\left(1, \frac{\gamma_n}{\zeta_n\|\Gamma^*_n\|}\right) \Gamma^*_n$ which concludes the proof. $\hfill\blacksquare$

\section{}\label{proof-LORL-convergence}
Proof of Proposition \ref{LORL-convergence}: In what follows, we show that none of the three steps in LORL Algorithm will increase the communication power $\mathcal{D}\left(\mathbf{P}, \mathbf{W}, \mathbf{S}\right)$. Note that the movement energy constraint in Eq. (\ref{lifetime-constraint}) does not depend on the cell partitioning and normalized flow matrix. Hence, it can be shown via the same argument as in Appendix \ref{proof-RL-convergence} that updating $\mathbf{W}$ and $\mathbf{S}$ according to the generalized Voronoi diagrams and Bellman-Ford Algorithm, respectively, does not increase $\mathcal{D}\left(\mathbf{P}, \mathbf{W}, \mathbf{S}\right)$. 
Note that for a fixed $\mathbf{W}$, $\mathbf{S}$ and $\left\{p_i \right\}_{i\neq n}$, according to Lemma \ref{geometric-locus-weighted-sum-square-distances}, the geometric locus of node $n$ for which the objective function $\mathcal{D}\left(\mathbf{P}, \mathbf{W}, \mathbf{S}\right)$ remains the same is a circle $\Phi_n$ centered at $z_n$ with radius $\|z_n - p_n\|$. Note that the update rule in Eq. (\ref{necessary-condition-network-lifetime-eq}) always keeps node $n$ in its \emph{valid region} determined by its limited movement energy, which is a circle centered at $\tilde{p}_n$ and radius $\frac{\gamma_n}{\zeta_n}$. A simple geometric reasoning indicates that by updating the position of node $n$ according to Eq. (\ref{necessary-condition-network-lifetime-eq}), node $n$ will either remain the same or move to the point inside its valid region that is closest to the point $z_n$, i.e., node $n$ will either remain on the circle $\Phi_n$ or move inside it, and the objective function $\mathcal{D}\left(\mathbf{P}, \mathbf{W}, \mathbf{S}\right)$ does not increase. Since the objective function has a lower bounded, i.e. $\mathcal{D}\left(\mathbf{P}, \mathbf{W}, \mathbf{S}\right) \geq 0$, and it is nonincreasing, LORL Algorithm is in iterative improvement algorithm and it converges. $\hfill\blacksquare$




\ifCLASSOPTIONcaptionsoff
  \newpage
\fi



%

\bibliographystyle{ieeetr}
\bibliography{main}




%








\end{document}